\documentclass[twoside]{IEEEtran}
\usepackage[hidelinks]{hyperref}
\usepackage[T1]{fontenc}
\usepackage{graphicx}

\usepackage{amssymb}
\usepackage{amsmath,amscd}
\usepackage{amsthm}
\usepackage{dsfont} 
\usepackage{paralist}
\usepackage[multiple]{footmisc} 
\usepackage{subfigure}
\usepackage{stackrel}
\usepackage{xr} 		
\usepackage{tikz}
\usepackage{pgfplots}

\usepackage{functan}
\usepackage{url}
\usepackage{float}
\usepackage{arydshln}
\usepackage{breakurl}

\floatstyle{ruled}
\newfloat{model}{thp}{lom}
\floatname{model}{Model}

\usepackage{amssymb}
\usepackage{amsfonts}
\usepackage{mathrsfs}
\usepackage{xspace}
\usepackage{bm}
\usepackage{upgreek}

\newcommand{\safemath}[2]{\newcommand{#1}{\ensuremath{#2}\xspace}}

\safemath{\bma}{\mathbf{a}}
\safemath{\bmb}{\mathbf{b}}
\safemath{\bmc}{\mathbf{c}}
\safemath{\bmd}{\mathbf{d}}
\safemath{\bme}{\mathbf{e}}
\safemath{\bmf}{\mathbf{f}}
\safemath{\bmg}{\mathbf{g}}
\safemath{\bmh}{\mathbf{h}}
\safemath{\bmi}{\mathbf{i}}
\safemath{\bmj}{\mathbf{j}}
\safemath{\bmk}{\mathbf{k}}
\safemath{\bml}{\mathbf{l}}
\safemath{\bmm}{\mathbf{m}}
\safemath{\bmn}{\mathbf{n}}
\safemath{\bmo}{\mathbf{o}}
\safemath{\bmp}{\mathbf{p}}
\safemath{\bmq}{\mathbf{q}}
\safemath{\bmr}{\mathbf{r}}
\safemath{\bms}{\mathbf{s}}
\safemath{\bmt}{\mathbf{t}}
\safemath{\bmu}{\mathbf{u}}
\safemath{\bmv}{\mathbf{v}}
\safemath{\bmw}{\mathbf{w}}
\safemath{\bmx}{\mathbf{x}}
\safemath{\bmy}{\mathbf{y}}
\safemath{\bmz}{\mathbf{z}}
\safemath{\bmzero}{\mathbf{0}}
\safemath{\bmone}{\mathbf{1}}

\bmdefine{\biad}{a}
\bmdefine{\bibd}{b}
\bmdefine{\bicd}{c}
\bmdefine{\bidd}{d}
\bmdefine{\bied}{e}
\bmdefine{\bifd}{f}
\bmdefine{\bigd}{g}
\bmdefine{\bihd}{h}
\bmdefine{\biid}{i}
\bmdefine{\bijd}{j}
\bmdefine{\bikd}{k}
\bmdefine{\bild}{l}
\bmdefine{\bimd}{m}
\bmdefine{\bind}{n}
\bmdefine{\biod}{o}
\bmdefine{\bipd}{p}
\bmdefine{\biqd}{q}
\bmdefine{\bird}{r}
\bmdefine{\bisd}{s}
\bmdefine{\bitd}{t}
\bmdefine{\biud}{u}
\bmdefine{\bivd}{v}
\bmdefine{\biwd}{w}
\bmdefine{\bixd}{x}
\bmdefine{\biyd}{y}
\bmdefine{\bizd}{z}

\bmdefine{\bixid}{\xi}
\bmdefine{\bilambdad}{\lambda}
\bmdefine{\bimud}{\mu}
\bmdefine{\bithetad}{\theta}
\bmdefine{\biphid}{\phi}
\bmdefine{\bideltad}{\delta}

\safemath{\bmia}{\biad}
\safemath{\bmib}{\bibd}
\safemath{\bmic}{\bicd}
\safemath{\bmid}{\bidd}
\safemath{\bmie}{\bied}
\safemath{\bmif}{\bifd}
\safemath{\bmig}{\bigd}
\safemath{\bmih}{\bihd}
\safemath{\bmii}{\biid}
\safemath{\bmij}{\bijd}
\safemath{\bmik}{\bikd}
\safemath{\bmil}{\bild}
\safemath{\bmim}{\bimd}
\safemath{\bmin}{\bind}
\safemath{\bmio}{\biod}
\safemath{\bmip}{\bipd}
\safemath{\bmiq}{\biqd}
\safemath{\bmir}{\bird}
\safemath{\bmis}{\bisd}
\safemath{\bmit}{\bitd}
\safemath{\bmiu}{\biud}
\safemath{\bmiv}{\bivd}
\safemath{\bmiw}{\biwd}
\safemath{\bmix}{\bixd}
\safemath{\bmiy}{\biyd}
\safemath{\bmiz}{\bizd}

\safemath{\bmxi}{\bixid}
\safemath{\bmlambda}{\bilambdad}
\safemath{\bmmu}{\bimud}
\safemath{\bmtheta}{\bithetad}
\safemath{\bmphi}{\biphid}
\safemath{\bmdelta}{\bideltad}

\safemath{\bA}{\mathbf{A}}
\safemath{\bB}{\mathbf{B}}
\safemath{\bC}{\mathbf{C}}
\safemath{\bD}{\mathbf{D}}
\safemath{\bE}{\mathbf{E}}
\safemath{\bF}{\mathbf{F}}
\safemath{\bG}{\mathbf{G}}
\safemath{\bH}{\mathbf{H}}
\safemath{\bI}{\mathbf{I}}
\safemath{\bJ}{\mathbf{J}}
\safemath{\bK}{\mathbf{K}}
\safemath{\bL}{\mathbf{L}}
\safemath{\bM}{\mathbf{M}}
\safemath{\bN}{\mathbf{N}}
\safemath{\bO}{\mathbf{O}}
\safemath{\bP}{\mathbf{P}}
\safemath{\bQ}{\mathbf{Q}}
\safemath{\bR}{\mathbf{R}}
\safemath{\bS}{\mathbf{S}}
\safemath{\bT}{\mathbf{T}}
\safemath{\bU}{\mathbf{U}}
\safemath{\bV}{\mathbf{V}}
\safemath{\bW}{\mathbf{W}}
\safemath{\bX}{\mathbf{X}}
\safemath{\bY}{\mathbf{Y}}
\safemath{\bZ}{\mathbf{Z}}

\safemath{\bZero}{\mathbf{0}}
\safemath{\bOne}{\mathbf{1}}
\safemath{\bDelta}{\mathbf{\Delta}}
\safemath{\bLambda}{\mathbf{\UpLambda}}
\safemath{\bPhi}{\mathbf{\Upphi}}
\safemath{\bSigma}{\mathbf{\Upsigma}}
\safemath{\bOmega}{\mathbf{\Upomega}}
\safemath{\bTheta}{\mathbf{\Uptheta}}

\bmdefine{\biAd}{A}
\bmdefine{\biBd}{B}
\bmdefine{\biCd}{C}
\bmdefine{\biDd}{D}
\bmdefine{\biEd}{E}
\bmdefine{\biFd}{F}
\bmdefine{\biGd}{G}
\bmdefine{\biHd}{H}
\bmdefine{\biId}{I}
\bmdefine{\biJd}{J}
\bmdefine{\biKd}{K}
\bmdefine{\biLd}{L}
\bmdefine{\biMd}{M}
\bmdefine{\biOd}{N}
\bmdefine{\biPd}{O}
\bmdefine{\biQd}{P}
\bmdefine{\biRd}{R}
\bmdefine{\biSd}{S}
\bmdefine{\biTd}{T}
\bmdefine{\biUd}{U}
\bmdefine{\biVd}{V}
\bmdefine{\biWd}{W}
\bmdefine{\biXd}{X}
\bmdefine{\biYd}{Y}
\bmdefine{\biZd}{Z}

\bmdefine{\biDelta}{\Delta}
\bmdefine{\biLambda}{\Lambda}
\bmdefine{\biPhi}{\Phi}
\bmdefine{\biSigma}{\Sigma}
\bmdefine{\biOmega}{\Omega}
\bmdefine{\biTheta}{\Theta}

\safemath{\bimA}{\biAd}
\safemath{\bimB}{\biBd}
\safemath{\bimC}{\biCd}
\safemath{\bimD}{\biDd}
\safemath{\bimE}{\biEd}
\safemath{\bimF}{\biFd}
\safemath{\bimG}{\biGd}
\safemath{\bimH}{\biHd}
\safemath{\bimI}{\biId}
\safemath{\bimJ}{\biJd}
\safemath{\bimK}{\biKd}
\safemath{\bimL}{\biLd}
\safemath{\bimM}{\biMd}
\safemath{\bimN}{\biNd}
\safemath{\bimO}{\biOd}
\safemath{\bimP}{\biPd}
\safemath{\bimQ}{\biQd}
\safemath{\bimR}{\biRd}
\safemath{\bimS}{\biSd}
\safemath{\bimT}{\biTd}
\safemath{\bimU}{\biUd}
\safemath{\bimV}{\biVd}
\safemath{\bimW}{\biWd}
\safemath{\bimX}{\biXd}
\safemath{\bimY}{\biYd}
\safemath{\bimZ}{\biZd}

\safemath{\bimDelta}{\biDelta}
\safemath{\bimLambda}{\biLambda}
\safemath{\bimPhi}{\biPhi}
\safemath{\bimSigma}{\biSigma}
\safemath{\bimOmega}{\biOmega}
\safemath{\bimTheta}{\biTheta}

\safemath{\setA}{\mathcal{A}}
\safemath{\setB}{\mathcal{B}}
\safemath{\setC}{\mathcal{C}}
\safemath{\setD}{\mathcal{D}}
\safemath{\setE}{\mathcal{E}}
\safemath{\setF}{\mathcal{F}}
\safemath{\setG}{\mathcal{G}}
\safemath{\setH}{\mathcal{H}}
\safemath{\setI}{\mathcal{I}}
\safemath{\setJ}{\mathcal{J}}
\safemath{\setK}{\mathcal{K}}
\safemath{\setL}{\mathcal{L}}
\safemath{\setM}{\mathcal{M}}
\safemath{\setN}{\mathcal{N}}
\safemath{\setO}{\mathcal{O}}
\safemath{\setP}{\mathcal{P}}
\safemath{\setQ}{\mathcal{Q}}
\safemath{\setR}{\mathcal{R}}
\safemath{\setS}{\mathcal{S}}
\safemath{\setT}{\mathcal{T}}
\safemath{\setU}{\mathcal{U}}
\safemath{\setV}{\mathcal{V}}
\safemath{\setW}{\mathcal{W}}
\safemath{\setX}{\mathcal{X}}
\safemath{\setY}{\mathcal{Y}}
\safemath{\setZ}{\mathcal{Z}}
\safemath{\emptySet}{\varnothing}

\safemath{\colA}{\mathscr{A}}
\safemath{\colB}{\mathscr{B}}
\safemath{\colC}{\mathscr{C}}
\safemath{\colD}{\mathscr{D}}
\safemath{\colE}{\mathscr{E}}
\safemath{\colF}{\mathscr{F}}
\safemath{\colG}{\mathscr{G}}
\safemath{\colH}{\mathscr{H}}
\safemath{\colI}{\mathscr{I}}
\safemath{\colJ}{\mathscr{J}}
\safemath{\colK}{\mathscr{K}}
\safemath{\colL}{\mathscr{L}}
\safemath{\colM}{\mathscr{M}}
\safemath{\colN}{\mathscr{N}}
\safemath{\colO}{\mathscr{O}}
\safemath{\colP}{\mathscr{P}}
\safemath{\colQ}{\mathscr{Q}}
\safemath{\colR}{\mathscr{R}}
\safemath{\colS}{\mathscr{S}}
\safemath{\colT}{\mathscr{T}}
\safemath{\colU}{\mathscr{U}}
\safemath{\colV}{\mathscr{V}}
\safemath{\colW}{\mathscr{W}}
\safemath{\colX}{\mathscr{X}}
\safemath{\colY}{\mathscr{Y}}
\safemath{\colZ}{\mathscr{Z}}

\safemath{\opA}{\mathbb{A}}
\safemath{\opB}{\mathbb{B}}
\safemath{\opC}{\mathbb{C}}
\safemath{\opD}{\mathbb{D}}
\safemath{\opE}{\mathbb{E}}
\safemath{\opF}{\mathbb{F}}
\safemath{\opG}{\mathbb{G}}
\safemath{\opH}{\mathbb{H}}
\safemath{\opI}{\mathbb{I}}
\safemath{\opJ}{\mathbb{J}}
\safemath{\opK}{\mathbb{K}}
\safemath{\opL}{\mathbb{L}}
\safemath{\opM}{\mathbb{M}}
\safemath{\opN}{\mathbb{N}}
\safemath{\opO}{\mathbb{O}}
\safemath{\opP}{\mathbb{P}}
\safemath{\opQ}{\mathbb{Q}}
\safemath{\opR}{\mathbb{R}}
\safemath{\opS}{\mathbb{S}}
\safemath{\opT}{\mathbb{T}}
\safemath{\opU}{\mathbb{U}}
\safemath{\opV}{\mathbb{V}}
\safemath{\opW}{\mathbb{W}}
\safemath{\opX}{\mathbb{X}}
\safemath{\opY}{\mathbb{Y}}
\safemath{\opZ}{\mathbb{Z}}
\safemath{\opZero}{\mathbb{O}}
\safemath{\identityop}{\opI}

\safemath{\veca}{\bma}
\safemath{\vecb}{\bmb}
\safemath{\vecc}{\bmc}
\safemath{\vecd}{\bmd}
\safemath{\vece}{\bme}
\safemath{\vecf}{\bmf}
\safemath{\vecg}{\bmg}
\safemath{\vech}{\bmh}
\safemath{\veci}{\bmi}
\safemath{\vecj}{\bmj}
\safemath{\veck}{\bmk}
\safemath{\vecl}{\bml}
\safemath{\vecm}{\bmm}
\safemath{\vecn}{\bmn}
\safemath{\veco}{\bmo}
\safemath{\vecp}{\bmp}
\safemath{\vecq}{\bmq}
\safemath{\vecr}{\bmr}
\safemath{\vecs}{\bms}
\safemath{\vect}{\bmt}
\safemath{\vecu}{\bmu}
\safemath{\vecv}{\bmv}
\safemath{\vecw}{\bmw}
\safemath{\vecx}{\bmx}
\safemath{\vecy}{\bmy}
\safemath{\vecz}{\bmz}

\safemath{\veczero}{\bmzero}
\safemath{\vecone}{\bmone}
\safemath{\vecxi}{\bmxi}
\safemath{\veclambda}{\bmlambda}
\safemath{\vecmu}{\bmmu}
\safemath{\vectheta}{\bmtheta}
\safemath{\vecphi}{\bmphi}
\safemath{\vecdelta}{\bmdelta}

\safemath{\matA}{\bA}
\safemath{\matB}{\bB}
\safemath{\matC}{\bC}
\safemath{\matD}{\bD}
\safemath{\matE}{\bE}
\safemath{\matF}{\bF}
\safemath{\matG}{\bG}
\safemath{\matH}{\bH}
\safemath{\matI}{\bI}
\safemath{\matJ}{\bJ}
\safemath{\matK}{\bK}
\safemath{\matL}{\bL}
\safemath{\matM}{\bM}
\safemath{\matN}{\bN}
\safemath{\matO}{\bO}
\safemath{\matP}{\bP}
\safemath{\matQ}{\bQ}
\safemath{\matR}{\bR}
\safemath{\matS}{\bS}
\safemath{\matT}{\bT}
\safemath{\matU}{\bU}
\safemath{\matV}{\bV}
\safemath{\matW}{\bW}
\safemath{\matX}{\bX}
\safemath{\matY}{\bY}
\safemath{\matZ}{\bZ}
\safemath{\matzero}{\bmzero}

\safemath{\matDelta}{\bDelta}
\safemath{\matLambda}{\bLambda}
\safemath{\matPhi}{\bPhi}
\safemath{\matSigma}{\bSigma}
\safemath{\matOmega}{\bOmega}
\safemath{\matTheta}{\bTheta}

\safemath{\matidentity}{\matI}
\safemath{\matone}{\matO}

\safemath{\rnda}{A}
\safemath{\rndb}{B}
\safemath{\rndc}{C}
\safemath{\rndd}{D}
\safemath{\rnde}{E}
\safemath{\rndf}{F}
\safemath{\rndg}{G}
\safemath{\rndh}{H}
\safemath{\rndi}{I}
\safemath{\rndj}{J}
\safemath{\rndk}{K}
\safemath{\rndl}{L}
\safemath{\rndm}{M}
\safemath{\rndn}{N}
\safemath{\rndo}{O}
\safemath{\rndp}{P}
\safemath{\rndq}{Q}
\safemath{\rndr}{R}
\safemath{\rnds}{S}
\safemath{\rndt}{T}
\safemath{\rndu}{U}
\safemath{\rndv}{V}
\safemath{\rndw}{W}
\safemath{\rndx}{X}
\safemath{\rndy}{Y}
\safemath{\rndz}{Z}

\safemath{\rveca}{\bimA}
\safemath{\rvecb}{\bimB}
\safemath{\rvecc}{\bimC}
\safemath{\rvecd}{\bimD}
\safemath{\rvece}{\bimE}
\safemath{\rvecf}{\bimF}
\safemath{\rvecg}{\bimG}
\safemath{\rvech}{\bimH}
\safemath{\rveci}{\bimI}
\safemath{\rvecj}{\bimJ}
\safemath{\rveck}{\bimK}
\safemath{\rvecl}{\bimL}
\safemath{\rvecm}{\bimM}
\safemath{\rvecn}{\bimN}
\safemath{\rveco}{\bomO}
\safemath{\rvecp}{\bimP}
\safemath{\rvecq}{\bimQ}
\safemath{\rvecr}{\bimR}
\safemath{\rvecs}{\bimS}
\safemath{\rvect}{\bimT}
\safemath{\rvecu}{\bimU}
\safemath{\rvecv}{\bimV}
\safemath{\rvecw}{\bimW}
\safemath{\rvecx}{\bimX}
\safemath{\rvecy}{\bimY}
\safemath{\rvecz}{\bimZ}

\safemath{\rvecxi}{\bmxi}
\safemath{\rveclambda}{\bmlambda}
\safemath{\rvecmu}{\bmmu}
\safemath{\rvectheta}{\bmtheta}
\safemath{\rvecphi}{\bmphi}

\safemath{\rmatA}{\bimA}
\safemath{\rmatB}{\bimB}
\safemath{\rmatC}{\bimC}
\safemath{\rmatD}{\bimD}
\safemath{\rmatE}{\bimE}
\safemath{\rmatF}{\bimF}
\safemath{\rmatG}{\bimG}
\safemath{\rmatH}{\bimH}
\safemath{\rmatI}{\bimI}
\safemath{\rmatJ}{\bimJ}
\safemath{\rmatK}{\bimK}
\safemath{\rmatL}{\bimL}
\safemath{\rmatM}{\bimM}
\safemath{\rmatN}{\bimN}
\safemath{\rmatO}{\bimO}
\safemath{\rmatP}{\bimP}
\safemath{\rmatQ}{\bimQ}
\safemath{\rmatR}{\bimR}
\safemath{\rmatS}{\bimS}
\safemath{\rmatT}{\bimT}
\safemath{\rmatU}{\bimU}
\safemath{\rmatV}{\bimV}
\safemath{\rmatW}{\bimW}
\safemath{\rmatX}{\bimX}
\safemath{\rmatY}{\bimY}
\safemath{\rmatZ}{\bimZ}

\safemath{\rmatDelta}{\bimDelta}
\safemath{\rmatLambda}{\bimLambda}
\safemath{\rmatPhi}{\bimPhi}
\safemath{\rmatSigma}{\bimSigma}
\safemath{\rmatOmega}{\bimOmega}
\safemath{\rmatTheta}{\bimTheta}

\usepackage{amssymb}
\usepackage{amsfonts}
\usepackage{mathrsfs}
\usepackage{xspace}
\usepackage{bm}
\usepackage{fancyref}
\usepackage{textcomp}

\usepackage{multirow}
\usepackage{stmaryrd}

\Macro{l2}{2}
\Macro{l1}{1}
\Macro{l0}{0}
\Macro{F}{\mathsf{F}}
\Macro{ln}{*}

\newcommand{\lefto}{\mathopen{}\left}

\DeclareMathOperator*{\argmin}{arg\;min}		

\newcommand{\abs}[1]{\lefto\lvert#1\right\rvert}		

\safemath{\dirac}{\delta}					
\safemath{\krond}{\dirac}					

\safemath{\upto}{\uparrow}
\safemath{\downto}{\downarrow}
\safemath{\iu}{j}							
\safemath{\ev}{\lambda}						
\safemath{\hilseqspace}{l^{2}}				
\newcommand{\banachfunspace}[1]{\setL^{#1}}	
\safemath{\hilfunspace}{\banachfunspace{2}}	
\newcommand{\floor}[1]{\lfloor #1 \rfloor}

\safemath{\SNR}{\text{\sc snr}} 				
\safemath{\No}{N_0}							
\safemath{\Es}{E_s}							
\safemath{\Eb}{E_b}							
\safemath{\EbNo}{\frac{\Eb}{\No}}
\safemath{\EsNo}{\frac{\Es}{\No}}

\DeclareMathOperator{\CHop}{\ensuremath{\opH}} 
\safemath{\tvir}{\rndh_{\CHop}}				
\safemath{\tvtf}{\rndl_{\CHop}}				
\safemath{\spf}{\rnds_{\CHop}}				
\safemath{\bff}{H_{\CHop}}					

\safemath{\ircf}{r_{h}}						
\safemath{\tftvcf}{r_{s}}					
\safemath{\tfcf}{r_{l}}						
\safemath{\bfcf}{r_{H}}						

\safemath{\tcorr}{c_h}						
\safemath{\scf}{c_{s}}						
\safemath{\tfcorr}{c_{l}}					
\safemath{\fcorr}{c_{H}}						

\safemath{\mi}{I}							
\safemath{\capacity}{C}						

\safemath{\normal}{\mathcal{N}}			
\safemath{\jpg}{\mathcal{CN}}			
\safemath{\mchain}{\leftrightarrow}		

\safemath{\dB}{\,\mathrm{dB}}
\safemath{\dBm}{\,\mathrm{dBm}}
\safemath{\Hz}{\,\mathrm{Hz}}
\safemath{\kHz}{\,\mathrm{kHz}}
\safemath{\MHz}{\,\mathrm{MHz}}
\safemath{\GHz}{\,\mathrm{GHz}}
\safemath{\s}{\,\mathrm{s}}
\safemath{\ms}{\,\mathrm{ms}}
\safemath{\mus}{\,\mathrm{\text{\textmu}s}}
\safemath{\ns}{\,\mathrm{ns}}
\safemath{\ps}{\,\mathrm{ps}}
\safemath{\meter}{\,\mathrm{m}}
\safemath{\mm}{\,\mathrm{mm}}
\safemath{\cm}{\,\mathrm{cm}}
\safemath{\W}{\,\mathrm{W}}
\safemath{\mW}{\, \mathrm{mW}}
\safemath{\J}{\,\mathrm{J}}
\safemath{\K}{\,\mathrm{K}}
\safemath{\bit}{\,\mathrm{bit}}
\safemath{\nat}{\,\mathrm{nat}}

\safemath{\define}{\triangleq}			

\safemath{\equivalent}{\sim}
\safemath{\distas}{\sim}					
\safemath{\sdiff}{\Delta}				

\safemath{\reals}{\mathbb{R}}
\safemath{\positivereals}{\reals_{+}}
\safemath{\integers}{\mathbb{Z}}
\safemath{\posint}{\integers_{+}}
\safemath{\naturals}{\mathbb{N}}
\safemath{\posnaturals}{\naturals_{+}}
\safemath{\complexset}{\mathbb{C}}
\safemath{\rationals}{\mathbb{Q}}

\newcommand*{\fancyrefapplabelprefix}{app}		
\newcommand*{\fancyrefthmlabelprefix}{thm}		
\newcommand*{\fancyreflemlabelprefix}{lem}		
\newcommand*{\fancyrefcorlabelprefix}{cor}		
\newcommand*{\fancyrefdeflabelprefix}{def}		
\newcommand*{\fancyrefproplabelprefix}{prop}		
\newcommand*{\fancyrefexmpllabelprefix}{exmpl}
\frefformat{vario}{\fancyrefseclabelprefix}{Section~#1}
\frefformat{vario}{\fancyrefthmlabelprefix}{Theorem~#1}
\frefformat{vario}{\fancyreflemlabelprefix}{Lemma~#1}
\frefformat{vario}{\fancyrefcorlabelprefix}{Corollary~#1}
\frefformat{vario}{\fancyrefdeflabelprefix}{Definition~#1}
\frefformat{vario}{\fancyreffiglabelprefix}{Fig.~#1}
\frefformat{vario}{\fancyrefapplabelprefix}{Appendix~#1}
\frefformat{vario}{\fancyrefeqlabelprefix}{(#1)}
\frefformat{vario}{\fancyrefproplabelprefix}{Property~#1}
\frefformat{vario}{\fancyrefexmpllabelprefix}{Example~#1}

 \newtheorem{theorem}{Theorem}
 \newtheorem{corollary}{Corollary}   
 \newtheorem{proposition}{Proposition}
 \newtheorem{definition}{Definition}
 \newtheorem{lemma}{Lemma}
 \newtheorem{example}{Example}
 \newtheorem{remark}{Remark}

\safemath{\low}{\matL}
\safemath{\sparse}{\matS}
\safemath{\obs}{\matM}

\safemath{\dima}{n_1}
\safemath{\dimb}{n_2}
\safemath{\mrank}{r}
\safemath{\msparse}{s}

\renewcommand{\define}{\triangleq}

\safemath{\cplxi}{\imath}
\safemath{\cplxj}{\jmath}

\safemath{\dict}{\matD}
\safemath{\inputdim}{n}		
\safemath{\outputdim}{m}		
\safemath{\sparsity}{s}	
\safemath{\inputdimA}{{n_a}}	
\safemath{\inputdimB}{{n_b}}	
\safemath{\elemA}{{n_a}}	
\safemath{\elemB}{{n_b}}	
\safemath{\resA}{\matR_a}	
\safemath{\resB}{\matR_b}	
\safemath{\subD}{\matS} 
\safemath{\subA}{\matS_a} 
\safemath{\subB}{\matS_b} 
\safemath{\dicta}{\matA} 	
\safemath{\dictb}{\matB} 	
\safemath{\hollowS}{H}
\safemath{\hollowA}{H_a}
\safemath{\hollowB}{H_b}
\safemath{\cross}{Z}
\safemath{\coh}{\mu}			
\safemath{\coha}{\mu_a}			
\safemath{\cohb}{\mu_b}			
\safemath{\mubs}{\nu}	
\safemath{\cohm}{\mu_m} 
\safemath{\dictset}{\setD}	
\safemath{\dictsetp}{\dictset(\coh,\coha,\cohb)}	
\safemath{\dictsetgen}{\dictset_\text{gen}}
\safemath{\dictsetgenp}{\dictsetgen(\coh)}
\safemath{\dictsetonb}{\dictset_\text{onb}}
\safemath{\dictsetonbp}{\dictsetonb(\coh)}

\safemath{\leftside}{U}
\safemath{\rightsideA}{R_a}
\safemath{\rightsideB}{R_b}

\safemath{\indexS}{\setI_S} 

\safemath{\na}{n_a}			
\safemath{\nb}{n_b}			
\safemath{\coeffa}{p_i}	
\safemath{\coeffb}{q_j}	
\safemath{\seta}{\setP}		
\safemath{\setb}{\setQ}     
\safemath{\setw}{\setW}	
\safemath{\setz}{\setZ}	
\safemath{\cola}{\veca}		
\safemath{\colb}{\vecb}		
\safemath{\cold}{\vecd}		
\safemath{\inputvec}{\vecx} 	
\safemath{\error}{\vece}	
\safemath{\noiseout}{\vecz} 	
\safemath{\inputvecel}{x}
\safemath{\inputveca}{\vecx_a}
\safemath{\inputvecb}{\vecx_b}
\safemath{\outputvec}{\vecy}	
\safemath{\lambdamin}{\lambda_{\mathrm{min}}}

\safemath{\elltwo}{\ell_2}
\safemath{\ellone}{\ell_1}
\safemath{\ellzero}{\ell_0}
\safemath{\ellinf}{\ell_\infty}
\safemath{\licard}{Z(\coh,\coha,\cohb)}
\safemath{\xsol}{\hat{x}}
\safemath{\xbord}{x_b}		
\safemath{\xstat}{x_s}		
\safemath{\xstatLone}{\tilde{x}_s}
\safemath{\order}{\mathcal{O}} 
\safemath{\scales}{\Theta} 
\safemath{\ones}{\mathbf{1}} 
\safemath{\zeroes}{\mathbf{0}} 
\safemath{\thlone}{\kappa(\coh,\cohb)} 
\safemath{\constoneA}{\delta} 
\safemath{\constoneB}{\epsilon} 
\safemath{\nlarge}{L}				   
\safemath{\sumlarge}{S_\nlarge}
\safemath{\maxlarger}{P_\nlarge}	   
\safemath{\Pzero}{\textrm{P0}}	
\safemath{\Pone}{\textrm{P1}}
\safemath{\vecfir}{\vecw}			 
\safemath{\vecsec}{\vecz}
\safemath{\elvecfir}{w}              
\safemath{\elvecsec}{z}				 
\safemath{\nlargefir}{n}
\safemath{\normout}{\gamma}
\safemath{\auxfun}{h}

\safemath{\indexa}{\ell}
\safemath{\indexb}{r}
\safemath{\indexc}{i}
\safemath{\indexd}{j}

\safemath{\project}{P}

\DeclareMathOperator{\diag}{diag}

\DeclareMathOperator{\Real}{Re}
\DeclareMathOperator{\Imag}{Im}

\newcommand{\udim}{\overline{d}}
\newcommand{\ldim}{\underline{d}}
\newcommand{\snr}{\mathsf{snr}}
\newcommand{\dof}{\mathsf{dof}}
\newcommand{\Dof}{\mathsf{DoF}}
\providecommand{\lk}{\langle}
\providecommand{\rk}{\rangle}
\providecommand{\abs}[1]{\left \lvert#1 \right \rvert}

\providecommand{\floor}[1]{\left \lfloor#1\right \rfloor}

\def\ba#1\ea{\begin{align*}#1\end{align*}}	
\def\ban#1\ean{\begin{align}#1\end{align}}	

\newcounter{MYtempeqncnt}

\makeatletter
\newcommand{\Biggg}{\bBigg@{3}}
\newcommand{\vast}{\bBigg@{4}}
\newcommand{\Vast}{\bBigg@{5}}
\newcommand{\Vvast}{\bBigg@{7}}
\makeatother

\begin{document}

\title{Degrees of Freedom in
 Vector  Interference Channels}

\author{David~Stotz 
		 and Helmut~B\"olcskei,~\IEEEmembership{Fellow,~IEEE}

\thanks{The material in this paper was presented in part in \cite{SB12Allerton} at the 50th Annual Allerton Conference on Communication, Control, and Computing, Monticello, IL, Oct.~2012. We point out that the statement  in \cite[Thm.~1]{SB12Allerton} needs to be restricted to  hold  true    for almost all $\mathbf H$ only. An Online Addendum \cite{SB14extended} to this paper is available at  \mbox{http://www.nari.ee.ethz.ch/commth/research/downloads/dof\_addendum.pdf}.
}
 
  \thanks{The authors are with the Department of Information Technology and Electrical Engineering, ETH Zurich, CH-8092  Zurich, Switzerland (e-mail: \mbox{\{dstotz,boelcskei\}@nari.ee.ethz.ch}).}

 
}

\maketitle


\begin{abstract}
This paper continues the Wu-Shamai-Verd\'u program \cite{WSV11draft} on characterizing the degrees of freedom (DoF) of interference channels (ICs) through R\'enyi information dimension.
Specifically, we find a single-letter formula for the DoF of vector ICs,  encompassing multiple-input multiple-output (MIMO) ICs, time- and/or frequency-selective ICs, and combinations thereof, as well as scalar ICs as considered in \cite{WSV11draft}.
The DoF-formula we obtain lower-bounds the DoF of \emph{all} channels---with respect to the choice of the channel matrix---and upper-bounds the DoF of \emph{almost all} channels. It  applies to a large class of noise distributions, and its proof is  based on an extension of a result  by Guionnet and Shlyakthenko \cite{GS07} to the vector case in combination with the Ruzsa triangle inequality for differential entropy introduced by Kontoyiannis and Madiman \cite{KM12}.
As in scalar ICs, achieving full DoF  requires the use of singular input distributions. 
Strikingly, in the vector case it suffices to enforce singularity on the joint distribution of each  transmit vector.
This can be realized through signaling in subspaces of the ambient signal space, which is in accordance with the idea of interference alignment, and, most importantly, allows the scalar entries of the transmit vectors to have non-singular distributions.
The DoF-formula for vector ICs we obtain enables a unified treatment of ``classical'' interference alignment \`a la Cadambe and Jafar \cite{CJ08}, and Maddah-Ali et al.\ \cite{MMK08}, and the number-theoretic schemes proposed in \cite{MM09,EO09}. Moreover, it allows to calculate the DoF achieved by new signaling schemes for vector ICs.
We furthermore recover the result by Cadambe and Jafar on the non-separability of parallel ICs \cite{CJ09}  and we show that almost all parallel ICs are separable in terms of DoF. 
Finally, our results  apply to complex vector ICs, thereby extending the main findings of \cite{WSV11draft} to the complex case.
\end{abstract}





\section{Introduction}
Sparked by the surprising finding of Cadambe and Jafar \cite{CJ08} stating that $K/2$ degrees of freedom (DoF) can be realized in $K$-user interference channels (ICs) through interference alignment, the study of DoF in wireless networks has seen significant activity in recent years. The essence of interference alignment is to exploit channel variations in time/frequency/space to align interference at the receiving terminals in low-dimensional subspaces. This is accomplished through a clever  vector-signaling scheme.

Following the discovery in \cite{CJ08} it was shown that the basic idea of aligning interference to realize DoF can be applied to numerous further settings \cite{Jaf11}. Perhaps the most surprising of these results is that $K/2$ DoF can be realized in $K$-user scalar ICs with constant channel matrix \cite{MM09,EO09}, i.e., in complete absence of channel variations. Moreover, it is shown in \cite{MM09} that $K/2$ DoF are achievable for almost all (with respect to the channel matrix) constant scalar ICs. The schemes introduced in \cite{MM09,EO09} use Diophantine approximation or lattice structures to design full DoF-achieving transmit signals.   

In a tour de force Wu et al.\  \cite{WSV11draft} discovered that R\'enyi information dimension\footnote{In the remainder of the paper we call ``R\'enyi information dimension'' simply ``information dimension''.} \cite{Rey59} is a suitable tool for systematically characterizing the DoF achievable in constant scalar ICs. Specifically, it follows from the results in \cite{WSV11draft} that the real interference alignment schemes 
proposed in \cite{MM09, EO09} correspond to the use of singular input distributions. What is more, Wu et al.\ \cite{WSV11draft} found that  $K/2$ DoF can be achieved in a $K$-user constant scalar IC only by input distributions that have  a non-trivial singular component. This is a strong, negative, result as input distributions with a singular component are 
difficult to realize, or, more specifically, to approximate in practice. 
On the other hand, it is well-known that full DoF can be realized in ICs with channel variations in time/frequency or in constant multiple-input multiple-output (MIMO)  ICs, even if the individual entries of the transmit vectors do not have singular components. Reconciling these two lines of results  is one of the central goals of this paper.

\paragraph*{Contributions}
We continue the Wu-Shamai-Verd\'u program \cite{WSV11draft} by showing how information dimension can be used to characterize the DoF of vector ICs. This extension 
is relevant as ``classical'' interference alignment \`a la Cadambe and Jafar \cite{CJ08}, and Maddah-Ali et al.\ \cite{MMK08} relies on vector-valued signaling. The vector IC we consider  contains, as special cases, the MIMO IC, time-~and/or frequency-selective ICs, and combinations thereof, as well as the constant scalar IC studied in \cite{WSV11draft, MM09, EO09}. The Wu-Shamai-Verd\'u theory builds on a little known but highly useful result by Guionnet and Shlyakthenko \cite[Thm.~2.7]{GS07}, stating that the DoF in a scalar additive noise channel achieved by a given input distribution equal the input distribution's information dimension. We present an extension of this result to the vector case and to more general noise distributions, and we give an information-theoretic proof  based  on (a slight generalization of) the Ruzsa triangle inequality for differential entropy as introduced by Kontoyiannis and Madiman \cite{KM12}. The (generalized) Ruzsa triangle inequality is key for the  extension to general noise distributions and, in addition,  allows for a considerable simplification of the existing proofs (for the scalar case) in \cite{Wu11, GS07}. 

Our main result is a single-letter formula for the DoF of  vector ICs. 
This formula has the same structure as the one for the constant scalar case  in \cite{WSV11draft}. It lower-bounds the DoF of \emph{all} channels---with respect to the channel matrix---and upper-bounds the DoF of (Lebesgue) \emph{almost all} channels,  including channel matrices with all entries rational. 
While we adopt the strategy underlying the proof of the DoF-formula for constant scalar ICs \cite[Sect.~V-B]{WSV11draft}, our extension of the result by Guionnet and Shlyakthenko \cite[Thm.~2.7]{GS07}---besides pertaining to the vector case---yields a DoF-formula that applies to a significantly larger class of noise distributions. Moreover, on a conceptual basis   the DoF-formula we obtain leads to  fundamentally new implications. Specifically, we find that
while input distributions with a singular component are still needed to achieve full DoF in vector ICs, it suffices to enforce singularity on the joint distribution of each  transmit vector. This form of singularity is  realized  by taking, e.g., the transmit vectors to live in lower-dimensional subspaces of the ambient signal space, as is, in fact, done in interference alignment \`a la Cadambe and Jafar \cite{CJ08}, and Maddah-Ali et al.\ \cite{MMK08}. 

We demonstrate that our DoF-formula for vector ICs allows for a unified treatment of ``classical'' interference alignment as introduced in \cite{CJ08,MMK08} and the ``number-theoretic'' interference alignment schemes for constant scalar ICs as proposed in \cite{MM09,EO09}. In addition, the formula constitutes a tool for evaluating the DoF achieved by new signaling schemes for vector ICs.

Furthermore, we recover the result by Cadambe and Jafar on the non-separability of parallel ICs \cite{CJ09}, and we show that almost all (again, with respect to the channel matrix) parallel ICs are separable in terms of DoF, i.e., for almost all parallel ICs independent coding across subchannels achieves full DoF. Finally, our results apply to  complex signals and channel matrices, thereby  extending the main result in \cite{WSV11draft} to the complex case.

\paragraph*{Notation}
Random vectors are represented by uppercase letters, deterministic vectors by lowercase letters, in both cases using letters from the end of the alphabet. 
Boldface uppercase letters are used to indicate matrices. The $n\times n$ identity matrix is $\mathbf I_n$ and the all-zeros matrix is $\mathbf 0$. The $nm\times nm$ block diagonal  matrix with blocks $\mathbf A_1, ... , \mathbf A_n\in \mathbb R^{m\times m}$ on its main diagonal is denoted by  $\diag \lefto (\mathbf A_1, ... , \mathbf A_n \right )$. $\| \mathbf A \|_\infty :=\max \{ |a_{i,j}| \!\mid \! 1\leqslant i \leqslant m, 1\leqslant j \leqslant n \}$ stands for the $\ell^\infty$-norm of the matrix $\mathbf A\in \mathbb R^{m\times n}$. For $x\in\mathbb R$, we write $\lfloor x \rfloor$ for the largest integer not exceeding  $x$. For ${k\in \mathbb N\! \setminus \! \{0\}}$, we set $\lk x \rk _k \! :=\!  \lfloor k x \rfloor / k$ and $[x]_k \! := \! \lk x \rk_{2^k}$. 
The notation conventions  $\lfloor \cdot \rfloor$, $\lk \cdot \rk_k$, and $[\cdot ]_k$ are extended to real vectors and matrices through application on an entry by entry basis.  
For $x\in \mathbb R^{n}$, let $|x |_{\mathbb Z}:=\min_{u\in \mathbb Z^{ n}}\| x-  u\|_\infty$. We define the minimum and maximum distance of a set $\mathcal W\subseteq \mathbb R^{ n}$, respectively,  as $\mathsf m(\mathcal W):=\inf_{w_1,w_2\in \mathcal W, w_1\neq w_2}\|  w_1 -  w_2\|_\infty$ and  $\mathsf M(\mathcal W):=\sup_{w_1,w_2\in \mathcal W, w_1\neq w_2}\|  w_1- w_2 \|_\infty$, and we extend these definitions to sets of matrices accordingly. 
 For a discrete random matrix $\mathbf X$, we let $H(\mathbf X)$  be the entropy of the vector obtained by stacking the columns of $\mathbf X$.
 The differential entropy $h(\mathbf X)$ is defined analogously.
All logarithms are to the base $2$. By ``$a>0$'' we mean that the real constant $a$ is positive \textit{and} finite. For sets $\mathcal A,\mathcal B\subseteq \mathbb R^n$, $m\in \mathbb N$, and a scalar $\alpha\in\mathbb R$, we let $\mathcal A\times \mathcal B$ denote the cartesian product, $\mathcal A^m$ the $m$-th cartesian power of $\mathcal A$, $\mathcal A + \mathcal B:=\{a+b \! \mid\!  a\in \mathcal A, b\in \mathcal B\}$, and $\alpha \mathcal A:=\{\alpha a  \!\mid\! a\in \mathcal A\}$. 
The dimension  of a subspace $\mathcal V\subseteq \mathbb R^n$ is denoted by $\dim \mathcal V$ and $\mathcal V^\perp$ represents the   orthogonal complement of $\mathcal V$ with respect to the Euclidean inner product.
 $\mathbb E[\cdot ]$ stands for the expectation operator and $\stackrel{D}=$ means equality in distribution. For a measurable real-valued function $f$ and a measure\footnote{Throughout the paper, the terms ``measurable'' and ``measure'' are to be understood with respect to the Borel $\sigma$-algebra.} $\mu$ on its domain, the pushforward measure is given by $(f_{\!\ast}\mu)(\mathcal A)=\mu(f^{-1}(\mathcal A))$ for  Borel sets $A$. Integration of $f$ with respect to $\mu$ is denoted by $\int \! f(x)\mu\lk x \rk$, where for the special case of the Lebesgue measure $\lambda$ we put $\mathrm dx:=\lambda\lk x\rk$. We say that an integral $\int \! f(x)\mu\lk x \rk$ exists if the integral of the positive part $\max \{f(x),0\}$  or the integral of the negative part $-\min \{f(x),0\}$ is finite. Furthermore, the
integral $\int \! f(x)\mu\lk x \rk$ is said to be finite if it exists and if $|\int \! f(x)\mu\lk x \rk|<\infty$. The differential entropy $h(X)$ of a random vector with density $f_X$ is said to exist (to be finite), if the integral $h(X)=\int_{\mathbb R^n} f_X(x)\log \lefto (\frac{1}{f_X(x)}\right ) \mathrm dx$ exists (is finite). For Borel sets $\mathcal A$, we write $\mathds{1}_\mathcal A(x)$ for the characteristic function on $\mathcal A$.

\section{Setup and Definitions}
We consider a memoryless $K$-user (with $K\geqslant 2$)  vector IC with input-output (I/O) relation
\begin{align}	Y_i={\sqrt{\snr}}\sum_{j=1}^K \mathbf H_{i,j} X_j +  W_i,	\quad i=1, ... ,K ,	\label{eq:channel} \end{align}
where $X_i=(X_i[1] \, ... \, X_i[M])^T\in\mathbb R^M$ and $Y_i=(Y_i[1] \, ... \, Y_i[M])^T\in\mathbb R^M$ is the transmit and receive vector, respectively, corresponding to user $i$, $M$ is the dimension of the I/O  signal space, $\mathbf H_{i,j}\in\mathbb R^{M\times M}$ denotes the channel matrix between transmitter $j$ and receiver $i$, and the $W_i$ are i.i.d.\ zero-mean Gaussian random vectors with identity covariance matrix. For simplicity of exposition, we treat the real case throughout and show in 
 Section~\ref{sec:complex} how our results  can be extended to the complex case. Defining the $KM\times KM$ matrix
 \begin{align}	\mathbf H := \begin{pmatrix} \mathbf H_{1,1} & \cdots & \mathbf H_{1,K}   \\[-.05cm]  \vdots &\ddots &\vdots \\ \mathbf H_{K,1} &\cdots &\mathbf H_{K,K}\end{pmatrix},	\end{align}
we can rewrite \eqref{eq:channel} as
\begin{align}
\begin{pmatrix} Y_1  \\[-.1cm]  \vdots \\ Y_K \end{pmatrix}=\sqrt{\snr} \, \mathbf H \! \begin{pmatrix} X_1  \\[-.1cm] \vdots \\ X_K \end{pmatrix} + \begin{pmatrix} W_1   \\[-.1cm]  \vdots \\ W_K \end{pmatrix}. \label{eq:stacked}
\end{align}

The channel matrix $\mathbf H$ is assumed to be known perfectly at all transmitters and receivers and remains constant across channel uses. 
For each user $i=1,...,K$, we impose the average power constraint
\begin{align}	\frac{1}{MN}\sum_{n=1}^N\sum_{m=1}^M \left (x_{i}^{(n)}[m]\right )^2\leqslant 1,	\label{eq:power constraint}\end{align}
on codeword matrices $\left (x_{i}^{(1)} \,  ... \;  x_{i}^{(N)}\right )$ of blocklength $N$, where $x_{i}^{(n)}=\left (x_{i}^{(n)}[1] \, ... \, x_{i}^{(n)}[M]\right )^T\in\mathbb R^M$.  In contrast to the setting in \cite{WSV11draft}, where the action of each link is represented through scaling by a single coefficient, here each link acts as a linear operator (represented by a finite-dimensional matrix) on the corresponding transmit vector. To differentiate clearly, we henceforth  refer to the setting in \cite{WSV11draft} as the ``scalar IC''. Note that this terminology includes the channel matrix being constant, as opposed to the classical interference alignment setup in \cite{CJ08}, which also deals with scalar ICs but has varying channel matrices.

The vector IC setting encompasses the MIMO IC (with $M$ antennas at each user and $\mathbf H_{i,j}$ denoting the MIMO channel matrix between  transmitter $j$ and receiver $i$), and time-frequency-selective ICs, transformed into memoryless vector ICs through the use of guard periods/bands. For  purely frequency-selective single-antenna ICs, e.g., the use of OFDM \cite{PR80} results in
a memoryless vector channel with  diagonal $\mathbf H_{i,j}$ matrices. If guard intervals (of sufficient length) filled with zeros are used instead of the cyclic prefix in OFDM, we get memoryless vector ICs with $\mathbf H_{i,j}$ matrices that are not diagonal. For purely time-selective single-antenna ICs, we obtain a memoryless vector IC, again with  diagonal $\mathbf H_{i,j}$ matrices, without using guard regions,  by simply identifying the main diagonal entries of the $\mathbf H_{i,j}$'s with the corresponding channel coefficients.
When all the $\mathbf H_{i,j}$ are diagonal we obtain an important special case of the vector IC in \eqref{eq:stacked}, namely a parallel IC (with $M$ subchannels) as studied in \cite{CJ09}.  Finally, the vector IC also covers time-frequency-selective MIMO ICs.

Let $C_\text{sum}(\mathbf H; \snr)$  
be the  sum-capacity\footnote{For a  definition of the sum-capacity of discrete ICs see \cite[Sect.~6.1]{GK12}; for  continuous alphabets with a cost constraint, as needed here, we refer to  \cite[Sections~3.3 and~3.4]{GK12}.} for a given $\snr$. The {degrees of freedom}\footnote{For motivation on why to study this quantity see \cite[App.~A]{Jaf11}.} of the IC \eqref{eq:channel} are then defined as
\begin{align}	\Dof (\mathbf H) := \limsup_{\snr\to\infty}\frac{C_\text{sum}(\mathbf H; \snr)}{\frac{1}{2}\log \snr}.	\label{eq:defdof}	\end{align}
We call $\Dof(\mathbf H)/M$ the {normalized DoF}.

\section{R\'enyi Information Dimension}

One of the main feats of \cite{WSV11draft} was to recognize that (R\'enyi) information dimension is a suitable tool for characterizing the DoF achievable in scalar ICs. The main conceptual contribution of the present paper is to show that information dimension is a natural tool for analyzing the DoF in vector ICs as well. This, in turn, leads to a unified framework for  real interference alignment \cite{MM09}, \cite{EO09}---hinging on number-theoretic  properties of transmit signals and channel coefficients---and  ``classical'' interference alignment  relying on channel variations and vector signaling \cite{CJ08, MMK08, Jaf11}.

 Throughout the paper we will deal with general distributions $\mu$ on $\mathbb R^n$, the nature of which can often be understood better by invoking the following decomposition.
\vspace{.1cm}
\begin{proposition}
Every distribution $\mu$ on $\mathbb R^n$ can be decomposed uniquely as
\begin{align}	\mu = \alpha \mu_{\text{ac}} + \beta \mu_\text{d} + \gamma \mu_{s}	\label{eq:decomp}	,\end{align}
where $\mu_{\text{ac}}$ is an absolutely continuous, $\mu_{\text{d}}$ a discrete, and $\mu_{\text{s}}$ a singular distribution,\footnote{``Absolutely continuous'' is to be understood with respect to Lebesgue measure. A discrete distribution  is supported on a countable set of points, whereas a singular distribution lives on a set of  Lebesgue measure zero and, in addition, does not have any point masses. An example of a singular distribution on $\mathbb R$ is the Cantor distribution \cite[Ex.~1.2.4]{Dur10}.} and $\alpha,\beta,\gamma\geqslant 0$ satisfy $\alpha+\beta+\gamma=1$.
\end{proposition}
\begin{IEEEproof}
See for example \cite[Thm.~2.7.19 combined with Exercise~2.9.14]{Tay97}.
\end{IEEEproof}
We begin by defining information dimension and collecting some of its basic properties  used in the remainder of the paper.

\begin{definition}\label{def:infdim} Let $X$ be a random vector. We define the lower and upper information dimension of $X$ as
\begin{align} \ldim(X) := \liminf_{k\to\infty}\frac{H(\lk X \rk _k)}{\log k} \;\; \text{and}  \;\; \udim(X):= \limsup_{k\to\infty}\frac{H(\lk X \rk _k)}{\log k},  \label{eq:infdimdef}\end{align}
respectively.\footnote{Note that due to the scaling invariance of entropy, we have $H(\lk X \rk _k)=H(\lfloor kX \rfloor)$ for all integers $k>0$. The motivation for using $\lk X\rk_k$ in Definition~\ref{def:infdim} is the pointwise convergence of $\lk X\rk_k$, as $k\to\infty$, to the original random vector $X$.} If  $\ldim(X) = \udim(X)$ (possibly $=\infty$), then we say that the information dimension $d(X)$ of $X$ exists and we set $d(X):= \underline d(X) = \overline d(X)$.
\end{definition}

At first sight  \eqref{eq:infdimdef} may suggest that information dimension depends on the specific way $X$ is quantized, namely through application of the floor-operation on an entry-wise basis. However, the following equivalent definition shows that this is, in fact, not the case. 

\vspace{.1cm}
\begin{lemma}\label{lem:alternative}
For a random vector $X$ with distribution $\mu$, we have
\begin{align*}	&\ldim(X)=\liminf_{\varepsilon\to 0}\frac{\mathbb E \lefto [ \log \mu \lefto (B(X;\varepsilon )\right )\right ]}{\log \varepsilon}	 \quad \text{and} \\  &\udim(X)= \limsup_{\varepsilon\to 0}\frac{\mathbb E \lefto [ \log \mu \lefto (B(X;\varepsilon )\right )\right ]}{\log \varepsilon}	,	\end{align*}
where $B(x;\varepsilon)\subseteq \mathbb R^n$ denotes the ball with center $x$ and radius $\varepsilon$ with respect to an arbitrary  norm on $\mathbb R^n$. 
\end{lemma}
\begin{IEEEproof}
See \cite[App.~I]{WV10}. A  (slightly) different proof can be found in the Online Addendum \cite[Sect.~III-A]{SB14extended}.
\end{IEEEproof}

Further basic properties of information dimension we shall need are summarized next.

\vspace{.1cm}
\begin{lemma}[{\cite{Rey59},\cite{WSV11draft},\cite{GS07}}]\label{lem:properties}
\begin{enumerate}\setlength{\itemindent}{-0.325cm}
	\item     Let $X$ be a random vector in $\mathbb R^n$. Then $\ldim (X),\udim(X)<\infty$ if and only if $H(\lfloor X \rfloor)<\infty$. Moreover, in this case we have
			\begin{align}		0 \leqslant \ldim (X)\leqslant  \udim(X)	\leqslant n . 	\label{eq:finite}\end{align}
\setlength{\itemindent}{-0.0cm}
	\item Let $X_1, ... , X_K$ be independent random vectors in $\mathbb R^n$, such that $d(X_i)$ exists for $i=1, ... ,K$. Then 
			\begin{align}	 d\lefto (\begin{matrix} X_1\\[-.1cm]  \vdots  \\ X_K \end{matrix} \right )= \sum_{i=1}^K d(X_i).		\label{eq:sumind}\end{align}
	\item Let $X$ and $Y$ be independent random vectors in $\mathbb R^n$. Then 
			\begin{align}	\max \{ \udim(X), \udim(Y) \}&\leqslant \udim(X+Y) 	\label{eq:sum1} \\&\leqslant \udim(X)+\udim(Y).	\label{eq:sum2}\end{align}
		The inequality \eqref{eq:sum1} also holds when $\udim(\cdot)$ is replaced by $\ldim(\cdot )$ throughout.
	\item Let $X$ be a random vector in $\mathbb R^n$, $F\colon \mathbb R^n\to \mathbb R^n$ a map, and  $\| \cdot \|$ a norm on $\mathbb R^n$. 
\begin{itemize} \item If $F$ is Lipschitz, i.e.,  there exists a constant $C>0$ such that $\| F(x) -F(y) \| \leqslant C \|   x-y\|$, for all $x,y\in\mathbb R^n$, then 
\ban \ldim (F(X)) \leqslant \ldim (X) \quad \text{and} \quad \udim(F(X)) \leqslant \udim (X).		\label{eq:isoleq}			\ean 
\item If $F$ is bi-Lipschitz, i.e., there exist constants $c,C>0$ such that $c\|   x-y  \|\leqslant \|   F(x) -F(y) \| \leqslant C \|   x-y\|$, for all $x,y\in\mathbb R^n$, then
			\begin{align}	\ldim (F(X)) = \ldim (X) \quad \text{and} \quad \udim(F(X)) = \udim (X).		\label{eq:iso}\end{align}\end{itemize}
\end{enumerate}
\end{lemma}
\begin{IEEEproof}
The statement in 1) follows directly from \cite[Lem.~4]{SB14extended} with $p=k$ and $q=1$. 
Eq.\ \eqref{eq:sumind} follows from $H\lefto (X_1, ..., X_K\right )=\sum_{i=1}^K H(X_i)$ for independent $X_i$. 
For the inequalities \eqref{eq:sum1} and \eqref{eq:sum2}, we first note that $|H(\lk X \rk_k+\lk Y\rk_k)- H(\lk X+Y\rk_k)|\leqslant n\log 3$, which follows from Lemma~\ref{lem:distanceentropy} (in Appendix~\ref{app:A}) taking into account that $\| \lfloor k(X+Y) \rfloor -\lfloor kX\rfloor - \lfloor kY\rfloor\|_\infty\leqslant 1$ and that the minimum distance of the value set of $\lk X \rk_k +\lk Y \rk_k$ is lower-bounded by $1/k$ by  definition of $\lk \cdot \rk_k$. The inequality \eqref{eq:sum1} then follows from \begin{align*} H(\lk X\rk_k)&=H(\lk X \rk_k+\lk Y\rk_k \! \mid \!  \lk Y\rk_k)\\  &\leqslant H(\lk X \rk_k+\lk Y\rk_k)\\ &\leqslant H(\lk X+Y\rk_k)+n\log 3 , \end{align*} \begin{align*} H(\lk Y\rk_k)&=H(\lk X \rk_k+\lk Y\rk_k \! \mid \!  \lk X\rk_k)\\ &\leqslant H(\lk X \rk_k+\lk Y\rk_k) \\ &\leqslant H(\lk X+Y\rk_k)+n\log 3.\end{align*} Similarly for \eqref{eq:sum2}, we note that the value set of $\lk X+Y\rk_k$  is lower-bounded by $1/k$, and we apply Lemma~\ref{lem:distanceentropy} to obtain\begin{align*} H(\lk X+Y\rk_k) &\leqslant H(\lk X \rk_k+\lk Y\rk_k)+ n\log 3\\  & \leqslant H(\lk X\rk_k) + H(\lk Y\rk_k) + n\log 3 ,\end{align*}
where the last inequality follows from $H(V) \geqslant H(V)- H(V\!\! \mid \! \!  U+V) = I(V;U+V) = H(U+V)-H(U+V\! \mid \!  V) = H(U+V) - H(U)$ for  discrete random vectors $U,V$. 
To prove \eqref{eq:isoleq} we use the equivalent characterization of information dimension from Lemma~\ref{lem:alternative}. Since  $F$ is continuous, it is measurable and we find that 
\begin{align} B\lefto (x;\frac \varepsilon C\right )\subseteq F^{-1}\lefto ( B(F(x); \varepsilon)\right )\! , \end{align} which implies
\ban  \mu \lefto (	B\lefto (x;\frac \varepsilon C\right )\right  ) \leqslant F_\ast \mu \lefto  (B(F(x); \varepsilon)\right )\! 	,  \label{eq:flip}  \ean where $\mu$ is the distribution of $X$. Dividing \eqref{eq:flip} by $\log \varepsilon $ and taking  $\liminf_{\varepsilon \to 0}$ yields
\begin{align}	\liminf_{\varepsilon\to 0}\frac{\mathbb E\lefto [\log F_{\!\ast}\mu(B(F(X);\varepsilon))\right ]}{\log \varepsilon}\leqslant \liminf_{\varepsilon\to 0}\frac{\mathbb E\lefto [\log \mu(B(X;\varepsilon))\right ]}{\log \varepsilon}.\label{eq:lem2a} \end{align}
As $F_\ast \mu$ is the distribution of $F(X)$, this implies $\ldim (F(X))\leqslant \ldim(X)$.
 If $F$ is  bi-Lipschitz, we additionally have 
\begin{align*} F^{-1}\lefto ( B(F(x); \varepsilon)\right )\subseteq B\lefto (x;\frac \varepsilon c\right )\! , \end{align*}
which  together with \eqref{eq:lem2a} implies that 
\begin{align}	\liminf_{\varepsilon\to 0}\frac{\mathbb E\lefto [\log F_{\!\ast}\mu(B(F(X);\varepsilon))\right ]}{\log \varepsilon}=\liminf_{\varepsilon\to 0}\frac{\mathbb E\lefto [\log \mu(B(X;\varepsilon))\right ]}{\log \varepsilon},\label{eq:lem2b}\end{align}
and hence $\ldim (F(X))=\ldim(X)$.
The statements in \eqref{eq:lem2a} and \eqref{eq:lem2b} also hold for ``$\limsup$'' in place of  ``$\liminf$''.
\end{IEEEproof}

\begin{remark}  \label{rem:two} \begin{enumerate} \item Since linear isomorphisms $F:\mathbb R^n\to \mathbb R^n$ are bi-Lipschitz with constants $c=\inf_{\|x\|=1} \|F(x)\|$ and $C=\sup_{\|x\|=1} \|F(x)\|$, an immediate consequence of \eqref{eq:iso} and \eqref{eq:sumind} is the following:
Suppose $(v_1, ... , v_n)$ are linearly independent vectors in some vector space and $X_1, ... ,X_n$ are independent random variables with $d(X_i)=1$, for all $i$.
Then, we have
\begin{align}		d\lefto ( X_1v_1 + \ldots + X_n v_n \right ) \stackrel{\eqref{eq:iso}}=  d\lefto (\begin{matrix} X_1   \\[-.1cm]  \vdots  \\ X_n \end{matrix} \right ) \stackrel{\eqref{eq:sumind}}=\sum_{i=1}^n d(X_i)=n.	\label{eq:change} \end{align}

\item For independent random variables $X,Y$ with $d(X)=d(Y)=1$, it follows by \eqref{eq:finite} and \eqref{eq:sum1} that $d(X+Y)$ exists and \ban d(X+Y)=1. \label{eq:two}\ean This fact will often  be used in the remainder of the paper. 

\end{enumerate}
\end{remark}

Information dimension is, in general, difficult to compute analytically. However, for distributions that are discrete-continuous mixtures, there is an explicit formula, which is a simple extension of \cite[Thm.~3]{Rey59} to the vector case.

\vspace{.1cm}
\begin{proposition} \label{prop:mixed1}
Let $X$ be a random vector in $\mathbb R^n$ with $H(\lfloor X \rfloor)<\infty$ and distribution $\mu$ that admits a decomposition  into an absolutely continuous part $\mu_\text{ac}$ and a discrete part $\mu_\text{d}$  according to $\mu=\alpha \mu_\text{ac} + (1-\alpha) \mu_\text{d}$, where  $\alpha\in [0,1]$. Then, we have
\begin{align}	d(X)=n\alpha	. \label{eq:mixture}\end{align}
\end{proposition}
\begin{IEEEproof}
The proof follows closely that  of the corresponding result for the scalar case reported in \cite[Thm.~3]{Rey59}, and is provided in the Online Addendum \cite[Sect.~III-B]{SB14extended}  for completeness.  
\end{IEEEproof}

Another class of distributions that is amenable to analytical expressions for information dimension and will turn out crucial in the proof of our main result, Theorem~\ref{thm:main}, is that of self-similar homogeneous distributions. 

A self-similar distribution $\mu$ is characterized as follows. Let $\{F_1, ... ,F_m\}$ be a  finite set of contractions  on $\mathbb R^n$ endowed with norm $\|\cdot\|$, i.e., $F_i:\mathbb R^n \to \mathbb R^n$ with $\| F_i(x) - F_i(y) \| < \|x-y\|$ for all $x,y\in\mathbb R^n$ with $x\neq y$. The set $\{F_1, ... ,F_m\}$ is called an iterated function system. Given a probability vector $(\rho_1,...,\rho_m)$ we let $F$  be the random contraction drawn from $\{ F_1, ... ,F_m\}$ according to  $(\rho_1,...,\rho_m)$, i.e., $F=F_i$ with probability $\rho_i$. It turns out (see \cite[Sect.~1: (2)]{Hut81}) that there is a unique distribution $\mu$ on $\mathbb R^n$, which is invariant under the random contraction $F$, i.e., $F(X)\stackrel{D}=X$ if $X$ has distribution $\mu$ and is independent of $F$. This distribution satisfies
\begin{align}	\mu= \sum_{i=1}^m \rho_i F_{i\ast}\mu 	\label{eq:invdist}	\end{align}
and is called the self-similar distribution associated with the pair  $\left ( \{F_1, ... ,F_m\} ,(\rho_1,...,\rho_m) \right )$.
We refer the interested reader to \cite{Hut81} for general facts on self-similar distributions.  Analytical expressions for the information dimension of self-similar distributions are available only under suitable ``regularity'' conditions, such as the open set condition \cite{BHR05}.
\begin{definition}\label{def:opensetcondition}
An iterated function system $\{ F_1, ... ,F_m\}$ is said to satisfy the open set condition if there exists a nonempty bounded open set $\mathcal U\subseteq\mathbb R^n$ such that $\bigcup_{i=1}^m F_i(\mathcal U)\subseteq \mathcal U$ and $F_i(\mathcal U)\cap F_j(\mathcal U)=\emptyset$ for all $i\neq j$.
\end{definition}

In the special case where all $F_i$ are of the form $F_i(x):=rx + w_i$ for a positive scalar $r<1$ and some set of vectors $\mathcal W:=\{w_1, ... ,w_m\}\subseteq \mathbb R^n$,\label{p:W} the associated measure $\mu$ in \eqref{eq:invdist} is said to be a self-similar \textit{homogeneous} distribution with similarity ratio $r$. In this case, we can give the following explicit expression for a random vector $X$ with the  distribution $\mu$ in \eqref{eq:invdist} 
\begin{equation}	X :=\sum_{i\geqslant 0}	r^{i}W_i,	\label{eq:ifs} \end{equation}
where $\{ W_i\}_{i\geqslant 0}$ is a set of i.i.d.\ copies of a random vector $W$  drawn from the set $\mathcal W$ according to the probability vector $(\rho_1,...,\rho_m)$. To see this simply note that with $\widetilde W$  another copy of $W$, independent of $\{W_i\}_{i\geqslant 0}$, we have
\ba 	F(X)=rX+\widetilde W=\sum_{i\geqslant 0}r^{i+1}W_i+\widetilde W\stackrel{D}=\sum_{i\geqslant 0}	r^{i}W_i=X.		\ea
By uniqueness of $\mu$ in \eqref{eq:invdist} it follows that $X$ must have distribution $\mu$. Random vectors of the form \eqref{eq:ifs} will be used later in the paper to construct full DoF-achieving transmit vectors. This ingenious idea was first described in \cite{WSV11draft} for scalar input distributions. 
The information dimension of random vectors with self-similar homogeneous distribution satisfying the open set condition takes on a particularly simple form.

\vspace{.1cm}
\begin{theorem}\label{thm:infdimifs}
Let $ X$ be defined by \eqref{eq:ifs} with the underlying iterated function system satisfying the open set condition. Then, we have
\begin{align}	d( X)=\frac{H( W)}{\log \frac{1}{r}}	.\label{eq:infdimifs}\end{align}
\end{theorem}
\begin{IEEEproof}
See \cite[Thm.~2]{GH89} and \cite[Thm.~4.4]{You82}. 
\end{IEEEproof}

We will see in the proof of Theorem~\ref{thm:main} that the open set condition can be satisfied by suitably restricting the contraction parameter $r$ as a function of $\mathsf m(\mathcal W)$ and  $\mathsf M(\mathcal W)$.

\section{Main Results}
\label{sec:main}

Our first result is an achievability statement and shows that information dimension plays a fundamental role in the characterization of the DoF of vector ICs.

\vspace{.1cm}
\begin{theorem}\label{thm:mainb}
Let $X_1, ... ,X_K$ be independent random vectors in $\mathbb R^M$ such that $H(\lfloor X_i \rfloor )<\infty$, for all $i$, and such that all information dimension terms appearing in
\begin{align} 	\dof(X_1, ... ,X_K ; \mathbf H) &:= \nonumber \\ \sum_{i=1}^K  &\left [ d\lefto (\sum_{j=1}^K \mathbf H_{i,j} X_j \right )-	 d \lefto ( \sum_{j\neq i}^K \mathbf H_{i,j} X_j \right )\right ],	 \label{eq:dofallexist}\end{align}
exist. Then, we have
\ban  \Dof (\mathbf H )\geqslant  \dof(X_1, ... ,X_K ; \mathbf H) . \label{eq:mainneu}\ean
\end{theorem}
\begin{IEEEproof}
See Appendix~\ref{app:A}.
\end{IEEEproof}
The existence of the information dimension terms in \eqref{eq:dofallexist} is not guaranteed for general input distributions. In particular, existence of $d(X_i)$, $i=1,...,K$, does not necessarily imply  existence of the information dimension terms in \eqref{eq:dofallexist}. However, when all input distributions are either discrete-continuous mixtures or  self-similar homogeneous, then the  distributions of $\sum_{j=1}^K \mathbf H_{i,j} X_j $ and $\sum_{j\neq i}^K \mathbf H_{i,j} X_j$ are of the same nature as the input distributions, which by Proposition~\ref{prop:mixed1} and Theorem~\ref{thm:infdimifs}
guarantees existence of the information dimension terms in \eqref{eq:dofallexist}.

Even though a single-letter characterization of the capacity region of the (vector) IC is not available in the literature, the next result shows that we can get a single-letter formula for the DoF, which  holds for almost all channel matrices $\mathbf H$.

\vspace{.1cm}
\begin{theorem}[DoF-formula]\label{thm:main}
For (Lebesgue) almost all channel matrices $\mathbf H$  we have
\begin{align}		\Dof (\mathbf H ) = \sup_{ X_1, ... , X_K}\dof(X_1, ... ,X_K ; \mathbf H)  ,	\label{eq:main}\end{align}
where the supremum is taken over all independent $X_1, ... ,X_K$ such that $H(\lfloor X_i \rfloor )<\infty$, for all $i$, and such that all information dimension terms appearing in \eqref{eq:dofallexist} exist.\footnote{It turns out (in the proof of Theorem~\ref{thm:main}) that it is not necessary to restrict the supremization to transmit vectors satisfying the power constraint \eqref{eq:power constraint}.}\footnote{We point out that the statement  in \cite[Thm.~1]{SB12Allerton} needs to be restricted to  hold  true only   for almost all $\mathbf H$.}
\end{theorem}
\begin{IEEEproof}
See Appendix~\ref{app:A}.
\end{IEEEproof}
This formula extends the DoF-formula for scalar ICs established by Wu et al.\ \cite{WSV11draft} to the case of vector ICs.
The quantity $\dof(X_1,...,X_K;\mathbf H)$ can be interpreted as the DoF achieved by the particular choice of transmit vectors $X_1,...,X_K$. The terms $d\lefto (\sum_{j=1}^K \mathbf H_{i,j} X_j \right )-	 d \lefto ( \sum_{j\neq i}^K \mathbf H_{i,j} X_j \right )$ in \eqref{eq:dofallexist} can be understood as the difference of the information dimensions of the noise-free receive  signal   and the noise-free interference at user $i$. The DoF of a given channel are obtained by maximizing  $\dof(\cdot )$ over all independent transmit vectors satisfying $H(\lfloor X_i\rfloor)<\infty$, $i=1,...,K$.\footnote{The expression $\dof(X_1, ... ,X_K ; \mathbf H) $ is defined only if all information dimension terms in \eqref{eq:dofallexist}  exist. For brevity, we will  not always mention this existence condition.} 

We hasten to add that Theorem~\ref{thm:main} applies to \textit{almost all} channel matrices $\mathbf H$ only, and that an explicit characterization of the set of these matrices is not available. For a given channel matrix, it is therefore not clear whether Theorem~\ref{thm:main} applies. The  set of exceptions  is, however, of Lebesgue measure zero. What we do know is that Theorem~\ref{thm:main} applies to channel matrices $\mathbf H$ that have all entries rational.

\vspace{.1cm}
\begin{theorem}\label{thm:maind}
If all entries of $\mathbf H$ are rational, then the DoF-formula \eqref{eq:main} holds.
\end{theorem}
\begin{IEEEproof}
See Appendix~\ref{app:A}.
\end{IEEEproof}
For scalar ICs \cite{SB14} provides an explicit condition on the channel matrix $\mathbf H$ to admit $K/2$ DoF. This condition is satisfied by almost all $\mathbf H$ and the DoF of the corresponding class of channels are shown in \cite{SB14} to be given by \eqref{eq:main}.
In addition,  for the scalar case,  channel matrices that have all entries non-zero and rational admit  strictly less than $K/2$ DoF as shown in \cite{EO09}, and their exact DoF are characterized by Theorem~\ref{thm:maind} through \eqref{eq:main}. Thus, for the scalar case we have explicit characterizations of channels where \eqref{eq:main} applies and where $\Dof(\mathbf H)=K/2$ or $\Dof(\mathbf H)<K/2$. 
In \cite{WSV11draft} it is furthermore shown that Theorem~\ref{thm:main} also applies in the scalar case if all entries of $\mathbf H$ are algebraic numbers. These arguments can be extended to the vector case. For conciseness, we do, however,  not pursue this extension here.

The channel model in \eqref{eq:channel} assumes Gaussian noise, as was done in \cite{WSV11draft}.
The results in Theorems~\ref{thm:mainb}--\ref{thm:maind}  extend, however, to a fairly general class of noise distributions, as shown next.
\vspace{.1cm}
\begin{theorem}\label{thm:mainc}
The results in Theorems~\ref{thm:mainb}--\ref{thm:maind} continue to hold if  the Gaussian noise distribution in the channel model \eqref{eq:channel} is replaced by an absolutely continuous noise distribution satisfying $h(W_1), ... , h(W_K)>-\infty$ and $H(\lfloor W_1 \rfloor), ... , H(\lfloor W_K\rfloor)<\infty$.
\end{theorem}
\begin{IEEEproof}
See Appendix~\ref{app:A}.
\end{IEEEproof}

Our Theorems~\ref{thm:mainb}--\ref{thm:mainc} above extend \cite[Thm.~4]{WSV11draft} from the scalar case to the vector case and to a broader class of noise distributions. 
Like the proof of \cite[Thm.~4]{WSV11draft},
the proof of  our Theorem~\ref{thm:main} builds on two results, whose extensions from the scalar to the vector case are provided next. We begin with the following multi-letter characterization of the vector IC sum-capacity. 
\vspace{.1cm}
\begin{proposition}\label{prop:limiting}
The sum-capacity of the channel \eqref{eq:channel}  is given by
\begin{align}	C_\text{sum}(\mathbf H; \snr) = \lim_{N\to\infty} \frac{1}{N} \sup_{\mathbf X_1^N,\ldots , \mathbf X_K^N}\sum_{i=1}^K I\lefto (\mathbf X_i^N; \mathbf Y_i^N \right ),	 \label{eq:limiting1}	\end{align}
where $\mathbf X_i^N= \left (X_{i}^{(1)}\, ... \; X_{i}^{(N)}\right )$ and the supremum is taken over all independent $\mathbf X_1^N, ... ,\mathbf X_K^N$ satisfying
\begin{align}	\frac{1}{MN} \sum_{m=1}^M\sum_{n=1}^N \mathbb E \lefto [\left(X_{i}^{(n)}[m] \right)^2\right ]\leqslant 1 ,	\label{eq:apc} \quad \text{for all $i$}.\end{align}
\end{proposition}
\begin{IEEEproof}
The proof is a straightforward extension of Ahlswede's limiting characterization of the sum-capacity of the discrete memoryless IC \cite[Sect.~2: Lem.~1]{Ahl71} to continuous alphabet channels under an average power constraint.
\end{IEEEproof}

For the scalar IC, Wu et al.\ \cite{WSV11draft} single-letterized \eqref{eq:limiting1} through an ingenious construction, which we extend to the vector case  in the proof of Theorem~\ref{thm:main}. We hasten to add that this extension is relatively straightforward.


The second result we shall need in the proof of Theorem~\ref{thm:main} relates information dimension to the high-$\snr$ asymptotics of mutual information in additive noise vector channels.

%
%
\vspace{.1cm}
\begin{theorem}\label{thm:guionnet}
Let $X$ and $W$ be  independent random vectors in $\mathbb R^n$ such that $W$ has an absolutely continuous distribution with $h(W)>-\infty$ and $H(\lfloor W\rfloor )<\infty$. Then
\begin{align}	\limsup_{\snr\to\infty}\frac{I(X;\sqrt{\snr} X+W)}{\frac{1}{2}\log\snr}=\udim(X).	 \label{eq:guionnet}	\end{align}
\end{theorem}
\begin{IEEEproof}
See Appendix~\ref{app:B}.
\end{IEEEproof}

\begin{remark}\label{rem:sufficient}
By Lemma~\ref{lem:linearbound} in Appendix~\ref{app:A} it follows that $\mathbb E[W^TW]<\infty$ implies $H(\lfloor W \rfloor )<\infty$. Finite second moment of $W$ and $h(W)>-\infty$ are therefore sufficient for the conditions in Theorem~\ref{thm:guionnet} to be satisfied.
\end{remark}

For $X$ and $W$ scalar, \eqref{eq:guionnet} was proved in \cite[Thm.~2.7]{GS07} under the following assumptions: $h(W)>-\infty$, $\mathbb E[\log (1+|W|)]<\infty$, and  $\mathbb E[\log (1+|X|)]<\infty$. This set of assumptions is  more restrictive than that in Theorem~\ref{thm:guionnet} as it includes a condition on $X$ as well and, in addition, $H(\lfloor W\rfloor)<\infty$ is slightly   weaker than $\mathbb E[\log (1+|W|)]<\infty$. In \cite[Thm.~9]{Wu11}, \eqref{eq:guionnet}  was shown in the scalar case for arbitrary $X$ and Gaussian  $W$. Theorem~\ref{thm:guionnet} above applies to the vector case and has weaker assumptions than both \cite[Thm.~9]{Wu11} and \cite[Thm.~2.7]{GS07}  as it does not impose restrictions on $X$ and also covers more general noise distributions $W$.   The proof of Theorem~\ref{thm:guionnet}, provided in Appendix~\ref{app:B}, proceeds by first showing the result for the case of uniform noise, similar to \cite{GS07},  and then uses the Ruzsa triangle inequality for differential entropy \cite[Thm.~3.1]{KM12} to generalize to noise $W$ satisfying $h(W)>-\infty$ and $H(\lfloor W\rfloor )<\infty$. Recognizing that the Ruzsa triangle inequality  for differential entropy can be applied considerably  simplifies the arguments  in \cite[Thm.~2.7]{GS07} (for the scalar case), which are based on explicit manipulations of the densities   involved.
\begin{remark}\label{rem:wu}
An alternative proof of Theorem~\ref{thm:guionnet} can be obtained as follows. As in the proof  provided in Appendix~\ref{app:B}, one starts by employing the
 Ruzsa triangle inequality for differential entropy to establish that 
\ban & | I(X;\sqrt{\snr} X+W)-I(X;\sqrt{\snr} X+Z) | \nonumber \\ &\leqslant  \max \{ I(Z;Z-W), I(W; Z-W) \}	,\label{eq:wu}	\ean
for  independent random vectors $X$, $W$, and $Z$, with $H(\lfloor X\rfloor )<\infty$,  $Z$  standard Gaussian, and $W$ of absolutely continuous distribution with $h(W)>-\infty$ and $H(\lfloor W\rfloor )<\infty$. By \cite[Thm.~6]{WV12a} it follows that $h(Z-W)<\infty$, which implies that the right-hand side (RHS) of \eqref{eq:wu} is  finite, and thus  dividing \eqref{eq:wu} by $\frac{1}{2}\log \snr$ and letting $\snr \to \infty$ yields zero. In particular, this means that the asymptote 
 on the left-hand side (LHS) of \eqref{eq:guionnet} remains unchanged if general noise $W$ (with $h(W)>-\infty$ and $H(\lfloor W\rfloor )<\infty$) is replaced by Gaussian noise $Z$.
The proof is then completed through  an extension of  \cite[Thm.~9]{Wu11} to the vector case.
While this route would lead to a shorter proof, the proof presented in Appendix~\ref{app:B} is more direct, completely self-contained, and
illuminates the link between  $I(X;\sqrt{\snr} X+W)$ for uniformly distributed $W$
and information dimension through its alternative definition in Lemma~\ref{lem:alternative}.
\end{remark}
We note that  the conditions $h(W)>-\infty$, $H(\lfloor W\rfloor)<\infty$ in Theorem~\ref{thm:guionnet} do not imply each other. Informally speaking, $h(W)>-\infty$ is a condition on the microscopic structure of the density of $W$ whereas $H(\lfloor W\rfloor)<\infty$ restricts the density in a global sense. 
The condition $h(W)>-\infty$  cannot be relaxed. To see this, suppose that $h(W)=-\infty$ and consider a random vector $X$ in $\mathbb R^n$ with
absolutely continuous distribution, and such that $h(X)>-\infty$ and $H(\lfloor X\rfloor)<\infty$ (these assumptions are satisfied, e.g., for Gaussian $X$). 
By  Proposition~\ref{prop:mixed1}, we have  $d(X)=n$. On the other hand, 
 $I(\sqrt{\snr}X+W;W)=h(\sqrt{\snr}X+W)- h(\sqrt{\snr} X)\geqslant 0$ and therefore $h(\sqrt{\snr}X+W)\geqslant  \frac{1}{2}\log \snr + h(X)>-\infty$ for all $\snr>0$, which implies
%
\begin{align*}		I(X;\sqrt{\snr}X+W)=h(\sqrt{\snr}X+W)-h(W)=\infty,	\end{align*}
and thus $\limsup_{\snr\to\infty}\frac{I(X;\sqrt{\snr} X+W)}{\frac{1}{2}\log\snr}=\infty$. This obviously violates \eqref{eq:guionnet}.





\section{Implications of the DoF-Formula}

We now show how  the results  in  Theorems~\ref{thm:mainb}--\ref{thm:maind} can be used to develop insights into the fundamental limits of interference alignment. Specifically, the DoF-formula  \eqref{eq:main} allows us to treat scalar \cite{MM09,EO09} and  vector interference alignment schemes \cite{CJ08,MMK08,Jaf11} in a unified fashion, and constitutes a tool for evaluating new interference alignment schemes both for the scalar and the vector case. 

We begin the discussion with an extension of \cite[Thm.~5]{WSV11draft} to vector ICs, which states that for almost all channel matrices no more than $K/2$ normalized DoF can be achieved. 


\vspace{.1cm}
\begin{proposition}\label{prop:upperbound}
Suppose that there is a permutation $\sigma$ of $\{1, ... , K\}$ such that $\sigma(i)\neq i$ and $\det \mathbf H_{ i,\sigma(i)}\neq 0$, for all $i$. Then, we have
\begin{align}	\frac{\Dof(\mathbf H)}{M} \leqslant \frac{K}{2}.	\end{align}
\end{proposition}
\begin{IEEEproof}
Let $R_1(\snr_k), ...,R_K(\snr_k)$ be achievable rates for  users $1,...,K$, respectively, and  consider a sequence $\snr_k\to \infty$ such that
\ban \lim_{k\to \infty} \frac{R_1(\snr_k)+ \ldots +R_K(\snr_k)}{\frac{1}{2}\log \snr_k}=	\Dof(\mathbf H). \label{eq:limsum}\ean
We set 
\ban d_i:=\limsup_{k\to \infty} \frac{R_i(\snr_k)}{\frac{1}{2}\log \snr_k} , \quad i=1,...,K.\label{eq:sumlim}\ean
A straightforward extension of the arguments in  \cite[Sect.~IV-A]{CJ08} to the vector case yields that 
 $d_i+d_j\leqslant M$ if  $\det \mathbf H_{i,j}\neq 0$. 

We thus have
\ban 2 \Dof(\mathbf H) &\leqslant 2\sum_{i=1}^{K}d_i \label{eq:erklaeren}\\
&=\sum_{i=1}^{K}d_i +\sum_{i=1}^{K}d_{\sigma(i)}  \\   &=\sum_{i=1}^{K}\underbrace{(d_i +d_{\sigma(i)})}_{\text{$\leqslant M$, by assumption}}\\ &\leqslant KM,\ean
where \eqref{eq:erklaeren} follows since the sum of the $\limsup$s \eqref{eq:sumlim} is larger than or equal to the limit of the sum in \eqref{eq:limsum}.
\end{IEEEproof}

\begin{remark}\label{rem:aa}
The conditions for Proposition~\ref{prop:upperbound} to hold are satisfied for almost all channel matrices $\mathbf H$. 
To see this first note that the determinant of a matrix is a polynomial in the entries of the matrix and the set of zeros of every non-trivial polynomial has Lebesgue measure zero. This allows us to conclude that for almost all $\mathbf H$ we have $\det \mathbf H_{i,j}\neq 0$, for all $i,j$.
Second, we can always find a permutation $\sigma$ with $\sigma(i)\neq i$, for $i=1,...,K$, as we assume (throughout the paper) that $K\geqslant 2$.
\end{remark}

For channel matrices $\mathbf H$ that satisfy the conditions in Proposition~\ref{prop:upperbound}, it follows from Theorem~\ref{thm:mainb} that \eqref{eq:main} holds if we can construct input distributions such that $\dof(X_1, ... , X_K;\mathbf H)/M=K/2$.

\subsection{Singular input distributions}
\label{ssec:sing}

For scalar ICs it was observed in \cite{WSV11draft} that input distributions with a singular component play a central role in achieving the supremum in \eqref{eq:main}. The natural extension of this result to the vector case holds, albeit with vastly different consequences as we shall see below. 

\vspace{.1cm}
\begin{proposition}\label{prop:sing}
Suppose $X_1, ... , X_K$ are  independent random vectors in $\mathbb R^M$, whose distributions are discrete-continuous mixtures\footnote{This means that $\gamma=0$ in the decomposition \eqref{eq:decomp}.} and are such that $H(\lfloor X_i\rfloor )<\infty$, for $i=1,...,K$. If $\det \mathbf H_{i,j}\neq 0$, for all $i,j$, then  
\begin{align}	\frac{\dof(X_1, ... , X_K;\mathbf H)}{M}	\leqslant 1.	\label{eq:sing} \end{align}
\end{proposition}
\begin{IEEEproof}
The proof is inspired by the proof of the corresponding result for the scalar case in \cite[Sect.~III-C]{WSV11draft}, and is detailed in the Online Addendum \cite[Sect.~III-C]{SB14extended}.
\end{IEEEproof}

The condition $\det \mathbf H_{i,j}\neq 0$, for all $i,j$, means that all transmit signals fully contribute to the  signal at each receiver. As explained in Remark~\ref{rem:aa}, this condition  is  satisfied for almost all $\mathbf H$.
Proposition~\ref{prop:sing} has far-reaching consequences as it says that restricting the transmit vectors to have distributions that do not contain a singular component, we can achieve no more than one normalized DoF with single-letter transmit vectors,\footnote{By Theorem~\ref{thm:main} single-letter inputs are DoF-optimal for almost all channels. This, however, leaves open the possibility  that on a set of channel matrices of (Lebesgue) measure zero multi-letter inputs are needed to achieve $\Dof(\mathbf H)$.} irrespectively  of the number of users and for almost all channels. This shows that the quantity $\dof(X_1, ... , X_K;\mathbf H)$ is highly sensitive to the nature of the input distributions, a surprising fact first discovered for scalar ICs in \cite{WSV11draft}. The implications of this result in the scalar case and in the vector case are, however, vastly different. In the scalar case singular input distributions are 
difficult to realize, or, more specifically, to approximate, 
as uncountable subsets of $\mathbb R$ with zero Lebesgue measure have a  complicated structure. While  Gaussian input distributions can  be efficiently  approximated by  discrete constellations   through shaping \cite{Ung02}, corresponding techniques for the approximation of singular scalar input distributions do not seem to be available in the literature.
%
In the case of vector ICs  the input symbols are vector-valued and a striking new feature appears. Singularity of the transmit symbols (vectors) can be realized by simply choosing input distributions that are (continuously) supported on lower-dimensional subspaces of the ambient signal space. These  subspaces are chosen to maximize $\dof(X_1, ... , X_K;\mathbf H)$, 
which is precisely the idea underlying interference alignment \`a la Cadambe and Jafar \cite{CJ08}, and Maddah-Ali et al.\ \cite{MMK08}. We illustrate this point by way of an example.


\begin{example}\label{ex:1}
Let $K=3$, $M=2$, and 
\begin{align*}	\mathbf H = \left (\begin{array}{cc:cc:cc}	1 & 0 & 1 & 0 & 1 & 0 \\ 1 & 1& 1 &1 &0 &1\\ \hdashline  1&0&1&0&1&0\\ 2&2&0&1&1&1\\ \hdashline  1&0&2&0&1&1\\ 0&1&0&1&0&1 \end{array}\right )	.	\end{align*}
Choose the transmit vectors as
\begin{align}
X_1:= \widetilde X_1 \lefto (\begin{matrix} 1 \\1 \end{matrix}\right ), \quad X_2:= \widetilde X_2 \lefto (\begin{matrix} 1 \\2 \end{matrix} \right ), \quad X_3:= \widetilde X_3 \lefto ( \begin{matrix} 1 \\3 \end{matrix} \right ),	\label{eq:inputs2a}
\end{align}
where $\widetilde X_1,\widetilde X_2$, and $\widetilde X_3$ are independent random variables with absolutely continuous distributions and $H(\lfloor \widetilde X_i \rfloor )<\infty$, for $i=1,2,3$. Note that the vectors $X_1,X_2,X_3$ are continuously distributed on lines in $2$-dimensional space and hence have distributions that are supported on sets of Lebesgue measure zero; this renders the input distributions singular. Therefore $\dof(X_1,X_2,X_3;\mathbf H)/2$ is not bound by \eqref{eq:sing}, even though $ \det \mathbf H_{i,j}\neq 0$, for $i,j=1,2,3$. Indeed, a simple calculation reveals that
\begin{align}
& \dof(X_1,X_2,X_3;\mathbf H)\label{eq:first}  \\  &=d\lefto ( \widetilde X_1\lefto (\begin{matrix} 1 \\ 2 \end{matrix}\right )+(\widetilde X_2 +\widetilde X_3) \lefto (\begin{matrix} 1 \\ 3 \end{matrix}\right ) \right ) - d\lefto ((\widetilde X_2 +\widetilde X_3) \lefto (\begin{matrix} 1 \\ 3\end{matrix}\right )\right )  \nonumber \\ & \phantom{=}\!\!  + d\lefto ( \widetilde X_2\lefto (\begin{matrix} 1 \\ 2 \end{matrix}\right ) +(\widetilde X_1 +\widetilde X_3) \lefto (\begin{matrix} 1 \\ 4 \end{matrix}\right )\right )-d\lefto ((\widetilde X_1 +\widetilde X_3) \lefto (\begin{matrix} 1 \\ 4 \end{matrix}\right )\right ) \nonumber \\ & \phantom{=} \!\!  +d\lefto (\widetilde X_3\lefto (\begin{matrix} 4 \\ 3 \end{matrix}\right )+ (\widetilde X_1 +2\widetilde X_2) \lefto (\begin{matrix} 1 \\ 1 \end{matrix}\right ) \right )-d\lefto ((\widetilde X_1 +2\widetilde X_2) \lefto (\begin{matrix} 1 \\ 1 \end{matrix}\right )\right ) \nonumber \\
&= 2-1 +2-1+2-1 = 3, \label{eq:lost}
\end{align}
and hence $\Dof(\mathbf H)/2\geqslant 3/2$. Here, we used \eqref{eq:change} and \eqref{eq:mixture} together with $d(\widetilde X_i +\alpha \widetilde X_j)=1$, for $i\neq j$ and $\alpha=1,2$, which follows from \eqref{eq:two} as $d(\alpha \widetilde X_j)=d(\widetilde X_j)=1$ by \eqref{eq:iso}. Since $ \det \mathbf H_{i,j}\neq 0$ for $i,j=1,2,3$, Proposition~\ref{prop:upperbound} implies that $\Dof(\mathbf H)/2\leqslant 3/2$, and thus the transmit vectors in \eqref{eq:inputs2a} achieve the supremum in \eqref{eq:main}. We emphasize that while the distributions of the transmit vectors in this example are singular, their marginals are not. What is more, the marginals have absolutely continuous distributions, as the $\widetilde X_i$ have absolutely continuous distributions. We could, e.g.,  take the $\widetilde X_i$ to be Gaussian which would render the marginals Gaussian. 
\end{example}

The reader can readily verify that the example above is in the spirit  of  interference alignment as put forward in \cite{CJ08, MMK08}. The underlying idea is to choose the inputs  continuously distributed in  $M/2$-dimensional subspaces and such that at each receiver the interference aligns within an $M/2$-dimensional subspace, while  the desired signal plus interference is continuously distributed in $\mathbb R^M$. We finally note that the transmission scheme discussed in this example only works for $M$ even.

\subsection{Classical interference alignment and parallel ICs}
Interference alignment as introduced by Cadambe and Jafar \cite{CJ08}  
was first applied to single-antenna ICs with channel coefficients that vary  across channel uses. The signal model in \cite{CJ08} differs from the one considered in the present paper as we take the channel coefficients to be constant across (vector-)channel uses.  We can, however, cast the signal model in \cite{CJ08} into the vector IC model used in this paper as follows. We take a finite block generated by the model in \cite{CJ08} and stack the corresponding input and output symbols into vectors whose dimension equals the blocklength. The submatrices $\mathbf H_{i,j}$ of the resulting channel matrix $\mathbf H$ are diagonal  and $\mathbf H$ is then assumed to be constant across (vector-)channel uses. 
We illustrate the procedure  by way of an example.
%
\begin{example}\label{ex:prev}
Consider  a time-varying $3$-user single-antenna IC whose channel matrix alternates between
 \begin{align} {\mathbf H}[1]=\begin{pmatrix} 1& 1 &-1 \\ -1 & 1 & 1\\ 1 &-1 & 1 \end{pmatrix} \;\; \text{and}	\;\; {\mathbf H}[2]=\begin{pmatrix} 1& -1 &1 \\ 1 & 1 & -1\\ -1 &1 & 1 \end{pmatrix}. \label{eq:encountered} \end{align}  
  As $\mathbf H[1]+\mathbf H[2]= 2 \mathbf I_2$, a simple transmit scheme where each  data symbol is sent  over two consecutive channel uses allows the receivers  interference-free access to the data symbols by simply adding the corresponding two consecutive receive symbols. Since each data symbol is repeated once, this scheme achieves $3/2$  normalized DoF.
Stacking the I/O relation induced by \eqref{eq:encountered} over blocks of length $2$ results in a vector IC of dimension $M=2$ with channel matrix
\begin{align}	\mathbf H = \left (\begin{array}{cc:cc:cc}	1 & 0 & 1 & 0 & -1 & 0 \\ 0 & 1& 0 &-1 &0 &1\\ \hdashline  -1&0&1&0&1&0\\ 0&1&0&1&0&-1\\ \hdashline  1&0&-1&0&1&0\\ 0&-1&0&1&0&1 
\end{array}\right )	.	\label{eq:bigmatrix}\end{align}
Reflecting that each data symbol is sent  over two consecutive channel uses, the corresponding transmit vectors are given by
\begin{align}
X_1:= \widetilde X_1 \lefto (\begin{matrix} 1 \\1 \end{matrix}\right ), \quad X_2:= \widetilde X_2 \lefto (\begin{matrix} 1 \\1 \end{matrix} \right ), \quad X_3:= \widetilde X_3 \lefto ( \begin{matrix} 1 \\1 \end{matrix} \right ),	\label{eq:inputs2}
\end{align}
where $\widetilde X_1,\widetilde X_2$, and $\widetilde X_3$ are independent random variables with absolutely continuous distributions and $H(\lfloor \widetilde X_i \rfloor )<\infty$, for $i=1,2,3$. 
A  calculation similar to the one in \eqref{eq:first}--\eqref{eq:lost} shows that $\dof(X_1,X_2,X_3;\mathbf H)/2 = 3/2$, in accordance with what was found above. 
\end{example}

The submatrices $\mathbf H_{i,j}$ in \eqref{eq:bigmatrix} are, indeed, all diagonal. This represents the structure of a parallel IC.
For general  parallel ICs with  $\mathbf H_{i,j}=\diag (h_{i,j}[1], ... , h_{i,j}[M] )$, $i,j=1,...,K$, we introduce the notation
\begin{align}
\mathbf H[m] := \begin{pmatrix} h_{1,1}[m] & \cdots & h_{1,K}[m]   \\[-.05cm]  \vdots &\ddots &\vdots \\ h_{K,1}[m] &\cdots &h_{K,K}[m]\end{pmatrix}, \quad m=1, ... ,M . \label{eq:parallelIC}
\end{align}
Here, $\mathbf H[m]$ is the interference matrix of the $m$-th (scalar) subchannel (as already encountered in \eqref{eq:encountered}), which acts on  $(X_1[m] \, ... \, X_K[m])^T$, i.e., the vector consisting of the $m$-th entry of each transmit vector. We note that for parallel ICs the conditions on $\mathbf H$ in Propositions~\ref{prop:upperbound} and \ref{prop:sing} are satisfied if all subchannel matrices $\mathbf H[m]$ are fully connected, i.e., $h_{i,j}[m]\neq 0$, for  $i,j=1,... ,K$, $m=1,..., M$, as this guarantees that
\ban \det \mathbf H_{i,j} = \prod_{m=1}^M h_{i,j}[m] \neq 0		 , \quad \text{$i,j=1,... ,K$}  \label{eq:dets} .\ean

Since parallel ICs describe the classical interference alignment setup \`a la  \cite{CJ08, MMK08}, it behooves us, in the next subsection, to particularize our main results to that case and to discuss the resulting implications. Besides recovering results known in the literature \cite{CJ08, CJ09}, we obtain a number of new insights. We emphasize, however, that the general vector IC setting is still highly relevant as MIMO ICs,  and general time-frequency-selective channels \cite{Dur11} usually lead to $\mathbf H_{i,j}$-matrices that are not diagonal.

\subsection{Parallel ICs}\label{sec:product}

An interesting question in the context of parallel ICs is that of separability.  Concretely, a parallel IC is said to be separable if its capacity region equals the (Minkowski) sum of the capacity regions of the individual subchannels. This means  that input distributions that are independent across subchannels are optimal for separable parallel ICs. Specifically, if a separable parallel IC is obtained from a purely time-selective channel as in Example~\ref{ex:prev}, then coding across time is not needed to achieve capacity. It was shown in \cite{CJ09} that  non-separable parallel ICs do exist. 
This phenomenon stands in stark contrast to the (Gaussian) multiple access and the (Gaussian) broadcast channel, both of which are always separable \cite{CJ09}.  

Here, we  shall deal with the weaker notion of separability in terms of DoF, i.e., the question of whether full DoF can be achieved by transmit vectors that are independent across subchannels. 
 The non-separability result in \cite{CJ09} was proved by showing that the channel with $K=3$, $M=2$, and the subchannel matrices in \eqref{eq:encountered}
is not separable. 
We discussed this very example in the previous subsection showing that, indeed,  transmit vectors with identical (and hence dependent) entries (cf.\  \eqref{eq:inputs2}) achieve full DoF. 

We turn to  analyzing the question of separability in more generality. First, we show how the  non-separability result in \cite{CJ09} (in terms of DoF) can be recovered through our DoF-formula.


\vspace{.1cm}
\begin{proposition}\label{prop:gain}
For a parallel IC with channel matrix $\mathbf H$ such  that \eqref{eq:main} is satisfied for $\mathbf H[m]$,  $m=1,...,M$, 
we have
\begin{align}	\Dof (\mathbf H ) \geqslant \sum_{m=1}^M\Dof (\mathbf H[m] ).	\label{eq:ineq}\end{align}
There exist cases where the inequality in \eqref{eq:ineq} is strict.
\end{proposition}
\begin{IEEEproof}
By Theorem~\ref{thm:mainb} we have $ \Dof (\mathbf H )\geqslant \sup_{X_1,...,X_K}  \dof(X_1, ... ,X_K ; \mathbf H)$ for all $\mathbf H$. The inequality in \eqref{eq:ineq} follows by restricting the supremization in \eqref{eq:main} to  vectors $X_i=(X_i[1]\, ... \, X_i[M])^T$ which, in addition to satisfying the conditions  in the supremization, have independent entries (across $m=1,...,M$) and for which all information dimension terms appearing in the expression \ban &\dof(X_1[m] , ... , X_K[m] ; \mathbf H[m]) \nonumber \\ &= \sum_{i=1}^K \left [ d\lefto (\sum_{j=1}^K h_{i,j}[m] X_j[m]  \right )-	 d \lefto ( \sum_{j\neq i}^K h_{i,j}[m] X_j[m] \right ) \right ] \ean exist, for $m=1, ... ,M$. 
Evaluating  $\dof(X_1 , ... , X_K ; \mathbf H)$ for this class of inputs yields
\begin{align}		&\dof(X_1 , ... , X_K ; \mathbf H) \label{eq:bythm2}\\ &= \sum_{i=1}^K\vast [ d\lefto (\begin{matrix} \sum_{j=1}^K h_{i,j}[1] X_j[1]  \\[-.05cm] \vdots \\ \sum_{j=1}^K h_{i,j}[M] X_j[M]  \end{matrix} \right )  \\ &\quad\quad\quad -	 d \lefto ( \begin{matrix} \sum_{j\neq i}^K h_{i,j}[1] X_j[1]  \\[-.05cm]  \vdots \\ \sum_{j\neq i}^K h_{i,j}[M] X_j[M]  \end{matrix} \right )\vast ]	
\end{align}
\begin{align}
&\stackrel{\eqref{eq:sumind}}= \sum_{m=1}^M
		 \sum_{i=1}^K \Biggg [ d\lefto (\sum_{j=1}^K h_{i,j}[m] X_j[m]  \right ) \\ &\quad\quad\quad\quad\quad  -	 d \lefto ( \sum_{j\neq i}^K h_{i,j}[m] X_j[m] \right ) \Biggg ] \label{eq:immediate}\\ &= \sum_{m=1}^M\dof(X_1[m] , ... , X_K[m] ; \mathbf H[m])  .	\label{eq:bythm}\end{align} 
Taking the supremum in \eqref{eq:bythm2}-\eqref{eq:bythm} over the class of input distributions specified above proves \eqref{eq:ineq}.

 A class of examples that renders the inequality in \eqref{eq:ineq} strict, and is different from the example reported in \cite{CJ09}, is obtained as follows. Take $K=3$ and $M=2$ with
\begin{align} {\mathbf H}[m]=\begin{pmatrix} 1& 0 & 0\\ 1 & \lambda[m] & 0\\ 1 &1 & 1 \end{pmatrix},	\quad m=1,2 ,\label{eq:prop6a}\end{align} 
	where $\lambda[1]$ and $\lambda[2]$ are nonzero, rational, and satisfy $\lambda[1]\neq\lambda[2]$. From \cite[Thm.~11]{WSV11draft}, we get $\Dof({\mathbf H}[m])<3/2$, for $m=1,2$. Now, consider the transmit vectors \begin{align}
X_1:= \widetilde X_1 \lefto (\begin{matrix} 1 \\1 \end{matrix}\right ), \quad X_2:= \widetilde X_2 \lefto (\begin{matrix} 1 \\1 \end{matrix}\right ), \quad X_3:= \widetilde X_3 \lefto (\begin{matrix} 1 \\0 \end{matrix}\right ),	\label{eq:inputs}
\end{align}
where $\widetilde X_1,\widetilde X_2,\widetilde X_3$ are independent random variables with $d(\widetilde X_i)=1$, for $i=1,2,3$. Note that the transmit vectors $X_1$ and $X_2$ have dependent, in fact identical, entries, which implies that we perform joint coding across subchannels. Inserting into \eqref{eq:dofallexist}, we find
\begin{align}
\nonumber &\dof(X_1,X_2,X_3;\mathbf H)\\ \nonumber    \,\;\; &= d\lefto ( \widetilde X_1\lefto (\begin{matrix} 1 \\ 1 \end{matrix}\right ) \right ) - d\lefto (\lefto (\begin{matrix} 0 \\ 0 \end{matrix}\right )\right ) \\ \nonumber & \phantom{=} + d\lefto ( \widetilde X_2\lefto (\begin{matrix} \lambda[1] \\ \lambda[2] \end{matrix}\right ) +\widetilde X_1 \lefto (\begin{matrix} 1 \\ 1 \end{matrix}\right )\right )-d\lefto (\widetilde X_1  \lefto (\begin{matrix} 1 \\ 1 \end{matrix}\right )\right )\\ \nonumber &\phantom{=}   +d\lefto (\widetilde X_3\lefto (\begin{matrix} 1 \\ 0 \end{matrix}\right )+ (\widetilde X_1 +\widetilde X_2) \lefto (\begin{matrix} 1 \\ 1 \end{matrix}\right ) \right )-d\lefto ((\widetilde X_1 +\widetilde X_2) \lefto (\begin{matrix} 1 \\ 1 \end{matrix}\right )\right ) \\ &\hspace{-.3cm}\stackrel{\eqref{eq:change},\eqref{eq:mixture}}= 1-0 +2-1+2-1 = 3, \label{eq:verweisen}
\end{align}
where we used  $d(\widetilde X_i + \widetilde X_j)=1$, for $i\neq j$, cf.~\eqref{eq:two}.
Applying Theorem~\ref{thm:mainb}, we therefore find that 
	\begin{align}	\frac{\Dof (\mathbf H )}{2}\stackrel{\eqref{eq:verweisen}}{\geqslant} \frac{3}{2}> \frac{\Dof({\mathbf H}[1])+\Dof({\mathbf H}[2])}{2}	.	\label{eq:explicit} \end{align}
\end{IEEEproof}

The channel matrices in \eqref{eq:prop6a} constitute an entire family of parallel ICs that are non-separable. We next show  that, however, almost all parallel ICs are separable (in terms of DoF).  This means that for almost all parallel ICs, we can achieve full DoF without coding  across subchannels or, in the setting of \cite{CJ08}, across time or frequency.


\vspace{.1cm}
\begin{proposition}\label{prop:sep}
For almost all parallel ICs, we have
\ban 	\Dof (\mathbf H ) = \sum_{m=1}^M\Dof (\mathbf H[m] ).	\label{eq:propind}\ean
\end{proposition}
\begin{IEEEproof}We will make repeated use of the fact that if two different properties hold individually   for almost all channel matrices, then they also hold simultaneously for almost all channel matrices. We may assume that all subchannel matrices $\mathbf H[m]$ are fully connected, since this holds for almost all parallel ICs.   Then, based on \eqref{eq:dets} we can apply Proposition~\ref{prop:upperbound} and find together with Proposition~\ref{prop:gain}  that
\begin{align}	\frac{\Dof(\mathbf H[1])+\ldots + \Dof(\mathbf H[M])}{M}\leqslant \frac{\Dof(\mathbf H)}{M}\leqslant \frac{K}{2},	\label{eq:sandwich}	\end{align}
for almost all parallel ICs.
Since we know that $\Dof (\mathbf H[m])=K/2$ for almost all $\mathbf H[m]$ \cite[Thm.~1]{MM09}, for $m=1,...,M$, it follows that for almost all parallel ICs we have $\Dof (\mathbf H[m])=K/2$, for $m=1,...,M$, as the cartesian product of sets of full (Lebesgue) measure (i.e., their complement has (Lebesgue) measure $0$) again has full (Lebesgue) measure.
From  \eqref{eq:sandwich} we thus obtain that $\Dof(\mathbf H)/M=K/2$ for almost all parallel ICs.
\end{IEEEproof}
The RHS  of \eqref{eq:propind} is achieved by transmit vectors with independent entries (across subchannels), where the individual entries are taken to achieve $\Dof(\mathbf H[m])$, for $m=1,..., M$.
This means that for almost all parallel ICs full DoF can be achieved without coding across subchannels. The downside  is, however, that the distributions of the scalar entries of the corresponding full DoF-achieving transmit vectors must have a singular component as each subchannel constitutes a scalar IC. 
If we allow joint coding across subchannels, on the other hand, there are ICs where full DoF can be achieved even with transmit vectors whose entries have distributions that do not contain a singular component as the example class in the proof of Proposition~\ref{prop:gain} shows. Specifically, the input vectors in \eqref{eq:inputs} have absolutely continuous scalar entries. The same interpretation applies to the example in  \eqref{eq:encountered} with the transmit vectors in \eqref{eq:inputs2}.





We  investigate the separability of parallel ICs and the nature of full DoF-achieving input distributions further by studying a parallel IC with $K=3$, $M=2$, and fully connected subchannel matrices $\mathbf H[1]$ and $\mathbf H[2]$. The fully connected scalar IC (i.e., $M=1$)  was studied in \cite{MM09}, where it was shown that $\Dof (\mathbf H )=K/2$ ($=3/2$ in this example)  for almost all channel matrices. In the following, we identify explicit conditions on $\mathbf H[1]$ and $\mathbf H[2]$ for $\Dof(\mathbf H)/M=K/2=3/2$ as well as conditions for $\Dof(\mathbf H)/M<3/2$. Note that Proposition~\ref{prop:sep} and the result in \cite{MM09} imply that $\Dof(\mathbf H)/M=3/2$ for almost all parallel $3$-user ICs with $2$ subchannels. 
We begin by stating a   result on the scaling invariance of $\Dof(\mathbf H)$, frequently used below.
\vspace{.1cm}

\begin{lemma}\label{lem:scaleinv}
Let $\mathbf D_1 =\diag (\mathbf D_{1,1}, ...  , \mathbf D_{1,K})$, $\mathbf D_2 =\diag (\mathbf D_{2,1}, ... , \mathbf D_{2,K})$, with $\mathbf D_{i,j}\in \mathbb R^{M\times M}$ and  $\det \mathbf D_{i,j} \neq 0$, for $i=1,2$ and $j=1,...,K$. Then, we have 
\ba 	\Dof(\mathbf H) = \Dof(\mathbf D_1 \mathbf H \mathbf D_2).		\ea
In particular, $\Dof(\mathbf H)$ is invariant with respect to scaling of a row or a column by a nonzero constant.
\end{lemma}
\begin{IEEEproof}
The proof follows the same line of arguments as the proof of the corresponding result for the scalar case \cite[Lem.~1]{EO09}. 
\end{IEEEproof}
Back to our example, we first note that, thanks to the scaling invariance in Lemma~\ref{lem:scaleinv}, it suffices to consider the ``standard $3$-user IC matrices'' from \cite[Sec.~VI-B]{MM09} 
\begin{align}	\mathbf H[m]=\begin{pmatrix} a[m] & 1 & 1\\ 1 & b[m] & 1\\ 1 &d[m] & c[m] \end{pmatrix}, \quad m=1,2, 	\label{eq:standard}\end{align}
where $a[m]$, $b[m]$, $c[m]$, and $d[m]$ are nonzero for $m=1,2$. 
 Since $\mathbf H[1]$ and $\mathbf H[2]$ are fully connected, we have $\det \mathbf H_{i,j}=\prod_{m=1}^2 h_{i,j}[m]\neq 0$, for all $i,j$, which, by Proposition~\ref{prop:upperbound}, implies that \ban \Dof(\mathbf H)/2\leqslant 3/2. \label{eq:always2}\ean 
\begin{enumerate}
	\item If, for $m=1,2$, $d[m]$ is rational and $a[m],b[m],c[m]$ are all irrational, then by \cite[Thm.~6]{WSV11draft} \eqref{eq:main} is satisfied for $\mathbf H[m]$ in place of $\mathbf H$, for $m=1,2$, and $\Dof (\mathbf H[1])=\Dof (\mathbf H[2])=3/2$, which combined with Proposition~\ref{prop:gain} and \eqref{eq:always2} yields
	\begin{align}	\frac{\Dof (\mathbf H )}{2}=\frac{3}{2}	. 	\label{eq:ex2a}\end{align}
Therefore, $\Dof (\mathbf H )=\Dof (\mathbf H[1])+\Dof (\mathbf H[2])$, which shows that  full DoF are achieved without coding across subchannels.
	\item If $d[1]=d[2]$ and $a[1]\neq a[2]$, $b[1]\neq b[2]$, $c[1]\neq c[2]$, a calculation similar to the one in Example~\ref{ex:1} (see \cite[Sect.~III-D]{SB14extended}) reveals that
	$ \dof(X_1,X_2,X_3;\mathbf H)= 3$ for independent transmit vectors each of which is continuously distributed along the direction $\begin{pmatrix}1 \\1 \end{pmatrix}$. Together with \eqref{eq:always2}, we hence get   
	\begin{align}	\frac{\Dof (\mathbf H )}{2}=\frac{3}{2}	.	\label{eq:ex2b}\end{align}
	This is achieved by joint coding across subchannels. Specifically, each transmit vector has  scalar entries that are dependent, in fact identical,   across subchannels,  which corresponds to repetition coding. Note that, here, to achieve full DoF we do not need the distributions of  the scalar entries of the transmit vectors to contain singular components.

	\item If all entries of the subchannel matrices are rational  and one of the entries on the main diagonal in \eqref{eq:standard} is identical across the two subchannels, then
	\begin{align}	\frac{\Dof (\mathbf H )}{2}<\frac{3}{2}	,\label{eq:ex2c}	\end{align}
	i.e., we have strictly less than $3/2$ normalized DoF.
	This follows from an extension of \cite[Thm.~8]{WSV11draft} to the vector case, detailed in the Online Addendum \cite[Sect.~III-E]{SB14extended}.
%
\end{enumerate}

To further demonstrate what we can get out of our DoF-formula,  we next analyze an example that was studied in \cite[Sect.~III-C]{CJ08}. Specifically, we recover the statement that $4/3$ normalized DoF can be achieved for almost all  $3$-user parallel ICs with $3$ subchannels using joint coding across subchannels.  


\begin{example}
We take $K=3$, $M=3$, and again assume that the subchannel matrices $\mathbf H[m]$ are in standard form as in \eqref{eq:standard}. We further assume that the channel coefficients $a[m],b[m],c[m],d[m]$, $m=1,2,3$, are chosen randomly  with respect to a joint (across $a,b,c,d,$ and $m$) absolutely continuous distribution. We take the transmit vectors as
\begin{align} \nonumber   X_1:= \widetilde X_{1,1} \begin{pmatrix} 1\\ 1 \\ 1\end{pmatrix}+ \widetilde X_{1,2} \begin{pmatrix} d[1]\\ d[2] \\ d[3]\end{pmatrix}, \quad  X_2:=\widetilde X_2 \begin{pmatrix} 1 \\ 1 \\ 1\end{pmatrix},  \\ \quad \text{and} \quad   X_3:= \widetilde X_3 \begin{pmatrix}1\\ 1 \\ 1\end{pmatrix}, \label{eq:transmit} \end{align}
where $\widetilde X_{1,1}, \widetilde X_{1,2}, \widetilde X_2$, and $\widetilde X_3$ are independent random variables with absolutely continuous distributions and $H(\lfloor \widetilde X_{1,1} \rfloor ),H(\lfloor \widetilde X_{1,2} \rfloor ),H(\lfloor \widetilde X_{2} \rfloor ),H(\lfloor \widetilde X_{3} \rfloor )<\infty$.
We then find \eqref{eq:as}--\eqref{eq:as3} displayed at the top of the next page, 
\setcounter{equation}{70}
where \eqref{eq:as2} holds almost surely with respect to the channel coefficients $a[m],b[m],c[m],d[m]$, as a consequence of  each of the sets \ban  \nonumber  \left \{ \begin{pmatrix} a[1]\\ a[2] \\ a[3]\end{pmatrix} , \begin{pmatrix} a[1]d[1]\\ a[2]d[2] \\ a[3]d[3]\end{pmatrix}, \begin{pmatrix}1\\1\\1 \end{pmatrix}\right \},    \\ \nonumber   \left \{ \begin{pmatrix}1\\1\\1 \end{pmatrix}, \begin{pmatrix} d[1]\\ d[2] \\ d[3]\end{pmatrix} , \begin{pmatrix} b[1]\\ b[2] \\ b[3]\end{pmatrix} \right \} ,  \nonumber \;\;\;    \left \{ \begin{pmatrix}1\\1\\1 \end{pmatrix}, \begin{pmatrix} d[1]\\ d[2] \\ d[3]\end{pmatrix} , \begin{pmatrix} c[1]\\ c[2] \\ c[3]\end{pmatrix} \right \} \ean being  linearly independent almost surely. In \eqref{eq:as3} we used \eqref{eq:two}. The transmit vectors in \eqref{eq:transmit} therefore achieve $4/3$ normalized DoF almost surely (with respect to the channel coefficients). 
\end{example}


%

\begin{figure*}[!t]
\normalsize
\setcounter{MYtempeqncnt}{\value{equation}}
\setcounter{equation}{66}
\begin{align}
\dof( X_1, X_2 , X_3 ; \mathbf H)    &= 
 d\lefto (\widetilde X_{1,1}\begin{pmatrix} a[1]\\ a[2] \\ a[3]\end{pmatrix}  \! +  \widetilde X_{1,2} \begin{pmatrix} a[1]d[1]\\ a[2]d[2] \\ a[3]d[3]\end{pmatrix} \! + (\widetilde X_2 + \widetilde X_3) \begin{pmatrix} 1 \\ 1 \\ 1\end{pmatrix} \right )  \! -		 d \lefto ( \! (\widetilde X_2+\widetilde X_3) \begin{pmatrix} 1 \\ 1 \\ 1\end{pmatrix} \right ) \nonumber  \\
& \phantom{=} 
+ d\lefto (\!  (\widetilde X_{1,1}+\widetilde X_3)\! \begin{pmatrix} 1\\ 1 \\ 1\end{pmatrix} \! +\widetilde X_{1,2} \begin{pmatrix} d[1]\\ d[2] \\ d[3]\end{pmatrix} \! +\widetilde X_2 \begin{pmatrix} b[1] \\ b[2] \\ b[3]\end{pmatrix} \right ) \! -	 d \lefto ( \!  (\widetilde X_{1,1}+\widetilde X_3) \! \begin{pmatrix} 1\\ 1 \\ 1\end{pmatrix} \! +\widetilde X_{1,2} \begin{pmatrix} d[1]\\ d[2] \\ d[3]\end{pmatrix}\right ) \nonumber \\
&\phantom{=}
 + d\lefto ( \widetilde X_{1,1}\begin{pmatrix} 1\\ 1 \\ 1\end{pmatrix} \! + (\widetilde X_{1,2}+\widetilde X_2)\! \begin{pmatrix} d[1]\\ d[2] \\ d[3]\end{pmatrix} \! +\widetilde X_3 \begin{pmatrix} c[1]\\ c[2] \\ c[3]	\end{pmatrix}\right ) \! -	  d \lefto (\widetilde X_{1,1}\begin{pmatrix} 1\\ 1 \\ 1\end{pmatrix} \! +(\widetilde X_{1,2} +\widetilde X_2)\! \begin{pmatrix} d[1]\\ d[2] \\ d[3]\end{pmatrix}  \right )  \label{eq:as} \\
&\stackrel{\eqref{eq:iso}}=d\lefto (\begin{pmatrix} \widetilde X_{1,1}\\ \widetilde X_{1,2} \\\widetilde X_2+\widetilde X_3\end{pmatrix}\right ) -d\lefto ( \widetilde X_2+\widetilde X_3\right )+d\lefto (\begin{pmatrix} \widetilde X_{1,1}+\widetilde X_3\\ \widetilde X_{1,2} \\ \widetilde X_2\end{pmatrix}\right ) -d\lefto (\begin{pmatrix} \widetilde X_{1,1} + \widetilde X_3\\ \widetilde X_{1,2}  \nonumber  \end{pmatrix}\right )\\ &\phantom{=} +d\lefto (\begin{pmatrix} \widetilde X_{1,1}\\ \widetilde X_{1,2}+\widetilde X_2 \\ \widetilde X_3\end{pmatrix}\right ) -d\lefto (\begin{pmatrix} \widetilde X_{1,1} \\ \widetilde X_{1,2}  +\widetilde X_2 \end{pmatrix}\right )    \label{eq:as2} \\&\stackrel{\eqref{eq:sumind}}=3 -1 +3 -2 +3-2 = 4 ,\label{eq:as3}
\end{align} 
\setcounter{equation}{\value{MYtempeqncnt}}
\hrulefill
\vspace{-.6cm}
\end{figure*}

\subsection{Verifying the MIMO interference alignment conditions in \cite{BCT11}} \label{ssec:MIMO}



We now show how our general DoF-formula can be used to verify the interference alignment conditions for MIMO ICs with constant channel matrices analyzed in \cite[Eqs.~(3)--(4)]{BCT11}. 
For  simplicity of exposition, we   assume $M$ antennas at each transmitter  and receiver. The interference alignment  conditions in \cite[Eqs.~(3)--(4)]{BCT11} are summarized as follows. To achieve $\ell$ DoF, find pairs of subspaces $(\mathcal U_1,\mathcal V_1), ... , (\mathcal U_K,\mathcal V_K)$ of $\mathbb R^M$ with $d_i:=\dim\mathcal U_i=\dim \mathcal V_i$ and $\sum_{i=1}^K d_i=\ell$, such that 
\begin{align}	\mathbf H_{i,j}\, \mathcal U_j &\subseteq  \mathcal V_i^\perp ,  \quad \text{ for all $i\neq j$,}		\label{eq:feasibility1} \\
			\dim(\pi_{\mathcal V_i}(\mathbf H_{i,i} \, \mathcal U_i )) &=\dim(\mathcal U_i),	\quad  \text{ for all $i=1,... ,K$},	\label{eq:feasibility2} \end{align}
where $\pi_{\mathcal V_i}$ denotes the orthogonal projection operator onto $\mathcal V_i$.
These pairs of subspaces are associated with a transmit-receive scheme as follows. 
We choose  $X_i$ to be continuously distributed in the subspace $\mathcal U_i$ (i.e., the distribution of $X_i$ has a density supported on $\mathcal U_i$) and such that $H(\lfloor X_i \rfloor )<\infty$. We furthermore take $X_1,...,X_K$ to be independent. The $i$-th receiver computes the orthogonal projection  onto $\mathcal V_i$.  By \eqref{eq:feasibility1} this results in  interference-free signals at all receivers and \eqref{eq:feasibility2} guarantees that in the process the desired signal at the $i$-th receiver does not experience  dimensionality reduction.


Next, we show how our achievability result \eqref{eq:mainneu} can be used to verify that the transmit-receive scheme summarized above, indeed, achieves $\ell$ DoF. To this end, we first organize the elements of a basis for $\mathcal V_i$ into the  first $d_i$ columns of an $M\times M$ matrix $\mathbf V_i= ( v_1 , ... , v_{d_i} , \tilde v_{1}, ..., \tilde v_{M-d_i})$ with the remaining columns consisting of a basis for the $(M-d_i)$-dimensional orthogonal complement of $\mathbf H_{i,i}\,  \mathcal U_i$. Note that by \eqref{eq:feasibility2} multiplication of   $\mathcal U_i$ by $\mathbf H_{i,i}$  must necessarily be  dimensionality-preserving, i.e., 
 $\mathbf H_{i,i}\,  \mathcal U_i$ must  be $d_i$-dimensional. 
We now show that $\det \mathbf V_i\neq 0$. To this end, suppose that $v\in\mathbb R^M$ lies in the orthogonal complement of the column space of $\mathbf V_i$. Then, $v$ has to be orthogonal to both $\mathcal V_i$ and the orthogonal complement of $\mathbf H_{i,i}\, \mathcal U_i$, and it follows that $\pi_{\mathcal V_i}(v)=0$ and  $v\in \mathbf H_{i,i}\,  \mathcal U_i$. But by \eqref{eq:feasibility2} this can only hold for $v=0$ as the application of $\pi_{\mathcal V_i}$ to the subspace $\mathbf H_{i,i}\,  \mathcal U_i$ can not result in dimensionality reduction.
The orthogonal complement of the column space of $\mathbf V_i$ is therefore trivial, and we can conclude that  $\det \mathbf V_i\neq 0$ for all $i=1,...,K$. 
This yields
\begin{align}	\nonumber	&\dof(X_1, ... ,X_K ; \mathbf H)	\\ &= \sum_{i=1}^K \left [ d\lefto (\sum_{j=1}^K \mathbf H_{i,j} X_j \right )-	 d \lefto ( \sum_{j\neq i}^K \mathbf H_{i,j} X_j \right )\right ]\\
										&\stackrel{\eqref{eq:iso}}=\sum_{i=1}^K \Bigg [d\Bigg (\sum_{j=1}^K \mathbf V_i^T \mathbf H_{i,j} X_j \Bigg )-	 d \Bigg ( \sum_{j\neq i}^K \mathbf V_i^T \mathbf H_{i,j} X_j \Bigg )\Bigg ] \label{eq:appiso}\\ 
&\stackrel{\eqref{eq:feasibility1}}=\sum_{i=1}^K \Vvast [d\begin{pmatrix} v_1^T\mathbf H_{i,i}X_i\\[-.1cm]  \vdots \\v_{d_i}^T\mathbf H_{i,i}X_i\\ \tilde v_1^T\sum_{j\neq i}^K\mathbf H_{i,j}X_j\\[-.05cm]  \vdots \\ \tilde v_{M-d_i}^T\sum_{j\neq i}\mathbf H_{i,j}X_j \end{pmatrix} \nonumber \\ & \hspace{3cm}-	 d\begin{pmatrix} 0\\[-.15cm]  \vdots \\0 \\ \tilde v_1^T\sum_{j\neq i}^K\mathbf H_{i,j}X_j\\[-.05cm]  \vdots \\\tilde v_{M-d_i}^T\sum_{j\neq i}\mathbf H_{i,j}X_j \end{pmatrix} \Vvast ]\\
								&\stackrel{\eqref{eq:sumind}}=\sum_{i=1}^K  d\!\begin{pmatrix} v_1^T\mathbf H_{i,i}X_i\\[-.05cm]  \vdots \\v_{d_i}^T\mathbf H_{i,i}X_i \end{pmatrix} 	\label{eq:unklar2} \\
										 &=\sum_{i=1}^K  d(\mathbf V_i^T \mathbf H_{i,i}X_i)	\label{eq:unklar0}\\
										&\stackrel{\eqref{eq:iso} }=\sum_{i=1}^K d_i \label{eq:unklar} \\
										&=\ell,	
\end{align}
where in the application of \eqref{eq:iso} in \eqref{eq:appiso} we used the fact that the map $\mathbf V_i$ is bi-Lipschitz as a consequence of $\det \mathbf V_i\neq 0$. Furthermore, in the application of \eqref{eq:iso}  in \eqref{eq:unklar} we used that $d(X_i)=d_i$ by Proposition~\ref{prop:mixed1} together with the fact that $\mathbf V_i^T\mathbf H_{i,i}$ induces a  bi-Lipschitz map on $\mathcal U_i$ since $\det \mathbf V_i\neq 0$ and   $\mathbf H_{i,i}$ acts isomorphically on $\mathcal U_i$ by \eqref{eq:feasibility2}.


Conditions on the existence of  pairs of subspaces that satisfy  \eqref{eq:feasibility1} and \eqref{eq:feasibility2} can be expressed in terms of  algebraic equations, which can be studied using methods from algebraic geometry, see \cite{BCT11}, \cite{Jaf11}. 
In particular, it is shown in \cite[Corollary~8]{BCT11} that if $d_1=\ldots = d_K$, then for almost all MIMO ICs no more than $\ell/M=2$ normalized DoF can be achieved with the interference alignment scheme described in this subsection.

\subsection{The complex case}
\label{sec:complex}
The results stated so far apply to real signals and channel matrices. We next describe how  Theorems~\ref{thm:mainb} and \ref{thm:main} can be extended to the complex case, thereby also providing an extension of the main result in \cite{WSV11draft} to the complex case. 

For a vector IC with transmit and receive signals in $\mathbb C^M$ and $\mathbf H_{i,j}\in\mathbb C^{M\times M}$, for $i,j=1,...,K$, simply stack the real and imaginary parts of the transmit, receive, and noise vectors in \eqref{eq:channel}, and stack the real and imaginary parts of the matrices $\mathbf H_{i,j}$ correspondingly according to
\ban \begin{pmatrix} \Real (\mathbf H_{i,j}) & - \Imag(\mathbf H_{i,j})\\ \Imag(\mathbf H_{i,j}) & \Real (\mathbf H_{i,j})\label{eq:complexstructure}\end{pmatrix}.\ean 
Then  apply Theorem~\ref{thm:mainb} 
to the resulting vector IC in $\mathbb R^{2M}$ and divide the  so obtained number of DoF by $2$ to account for the fact that we would like to think in terms of complex dimensions (degrees of freedom) for the complex IC.
As the scalar complex IC is turned into a $2$-dimensional real vector IC, we see that having a characterization of the DoF for vector ICs constitutes the basis for extending the main result in \cite{WSV11draft} to the complex case. The DoF-formula in Theorem~\ref{thm:main} can analogously  be shown  to hold for
 the complex case. The achievability part of the corresponding proof is straightforward,  the converse requires certain modifications taking into account  
 the specific structure of the effective channel matrix in \eqref{eq:complexstructure}. For brevity, we do not present  the   proof of this extension.

The  extensions of Theorems~\ref{thm:mainb} and~\ref{thm:main} to the complex case  allow us to treat  interference alignment  schemes for  complex channel matrices $\mathbf H$. This is illustrated next  by analyzing  the following well-known example using Theorem~\ref{thm:mainb}.

\begin{example}[{\cite{CJ08}}]\label{ex:dft}
We consider a cyclic variant of the example in \cite[App.~1]{CJ08}, whose main idea is to exploit channel-induced differences in  propagation delays to perform interference alignment. Let $M$ be even, take $K\geqslant 3$, and consider the  complex channel with I/O relation 
\begin{align}	Y_{k}[m] = 	\sqrt{\snr}\left (X_{k}[m] + \sum_{\ell\neq k}^K X_\ell[m-1]\right ) + W_k[m],	\label{eq:iocomplex} \end{align}
for $k=1, ... ,K$, $m=1,... , M$, where the argument in square brackets, e.g., in $X_k[m]$,  is to be understood modulo $M$. The corresponding channel matrices are given by $\mathbf H_{k,k}=\mathbf I_{M}$, and
\ba  \mathbf H_{k,\ell}=\begin{pmatrix}	0 & \cdots& \cdots & 0 & 1\\
											1 & 0 & \cdots &\cdots & 0\\[-.05cm] 
											0 & \ddots & \ddots &  &\vdots\\[-.05cm] 
											\vdots &\ddots & \ddots &\ddots &\vdots\\
											0 & \cdots  &0 & 1 &0 	\end{pmatrix}, \quad k\neq \ell, \quad k,\ell=1,...,K	.\ea

Denoting the $m$-th  vector of the standard basis in $\mathbb R^M$ by $e_m$, we choose the transmit vectors $X_k$ to be continuously distributed in the subspace spanned by $\{e_m \! \mid \! 1\leqslant m \leqslant M,\; \text{$m$ even}\}$ such that $H(\lfloor X_k \rfloor)<\infty$. This accounts for the fact that the transmitters remain silent in odd time slots. The vectors $\mathbf H_{k,\ell}X_\ell$,  $k\neq \ell$, are then continuously distributed in the subspace spanned by $\{e_m \! \mid \!   1\leqslant m \leqslant M,\; \text{$m$ odd}\}$, which reflects the fact that interference at each receiver is confined to odd time slots. 
Inserting into \eqref{eq:dofallexist}, we find that
\ba 	 &\dof(X_1,...,X_K;\mathbf H)\\ &=\sum_{k=1}^K \left [ d\lefto (\sum_{\ell=1}^K \mathbf H_{k,\ell} X_\ell \right )-	 d \lefto ( \sum_{\ell\neq k}^K \mathbf H_{k,\ell} X_\ell \right )\right ]\\ &=\sum_{k=1}^K \left[M - \frac{M}{2}\right ]\\ &=KM/2,	\ea
where we used $d\lefto (\sum_{\ell=1}^K \mathbf H_{k,\ell} X_\ell \right )= d \lefto ( \mathbf H_{k,k}X_k + \sum_{\ell\neq k}^K \mathbf H_{k,\ell} X_\ell \right ) =M$ and  $d \lefto ( \sum_{\ell\neq k}^K \mathbf H_{k,\ell} X_\ell \right )=M/2$, both thanks to \eqref{eq:change}.
 Moreover, since $\det \mathbf H_{k,\ell}\neq 0$, for all $k,\ell$, we can apply\footnote{In fact, we apply the extension of Proposition~\ref{prop:upperbound}---obtained mutatis mutandis---to  complex ICs.}  Proposition~\ref{prop:upperbound} to find that $\Dof (\mathbf H )/M \leqslant K/2$ and therefore
\begin{align}		\frac{\Dof (\mathbf H )}{M} = \frac{K}{2}.	\label{eq:fourier}\end{align}
This recovers the result in \cite[App.~1]{CJ08}, showing that the scheme above achieves full DoF.

The example can equivalently be analyzed in the frequency domain. We pursue this in the following as it  reveals interesting properties of the full DoF-achieving transmit vectors in the frequency domain.
First, we note that the cyclic channel matrices $\mathbf H_{k,\ell}$ are simultaneously diagonalized by the discrete Fourier transform (DFT). 
With $\widehat X[m]$, $\widehat Y[m]$, and $\widehat W[m]$ denoting the $M$-point DFT of the transmit, receive, and noise vectors, respectively, we obtain from  \eqref{eq:iocomplex} the following transformed I/O relation 
\begin{align}	&\widehat Y_k[m]=\sqrt{\snr}\left (\widehat X_k[m]+e^{-2\pi i \frac{m}{M}}\sum_{\ell\neq k}^K \widehat X_\ell[m]	\right )+\widehat W_k[m], \label{eq:iofourier}\end{align}
for $m=1,... , M$ and $k=1,... ,K$. This is a parallel IC with subchannel matrices
\ban 	\widehat{\mathbf H}[m]=\begin{pmatrix}	1 & e^{-2\pi i \frac{m}{M}}& \cdots & e^{-2\pi i \frac{m}{M}}\\[-.05cm] 
											e^{-2\pi i \frac{m}{M}} & \ddots & \ddots &\vdots \\
											\vdots & \ddots & \ddots &  e^{-2\pi i \frac{m}{M}}\\
											e^{-2\pi i \frac{m}{M}} &\cdots & e^{-2\pi i \frac{m}{M}} &1\\
												\end{pmatrix}	.	\label{eq:theparallel}\ean 
As the DFT applied at each transmitter and receiver is unitary, \eqref{eq:iofourier} constitutes an equivalent channel,
 i.e., every code for the original channel corresponds to a code for the modified channel \eqref{eq:iofourier} that achieves the same rate.
Moreover, since the DFT is invertible  the transformed  noise distributions meet the conditions in Theorem~\ref{thm:mainc}  if  $W_1,...,W_K$  in \eqref{eq:iocomplex} do. Indeed, since the DFT is unitary we have $h(\widehat{W}_k)=h(W_k)>-\infty$, for $k=1,...,K$. Moreover, by the first property in Lemma~\ref{lem:properties}  $H(\lfloor W_k \rfloor)<\infty$ is equivalent to $\ldim (W_k),\udim(W_k)<\infty$,   and we have $\ldim (W_k)=\ldim(\widehat{W}_k)$, $\udim (W_k)=\udim(\widehat{W}_k)$ by \eqref{eq:iso}, which, since   $\ldim (W_k),\udim(W_k)<\infty$ by assumption,  again using the  first property in Lemma~\ref{lem:properties},   establishes  that $H(\lfloor \widehat{W}_k \rfloor)<\infty$, for $k=1,...,K$.
We next analyze  the parallel IC in the frequency domain with respect to its separability.
The matrices $\widehat{\mathbf H}[m]$ are fully connected; we can therefore (cf.\ \eqref{eq:dets}) apply  Proposition~\ref{prop:upperbound} to the individual subchannels and find that
\begin{align}	\Dof (\widehat{\mathbf H}[m])\leqslant \frac{K}{2},	\quad m=1,...,M.\label{eq:strictineq}\end{align}
For $m=\frac{M}{2}$ and $m =M$ the channel matrix $\widehat{\mathbf H}[m]$ has real, in fact rational, entries. We can therefore apply   \cite[Thm.~2]{EO09}\footnote{Although \cite[Thm.~2]{EO09} is stated for real scalar ICs, it also applies to  scalar ICs with real channel matrix and complex inputs and outputs, as is the case here. To see this, simply stack the real and imaginary parts of input and output to obtain a parallel IC with $2$ subchannels and identical subchannel matrices. 
Since the overall channel matrix has rational entries only, we know by Theorem~\ref{thm:maind} that single-letter inputs are DoF-optimal.
Therefore transmit vectors with   independent  entries  achieving full DoF of the real scalar IC are DoF-optimal for the induced  parallel IC and we have equality in \eqref{eq:ineq}.
This implies that the DoF of the channel with complex inputs and outputs equal twice the DoF of the real scalar channel. By \cite[Thm.~2]{EO09} the DoF of each of the real scalar ICs are strictly less than $K/2$, and so the same holds true for the normalized DoF of the  IC with complex inputs and outputs.  } to conclude that the inequality \eqref{eq:strictineq} is strict for $m=\frac{M}{2}$ and $m =M$. 
This implies that the inequality \eqref{eq:ineq} is strict, which means that the channel \eqref{eq:iofourier} is a non-separable parallel IC and  joint coding across subchannels is mandatory to achieve full DoF.
We next show how the transmit vectors we chose in  the time domain amount to coding across subchannels in the frequency domain. 
Noting that
\begin{align}	\widehat X_k\lefto [m+\frac{M}{2}\right ] 	&= \frac{1}{\sqrt M}\sum_{m'\text{ even}}^M  X_k[m']e^{-2\pi i\frac{m m'}{M}}\underbrace{e^{-\pi i m'}}_{=\, (-1)^{m'}} \\	
&=
\widehat X_k[m], \label{eq:slots} \end{align}
for all $m$ and $k$, it follows that each user sends the same symbol over two different frequency slots, implementing repetition coding across subchannels. The receiver computes $\widehat Y_k[m]+\widehat Y_k\lefto [m+\frac{M}{2}\right]=2\sqrt{\snr}\widehat X_k[m]+\widehat W_k[m]+\widehat W_k\lefto [m+\frac{M}{2}\right]$ to recover the transmitted data symbols $\widehat X_k[m]$.  We see that  interference is canceled by  the phase difference induced by the channel across the frequency slots $m$ and $m+M/2$, i.e., for  $\ell\neq k$, we have \ba &e^{-2\pi i \frac{m}{M}}\widehat X_\ell[m] + e^{-2\pi i \frac{m+\frac{M}{2}}{M}}\widehat X_\ell\lefto [m+\frac{M}{2}\right ] \\ &=e^{-2\pi i \frac{m}{M}}\left (\widehat X_\ell[m] - \widehat X_\ell\lefto [m+\frac{M}{2}\right ]\right )\\ &\hspace{-.05cm}\stackrel{\eqref{eq:slots}}= 0. \ea  


\end{example}

\section{Open problems}


It would be desirable to prove that  the DoF-formula \eqref{eq:main} holds for \emph{all} rather than only \emph{almost all} channel matrices. This appears to be a very hard problem, whose solution may require novel results in number theory. We believe that the main obstacle in extending \eqref{eq:main} to  all channel matrices  resides in the approximation of the entropy of the noiseless output signals by the entropy of their quantized versions, see \eqref{eq:collecterror0}--\eqref{eq:collecterror2}. These approximations depend crucially on results from Diophantine approximation theory, which guarantee an approximation behavior  as in \cite[(112)]{WSV11draft}, and are known to hold for almost all $\mathbf H$ including special cases such as $\mathbf H$ with all entries rational or algebraic numbers. 

Despite the fact that our DoF-formula holds for almost all $\mathbf H$ only, it provides a comprehensive framework for studying the DoF in vector ICs. Knowing that the DoF-formula holds for \emph{all} channel matrices could be useful in finding sharper upper bounds on the DoF of ICs that have strictly less than $K/2$ DoF.

\section*{Acknowledgments}

The authors would like to thank Y.\ Wu, S.\ Shamai (Shitz), and S.\ Verd\'u  for interesting discussions and A.\ Guionnet for advice on the proof of  \cite[Sect.~2]{GS07}. We gratefully acknowledge an anonymous reviewer for pointing out the alternative proof of Theorem~\ref{thm:guionnet} explained in Remark~\ref{rem:wu}.

\appendices

\section{Proof of Theorem~\ref{thm:guionnet}}\label{app:B}
Since by Lemma~\ref{lem:properties}, $\udim(X)<\infty$ if and only if $H(\lfloor X\rfloor)<\infty$, we distinguish the two cases $H(\lfloor X\rfloor)=\infty$ and $H(\lfloor X\rfloor)<\infty$.
First, we show that $H(\lfloor X\rfloor )=\infty$ implies 
\ba \limsup_{\snr\to\infty}\frac{I(X;\sqrt{\snr} X+W)}{\frac{1}{2}\log\snr}=\infty.\ea
 For $\snr>0$, we have
\begin{align}		I(X;\sqrt{\snr}X+W)&\geqslant I(\lfloor X\rfloor ;  \sqrt{\snr}X+W)\label{eq:gray1}\\
&\geqslant I(\lfloor X\rfloor ; \lfloor \sqrt{\snr}X+W\rfloor)\label{eq:gray}\\&= H(\lfloor X\rfloor )-H(\lfloor X\rfloor \! \mid \!  \lfloor \sqrt{\snr}X+W\rfloor) 	,\label{eq:thm3e}	\end{align}
where \eqref{eq:gray1} and \eqref{eq:gray} are by the data processing inequality applied to  $\lfloor X \rfloor \text{ --- }  X \text{ --- } \sqrt{\snr}X+W$  and $\lfloor X \rfloor \text{ --- }  \sqrt{\snr}X+W \text{ --- } \lfloor \sqrt{\snr}X+W \rfloor $, respectively.
Now 
\ban &H(\lfloor X\rfloor \! \mid \!  \lfloor \sqrt{\snr}X+W\rfloor)\\ &  \! \leqslant H(\lfloor X\rfloor , \lfloor W\rfloor \! \mid \!  \lfloor \sqrt{\snr}X+W\rfloor) \label{eq:thm3a}\\ &  \! =H(\lfloor W\rfloor \!\mid \!\lfloor \sqrt{\snr}X+W\rfloor) \! +  \! H(\lfloor X\rfloor   \! \mid \!  \lfloor \sqrt{\snr}X+W\rfloor,\lfloor W\rfloor)\label{eq:thm3b}
\\ &  \! \leqslant  H(\lfloor W\rfloor )+ H(\lfloor X\rfloor \! \mid \!  \lfloor \sqrt{\snr}X+W\rfloor ,\lfloor W\rfloor) \label{eq:thm3c}\\ & \! <\infty,\label{eq:thm3d}\ean 
where \eqref{eq:thm3a} is by the chain rule and the non-negativity of entropy, \eqref{eq:thm3b} is again by the chain rule, 
and \eqref{eq:thm3d} holds since by assumption $H(\floor{W})<\infty$ and since the second term in \eqref{eq:thm3c}  is finite, which is seen as follows.
We show that given $\lfloor \sqrt{\snr}X+W\rfloor $ and $\lfloor W\rfloor$ the number of possible values of $\floor{ X }$ is bounded independently of $\snr$ when  $\snr$ is sufficiently large.
 First, we note that $\floor{ \sqrt{\snr} X }$ has integer-valued entries which are sandwiched according to 
%
\ban \floor{\sqrt {\snr}X_i+W_i}-\floor{W_i}  -2 &< \sqrt{\snr}X_i + W_i - W_i -1  \label{eq:combining0}\\&< \floor{ \sqrt{\snr} X_i }\\ &\leqslant \sqrt{\snr}X_i +W_i -W_i\\ &< \floor{\sqrt{\snr} X_i +W_i}+1 - \floor{W_i}. \label{eq:combining1}	\ean 
Here $X_i$ and $W_i$ denote the $i$-th entry of $X$ and $W$, respectively. 
Second,  $\floor{X}$ can be sandwiched according to
\ban \frac{1}{\sqrt{\snr}}\floor{\sqrt{\snr}X_i} -1 &\leqslant X_i -1 \\ &< \floor{X_i}\\ & \leqslant \frac{1}{\sqrt{\snr}}\left (\floor{\sqrt{\snr}X_i}+1 \right ).\label{eq:combining2}\ean
Combining \eqref{eq:combining0}--\eqref{eq:combining1} with \eqref{eq:combining2} we obtain
\ban &\frac{1}{\sqrt{\snr}}(\floor{\sqrt {\snr}X_i+W_i}-\floor{W_i}  -2) -1 \label{eq:combiningobtain0} \\ &< \floor{X_i}\\ & < \frac{1}{\sqrt{\snr}} (\floor{\sqrt{\snr} X_i +W_i}+2 - \floor{W_i}). \label{eq:combiningobtain1} \ean
From \eqref{eq:combiningobtain0}--\eqref{eq:combiningobtain1} we  conclude that the number of values $\lfloor X\rfloor$ can take on given $\lfloor \sqrt{\snr}X+W\rfloor $ and $\lfloor W\rfloor$ is bounded by $(4/\sqrt{\snr}+1)^n$. If we take $\snr$ sufficiently large, e.g., $\snr>1$, we obtain a bound that is independent of $\snr$, which concludes the argument.
 Finally, using \eqref{eq:thm3e} and \eqref{eq:thm3d} yields $I(X;\sqrt{\snr}X+W)=\infty$ for $H(\lfloor X\rfloor)=\infty$ and hence
\ban		\limsup_{\snr\to\infty}\frac{I(X;\sqrt{\snr} X+W)}{\frac{1}{2}\log\snr} = \infty,	\ean 
as was to be shown.

It remains to establish the result for $H(\lfloor X\rfloor )<\infty$. We begin by decomposing
\begin{align*} I(X;\sqrt{\snr} X+W)=h(\sqrt{\snr}X+W)-h(W)	,\end{align*}
which holds provided that $h(\sqrt{\snr}X+W)$ exists. Note that $h(W)$ exists  and satisfies $h(W)>-\infty$ by assumption. The existence of $h(\sqrt{\snr}X+W)$ is established as part of Proposition~\ref{prop:existence} below. Setting $t:=\frac{1}{\sqrt{\snr}}$ and using the scaling property for differential entropy, we have
\begin{align}	h(\sqrt{\snr}X+W)=h(X+tW) - n \log t.	\label{eq:existencesum} \end{align}
To finalize the proof of Theorem~\ref{thm:guionnet} it therefore suffices to establish the following:

\vspace{.1cm}
\begin{theorem}\label{thm:guionnet2}
Let $X$ and $W$ be independent random vectors in $\mathbb R^n$ such that $H(\lfloor X\rfloor )<\infty$  and $W$ has an absolutely continuous distribution with $h(W)>-\infty$ and $H(\lfloor W\rfloor )<\infty$. Then, we have
\begin{align*}	\limsup_{t\to 0}\frac{h(X+tW)}{\abs{\log t}}=\udim(X) - n.		\end{align*}
\end{theorem}

The proof of  Theorem~\ref{thm:guionnet2} proceeds as follows. We first state   a few auxiliary results, then prove the statement for the special case of uniformly distributed $W$, and finally extend the proof to the case where $W$ has general absolutely continuous distribution  with $h(W)>-\infty$ and $H(\lfloor W\rfloor )<\infty$.

\subsection{Auxiliary results}

We begin with the following result on the existence of differential entropy.

\begin{lemma}\label{lem:diffent}
Let $f,g$ be probability densities on $\mathbb R^n$. 
\begin{enumerate}\item  If 
$	\int_{\mathbb R^n}f(x) \log\lefto (\frac{1}{g(x)}\right ) \mathrm dx 	$
exists and does not  equal $+\infty$ (but possibly $-\infty$), then \linebreak $	\int_{\mathbb R^n}f(x) \log\lefto (\frac{1}{f(x)}\right ) \mathrm dx $
 exists and
\begin{align} 	\int_{\mathbb R^n}f(x) \log\lefto (\frac{1}{f(x)}\right ) \mathrm dx  \leqslant 	\int_{\mathbb R^n}f(x) \log\lefto (\frac{1}{g(x)}\right ) \mathrm dx .\label{eq:holds}\end{align}\\[-.7cm]
\item   If 
$	\int_{\mathbb R^n}f(x) \log\lefto (\frac{1}{f(x)}\right ) \mathrm dx 	$
is finite, then 
$	\int_{\mathbb R^n}f(x) \log\lefto (\frac{1}{g(x)}\right ) \mathrm dx 	$
exists and \eqref{eq:holds} holds.
\end{enumerate}

\end{lemma}
\begin{IEEEproof}
See \cite[Lem.~8.3.1]{Ash90}.
\end{IEEEproof}
\vspace{.1cm}
\begin{corollary}\label{cor:diffent}
Suppose that $W$ is a random vector in $\mathbb R^n$ with absolutely continuous distribution. If $H(\lfloor W \rfloor ) <\infty$, then $h(W)$ exists and 
\begin{align*}	h(W) \leqslant H(\lfloor W \rfloor ).	\end{align*}
\end{corollary}
\begin{IEEEproof}
Let $f$ be the density of $W$. For $x\in\mathbb R^n$, we define a  function $g$ according to
\begin{align*}	g(x) := \int_{\mathcal Q(\ell)} f(y) \mathrm d y ,	\end{align*}
where $\mathcal Q(\ell):= [\ell_1,\ell_1+1)\times ... \times [\ell_n,\ell_n+1)$, with $\ell=(\ell_1,... ,\ell_n)\in\mathbb Z^n$, is the $n$-dimensional cube of sidelength $1$ containing $x$. Since $g$ is  constant on each cube $\mathcal Q(\ell)$, and each $\mathcal Q(\ell)$ is a Borel set, $g$ is measurable, and  $\int_{\mathbb R^n} g(x)\mathrm d x= \int_{\mathbb R^n} f(x)\mathrm d x=1$ implies that $g$  is a probability density. Decomposing $\mathbb R^n$ into the union of all cubes $\mathcal Q(\ell)= [\ell_1,\ell_1+1)\times ... \times [\ell_n,\ell_n+1)$, for $\ell=(\ell_1,... ,\ell_n)\in\mathbb Z^n$, we get
\begin{align*}	&\int_{\mathbb R^n}f(x) \log\lefto (\frac{1}{g(x)}\right ) \mathrm dx\\ &= \sum_{\ell\in\mathbb Z^n}\int_{\mathcal Q(\ell)} f(x)\log\lefto (\frac{1}{\int_{\mathcal Q(\ell)}f(y)\mathrm dy}\right )\mathrm dx\\ &= H(\lfloor W \rfloor ) < \infty,	\end{align*}
and by the first part of Lemma~\ref{lem:diffent}, the proof is complete.
\end{IEEEproof}

We will make repeated use of the fact that the sum $X+W$ of independent random vectors $X$ and $W$ in $\mathbb R^n$ has  absolutely continuous distribution if  $X$ or $W$ is of absolutely continuous distribution \cite[p.~175]{Rud87}. Assuming that  $W$ has  density $f_W$, the density $f_{X+W}$ of $X+W$ is given by 
\ban f_{X+W}(y)=\mathbb E[f_W(y-X)]=\int_{\mathbb R^n} f_W(y-x) \mu\lk x\rk,  \label{eq:abscont0}\ean
where $\mu$ is the distribution of $X$.
Next we establish an existence result for the differential entropy term appearing on the RHS in \eqref{eq:existencesum}.

\vspace{.1cm}
\begin{proposition}\label{prop:existence}
Suppose that $X$ and $W$ are independent random vectors in $\mathbb R^n$ such that $W$ has an absolutely continuous distribution and $H(\lfloor X \rfloor ), H(\lfloor W \rfloor )<\infty$. Then $H(\lfloor X+tW\rfloor)<\infty$, for all $t>0$, and  $h(X+tW)$ exists and satisfies $h(X+tW)\leqslant H(\lfloor X+tW\rfloor)$ for all $t>0$. Moreover, we have
\begin{align}		\limsup_{t\to 0}\frac{h(X+tW)}{\abs{\log t}}\leqslant 0.	\label{eq:establ}\end{align}
\end{proposition}
\begin{IEEEproof}
Applying the chain rule, we get
\begin{align}	 \nonumber &H(\lfloor  X+tW \rfloor )\\ &\leqslant H(\lfloor X+t  W\rfloor,\lfloor   X\rfloor ) \nonumber \\ &= H(\lfloor   X \rfloor)+H(\lfloor   X+t  W\rfloor \! \mid \!  \lfloor   X\rfloor )\nonumber\\
&\leqslant H(\lfloor   X\rfloor )+H(\lfloor t  W\rfloor)+H(\lfloor   X+t  W\rfloor \! \mid \!  \lfloor   X \rfloor ,\lfloor t  W\rfloor ) .\label{eq:last}\end{align}
For $i=1,... ,n$, let $X_i$ and $W_i$ be the $i$-th entry of $X$ and $W$, respectively. We next show that the three terms on the RHS of \eqref{eq:last} are finite for all $t>0$ and, moreover,  bounded uniformly for  $t\leqslant 1$, which establishes  the existence of $h(X+tW)$ by Corollary~\ref{cor:diffent} and also proves \eqref{eq:establ}. The term $H(\lfloor   X\rfloor )$ in \eqref{eq:last} is bounded by assumption. For the last term in \eqref{eq:last}, we have $H(\lfloor   X+t  W\rfloor \! \mid \!  \lfloor   X \rfloor ,\lfloor t  W\rfloor )\leqslant n \log 2$, since 
\ba \floor{X_i}+\floor{tW_i} \leqslant \floor {X_i+tW_i} \leqslant \floor{X_i}+\floor{tW_i} +1 \ea
for $i=1,..., n$. 
To upper-bound the remaining term $H(\lfloor t  W \rfloor)$, we apply the chain rule and find
\begin{align}	H(\lfloor t  W \rfloor)&\leqslant H(\lfloor t\lfloor   W\rfloor\rfloor )+ H(\lfloor t  W \rfloor \! \mid \!  \lfloor t\lfloor   W\rfloor\rfloor )\\	
&\leqslant H(\lfloor   W\rfloor ) + H(\lfloor t  W \rfloor \! \mid \!  \lfloor t\lfloor   W\rfloor\rfloor )  , \label{eq:prop7a}	 \end{align}	
where \eqref{eq:prop7a} follows from the fact that $\lfloor t\lfloor   W\rfloor\rfloor$ is a function of $\lfloor   W\rfloor$ and the application of a function to a discrete random variable cannot increase entropy. Since
\ba \floor{t\floor{ W_i}}\leqslant \floor{tW_i} &\leqslant t\underbrace{(W_i - \floor{W_i})}_{<1} + t\floor{W_i} \\&< t+\floor{t\floor{W_i}}+1,  	\ea 
we have $H(\lfloor t  W \rfloor \! \mid \!  \lfloor t\lfloor   W\rfloor\rfloor )<\infty$ for all $t>0$. Moreover, $H(\lfloor t  W \rfloor \! \mid \!  \lfloor t\lfloor   W\rfloor\rfloor )$ is uniformly bounded by $n\log 2$ if $t\leqslant 1$. The sum  $X+tW$ has  absolutely continuous distribution as a consequence of $tW$ being of  absolutely continuous distribution. It thus follows from Corollary~\ref{cor:diffent} and $H(\lfloor  X+tW \rfloor )<\infty $ that $h(X+tW)$ exists and $h(X+tW)\leqslant H(\lfloor  X+tW \rfloor )$. Moreover, since $H(\lfloor  X+tW \rfloor )$ is uniformly bounded for all $t\leqslant 1$, we have 
\begin{align}		\limsup_{t\to 0}\frac{h(X+tW)}{\abs{\log t}}\leqslant 0.	\end{align}
\end{IEEEproof}

\subsection{Proof of Theorem~\ref{thm:guionnet2}}
The proof of Theorem~\ref{thm:guionnet2} constitutes one of the main technical difficulties in this paper. We proceed in two steps. First, we prove the statement for uniformly distributed noise $W$. Then, by employing a slight generalization of the Ruzsa triangle inequality for differential entropy \cite{KM12}, presented in Proposition~\ref{prop:ruzsa} below, we extend it to general noise distributions satisfying $h(W)>-\infty$ and $H(\lfloor W\rfloor )<\infty$. This considerably simplifies the proof of \cite[Thm.~2.7]{GS07} (for the scalar case), which is based on explicit manipulations of the densities   involved.
We begin by showing the statement for  uniformly distributed noise with the corresponding proof  inspired by \cite[Thm.~3.1]{GS07}.

\vspace{.1cm}
\begin{proposition}\label{prop:uni}
Let $X$ and $U$ be independent random vectors in $\mathbb R^n$ such that $H(\lfloor X\rfloor )<\infty$ and $U$ is uniformly distributed on $B(0;1)$, where the ball is taken with respect to the $\ell^\infty$-norm. Then, we have
\begin{align*}	\limsup_{t\to 0}\frac{h(X+tU)}{\abs{\log t}}=\udim(X) - n.		\end{align*}
\end{proposition}
\begin{IEEEproof}
Note first that the distribution of $U$ is absolutely continuous and $H(\lfloor U\rfloor)=0<\infty$. Thus, by Proposition~\ref{prop:existence}, $H(\lfloor X+tU\rfloor)<\infty$  and $h(X+tU)$ exists    and satisfies  $h(X+tU)\leqslant H(\lfloor X+tU\rfloor)$, for all $t>0$. 
The density of $X+tU$ is given by
\ban f_t(x):=\frac{1}{t^n}\mu \lefto (B(x;t)\right)\! , \label{eq:dens}\ean
where $\mu$ is the distribution of $X$. As $\log$ is an increasing function we have $\mathbb E \lefto [\log \mu(B(X+tU;t))\right ]\leqslant \mathbb E [ \log 1]=0$ and thus 
\ban h(X+tU)=\mathbb E \lefto [\log \frac{t^n}{\mu(B(X+tU;t))}\right ] \geqslant n \log t, \label{eq:exfin}\ean
for all $t>0$.
 Moreover, since  $U$ is uniformly distributed on $B(0;1)$, we have $\mu(B(X+tU;t))\leqslant \mu(B(X;2t))$ almost surely, and hence
\ban \limsup_{t\to 0}\frac{h(X+tU)}{\abs{\log t}} &\geqslant \limsup_{t\to 0}\frac{n \log t-\mathbb E \lefto [\log \mu(B(X;2t))\right ]}{\abs{\log t}} \nonumber \\ &=\limsup_{t\to 0}\frac{\mathbb E \lefto [\log \mu(B(X;2t))\right ]}{\log t} - n \nonumber \\ &=\limsup_{t\to 0}\frac{\mathbb E \lefto [\log \mu(B(X;2t))\right ]}{\log (2t)}
- n\nonumber \\ &=\udim (X) -n, \label{eq:prop10e}\ean
where we used  Lemma~\ref{lem:alternative} in the last step. 

To conclude the proof we need to show that  $\limsup_{t\to 0}\frac{h(X+tU)}{\abs{\log t}}\leqslant \udim (X) -n$. To this end, we take $\widetilde U$ uniformly distributed on $B(0,2)$,  independent of $X$, and note that  $X+t\widetilde U$ has  density
\ba g_t(x):=\frac{1}{(2t)^n}\mu \lefto (B(x;2t)\right). \ea
By \eqref{eq:exfin} it follows that $h(X+tU)>-\infty$. Combined with $h(X+tU)<\infty$ from above this means that $h(X+tU)$ is finite for all $t>0$. 
Therefore, we can apply the second part of Lemma~\ref{lem:diffent} to find  that 
\begin{align} 	h(X+tU) \leqslant 	\int_{\mathbb R^n}f_t(x) \log\lefto (\frac{1}{g_t(x)}\right ) \mathrm dx .\label{eq:prop10a}\end{align}
Writing the density of $X+tU$ as $f_t(x)=\int_{\mathbb R^n}\frac{1}{t^n}	\mathds{1}_{B(x;t)}(y)\mu\lk y\rk$, we find that 
\ban &\int_{\mathbb R^n}f_t(x) \log\lefto (\frac{1}{g_t(x)}\right ) \mathrm dx \\ &= \int_{\mathbb R^n}\int_{\mathbb R^n}\frac{1}{t^n}	\mathds{1}_{B(x;t)}(y)\mu\lk y\rk \log\lefto (\frac{(2t)^n}{\mu \lefto (B(x;2t)\right)}\right ) \mathrm dx  \\ &= \int_{\mathbb R^n}\int_{\mathbb R^n}\frac{1}{t^n}	\mathds{1}_{B(x;t)}(y)\log\lefto (\frac{(2t)^n}{\mu \lefto (B(x;2t)\right)}\right ) \mathrm dx\mu\lk y\rk 	\label{eq:prop10fub} \\ &= \int_{\mathbb R^n}\int_{\mathbb R^n}\frac{1}{t^n}	\mathds{1}_{B(0;t)}(x')\log\lefto (\frac{(2t)^n}{\mu \lefto (B(x'+y;2t)\right)}\right ) \mathrm dx'\mu\lk y\rk \label{eq:varsub}\\ &\leqslant \int_{\mathbb R^n}\int_{\mathbb R^n}\frac{1}{t^n}	\mathds{1}_{B(0;t)}(x')\log\lefto (\frac{(2t)^n}{\mu \lefto (B(y;t)\right)}\right ) \mathrm dx'\mu\lk y\rk \label{eq:contained}\\&= \int_{\mathbb R^n} \log\lefto (\frac{(2t)^n}{\mu \lefto (B(y;t)\right)}\right ) \mu\lk y\rk \label{eq:ttothen}\\ &=\mathbb E\lefto [\log \frac{1}{\mu \lefto (B(X;t)\right)}\right ]+ n\log (2t), \label{eq:lastline}\ean
where  \eqref{eq:prop10fub} follows from Fubini's Theorem, whose conditions are satisfied since the function in \eqref{eq:prop10fub} is integrable which in turn follows from \eqref{eq:lastline}, in \eqref{eq:prop10fub} we substituted $x':=x-y$,
\eqref{eq:contained} is thanks to $B(y;t)\subseteq B(x'+y;2t)$ for all $x'\in B(0;t)$, and to get \eqref{eq:ttothen} we used $\int_{\mathbb R^n}\frac{1}{t^n}	\mathds{1}_{B(0;t)}(x')dx'=1$. Together with \eqref{eq:prop10a} this yields 
\ban 	\limsup_{t\to 0}\frac{h(X+tU)}{\abs{\log t}} &\leqslant  	\limsup_{t\to 0}	\frac{ n\log (2t)-\mathbb E\lefto [\log \mu \lefto (B(X;t)\right)\right ]}{\abs{\log t}}	\nonumber \\ &=\limsup_{t\to 0}\frac{\mathbb E\lefto [\log \mu \lefto (B(X;t)\right)\right ]}{\log t}-n\\ &=\udim(X)-n,	\label{eq:prop10d} \ean
where we used  Lemma~\ref{lem:alternative} in the last step. This completes the proof.
\end{IEEEproof}

We  turn to the second step in the proof of Theorem~\ref{thm:guionnet2}, namely the extension of Proposition~\ref{prop:uni} to noise $W$ with general (absolutely continuous) distribution satisfying $h(W)>-\infty$ and $H(\lfloor W\rfloor )<\infty$. To this end, we need a slight generalization of the Ruzsa triangle inequality for differential entropy stated in \cite[Thm.~3.1]{KM12} for scalar random variables with absolutely continuous distributions.  The generalization that we state below i) applies to random vectors, and ii) allows for a general distribution for $X$ as long as $H(\lfloor X\rfloor)<\infty$.

\vspace{.1cm}
\begin{proposition}\label{prop:ruzsa}
Let $X$, $Y$, and $Z$ be independent random vectors in $\mathbb R^n$ such that \linebreak $H(\lfloor X\rfloor),H(\lfloor Y\rfloor), H(\lfloor Z \rfloor)<\infty$ and $Y$ and $Z$ have absolutely continuous distribution  with $h(Y),h(Z)>-\infty$. Then, we have
\ban 	h(X+Z)-h(X+Y) \leqslant h(Y-Z) - h(Y).		\label{eq:ruzsa} \ean
\end{proposition}
\begin{IEEEproof}
First,  note that, by Proposition~\ref{prop:existence}, the differential entropies $h(X+Z)$, $h(X+Y)$ and $h(Y-Z)$  exist.\footnote{Note that $h(Y), h(Z)<\infty$ is not sufficient for $h(Y-Z)<\infty$ to hold. In fact, \cite[Prop.~V.8]{BC15} shows that there exist  $Y$ with $h(Y)$ finite such that for every $Z$ with $h(Z)$ finite, we have $h(Y-Z)=\infty$.} We have 
\ban 		&h(X+Z)-h(Z)\\ &=I(X;X+Z) \label{eq:prop11a}\\ &\leqslant I(X;(X+Y,Y-Z))\label{eq:prop11b}\\ &= h(X+Y, Y-Z) - h(Y, Y-Z)\label{eq:prop11c}\\ 
							&= h(X+Y,Y-Z)-h(Y,Z) \label{eq:prop11d}\\ &= h(X+Y,Y-Z) - h(Y) - h(Z) \label{eq:prop11e} \\ 
 &\leqslant h(X+Y) + h(Y-Z) - h(Y) - h(Z), \label{eq:prop11f}\ean						
							
where \eqref{eq:prop11b} is by the data processing inequality noting that $X \text{ --- } (X+Y,Y-Z) \text{ --- }  X+Z$ forms a Markov chain, for \eqref{eq:prop11d} we applied the coordinate transformation $(x,y)\mapsto (x,x-y)$ to the vector in the second term in \eqref{eq:prop11c} and used the fact that the determinant of the corresponding Jacobian has absolute value $1$ together with \cite[(8.71), p.~254]{CT06},  \eqref{eq:prop11e} follows from the independence of $Y$ and $Z$, and \eqref{eq:prop11f} is by the chain rule. 
 Adding $h(Z)-h(X+Y)$ to both sides  yields \eqref{eq:ruzsa}.
\end{IEEEproof}

We are now ready to complete the proof of Theorem~\ref{thm:guionnet2}.
Let $U$ be uniformly distributed on the $\ell^\infty$-ball $B(0;1)$ and independent of $(X,W)$ for $X$ and $W$ as in the statement of Theorem~\ref{thm:guionnet2} and let $t>0$. Noting that $U$ and $W$ have absolutely continuous distributions, $h(W)>-\infty$ and $H(\lfloor W \rfloor)<\infty$ by assumption, and $h(U)=H(\lfloor U \rfloor)=0$, we apply Proposition~\ref{prop:ruzsa} twice to get 
\ban 	- h(tU-tW) +h(tU)& \leqslant 	h(X+tU) - h(X+tW)\label{eq:diffallexist1}\\ & \leqslant h(tW-tU) - h(tW).	\label{eq:diffallexist2}\ean
Corollary~\ref{cor:diffent} and Proposition~\ref{prop:existence} imply that all differential entropies appearing in \eqref{eq:diffallexist1} and \eqref{eq:diffallexist2} exist and $h(tU-tW),h(tW-tU)<\infty$ for  all $t>0$. Furthermore, by the scaling property of differential entropy we have $h(tU-tW)=h(tW-tU)$. Moreover, since the density $f_U$ of $U$ is bounded, i.e., $f_U(x)\leqslant C$ for all $x\in \mathbb R^n$, it follows that the density $f_{tU-tW}$ of $tU-tW$ is bounded as well for all $t>0$, since
\ba 	f_{tU-tW}(x)=\frac{1}{t^n}\int_{\mathbb R^{n}}f_U(u)f_W\lefto (u-\frac{x}{t}\right)\mathrm du\leqslant \frac{C}{t^n},	\ea 
 and thus 
\ba h(tU-tW)&=h(tW-tU)\\ &\geqslant \int_{\mathbb R^n}f_{tU-tW}(x)\log\lefto (\frac{t^n}{C}\right )\mathrm dx\\ &=\log\lefto (\frac{t^n}{C}\right ) >-\infty, 	\ea 
for all $t>0$.
Since  $h(U)=0$ and  $h(W)$ is finite by Corollary~\ref{cor:diffent} and the assumption $h(W)>-\infty$, we find that $h(U-W)$ and $h(W-U)$ are finite, which yields
\ba 	&\lim_{t\to 0} \frac{- h(tU-tW) +h(tU)}{\abs{\log t}}\\ &=\lim_{t\to 0} \frac{- h(U-W) -n\log t  +h(U)+n\log t}{\abs{\log t}}=0\\ &\hspace{-.14cm} \stackrel{\eqref{eq:diffallexist1}}{\leqslant} \lim_{t\to 0} \frac{ h(X+tU) - h(X+tW)}{\abs{\log t}}\\ &\hspace{-.14cm}\stackrel{\eqref{eq:diffallexist2}}{\leqslant} \lim_{t\to 0} \frac{ h(tW-tU) -h(tW)}{\abs{\log t}}\\ &=\lim_{t\to 0} \frac{ h(W-U) +n\log t-h(W)-n\log t}{\abs{\log t}}=0.\ea 
 We can therefore conclude that 
\ba 	\limsup_{t\to 0}\frac{h(X+tW)}{\abs{\log t}}=\limsup_{t\to 0}\frac{h(X+tU)}{\abs{\log t}}\stackrel{\text{Prop.~\ref{prop:uni}}}{=}\udim(X) -n,		\ea 
which completes the proof.
 \hfill \endIEEEproof



\section{Proofs of  Theorems~\ref{thm:mainb}--\ref{thm:mainc}}
\label{app:A}

The proofs build on the arguments  in \cite[Sect.~V-B]{WSV11draft}, which need to be extended to the vector case. To keep this paper as self-contained as possible, we detail the corresponding extensions and also include the steps of \cite{WSV11draft} that remain unchanged. We hasten to add that the main conceptual components in the proofs in this Appendix should be attributed  to Wu et al.\ \cite{WSV11draft}.

\subsection{Auxiliary results}
We begin with a number of auxiliary results. 


\vspace{.1cm}
\begin{lemma}\label{lem:moreaccurate}
Let $\mathbf X$ be a random matrix in $\mathbb R^{M\times N}$  and let $k\in\mathbb N$. Then, we have
\begin{align*}	0\leqslant H([ \mathbf X]_{k})-	\mathbb E\lefto [\log \frac{1}{\mu(B(\mathbf X;2^{-k}))}\right ]	\leqslant	MN\log 3, \end{align*}
where $\mu$ is the distribution of $\mathbf X$ and $B(\mathbf X;\varepsilon)$ is the $\ell^\infty$-ball\footnote{$B(\mathbf A;\varepsilon):=\{\mathbf B\in \mathbb R^{M\times N} \; \mid \; \|\mathbf A- \mathbf B\|_\infty <\varepsilon\}$  where $\| \cdot \|_\infty$ is defined in the Notation paragraph.}  with center $\mathbf X$ and radius $\varepsilon$.
\end{lemma}
\begin{IEEEproof}
For $\mathbf A=(a_{i,j})_{\substack{1\leqslant i \leqslant M\\ 1\leqslant j \leqslant N}}$ we define the ``cube'' in $\mathbb R^{M\times N}$ of sidelength $2^{-k}$ containing $\mathbf A$ as 
\begin{align}	 \nonumber \mathcal Q_k(\mathbf A):= &\left [\,[ a_{1,1}]_{k},[a_{1,1}]_{k}+2^{-k}\, \right )\times \ldots \\ &\times \left [\,[ a_{M,N}]_{k},[a_{M,N}]_{k}+2^{-k} \,\right )\label{eq:lem4a}	.\end{align}
Consider the decomposition of $\mathbb R^{M\times N}$ into cubes with sidelength $2^{-k}$ according to
\begin{align}	\mathbb R^{M\times N}=\bigcup_{\mathcal Q\in\{\mathcal Q_k(  \mathbf A) \mid \mathbf A\in \mathbb R^{M\times N}\}} \mathcal  Q .\end{align}
We will also need cubes of sidelength $3\cdot 2^{-k}$, defined according to
\ba  \nonumber \widehat{\mathcal  Q}_k(  \mathbf A) := & \left [\,[ a_{1,1}]_{k}-2^{-k},[a_{1,1}]_{k}+ 2^{-k+1}\, \right )\times \ldots \\ & \times \left [\,[ a_{M,N}]_{k}-2^{-k},[a_{M,N}]_{k}+ 2^{-k+1} \,\right ) .\ea 
Then, we have $\mathcal Q_k(\mathbf A)\subseteq B(\mathbf A;2^{-k})\subseteq \widehat{\mathcal  Q}_k(\mathbf A)$ and since $H([ \mathbf  X ]_{{k}} )=\mathbb E\lefto [\log \frac{1}{\mu(\mathcal Q_k(\mathbf X))}\right ]$, we get
\begin{align} 0& \leqslant H([\mathbf  X]_{k})-	\mathbb E\lefto [\log \frac{1}{ \mu(B(\mathbf X;2^{-k}))}\right ]\\ &=\mathbb E\lefto [\log\frac{\mu(B(\mathbf X;2^{-k}))}{\mu(\mathcal Q_k(\mathbf X))}\right ] \\ &\leqslant \mathbb E\lefto [\log\frac{\mu(\widehat{\mathcal  Q}_k(\mathbf X))}{\mu(\mathcal Q_k(\mathbf X))}\right ]  \\&\hspace{-.22cm}\stackrel{\text{Jensen}}{\leqslant} \log \mathbb E\lefto [\frac{\mu(\widehat{\mathcal  Q}_k(\mathbf X))}{\mu(\mathcal Q_k(\mathbf X))}\right ]  \\
& =MN\log 3 , \label{eq:lemend}\end{align} 
where in  \eqref{eq:lemend} we used the fact that for each $\mathbf A\in\mathbb R^{M\times N}$ the function $\mu(\mathcal Q_k(\mathbf X))$ is constant on the event $\mathbf X\in \mathcal Q_k(\mathbf A)$  and that each $\widehat{\mathcal  Q}_k(  \mathbf A)$ is the union of $3^{MN}$  cubes of the form \eqref{eq:lem4a}. 
\end{IEEEproof}

\vspace{.1cm}
\begin{lemma}\label{lem:expsub}
Let $  X$ be a random vector in $\mathbb R^n$. For $a>1$ we have 
\begin{align*}	\underline d(  X) = \liminf_{\ell\to\infty}\frac{H(\lk   X \rk_{a^\ell})}{\log(a^\ell)}	\;\; \text{and} \;\;  \overline d(  X) = \limsup_{\ell\to\infty}\frac{H(\lk    X \rk_{a^\ell})}{\log(a^\ell)}	.	\end{align*}
\end{lemma}
\begin{IEEEproof}
The proof follows closely that of the corresponding result for the scalar case reported in \cite[Prop.~2]{WV10}, and is provided in the Online Addendum \cite[Lem.~5]{SB14extended}  for completeness. 
\end{IEEEproof}

\vspace{.1cm}
\begin{lemma}\label{lem:neu}
Let $\mathbf X$ and $\mathbf W$ be independent random matrices in $\mathbb R^{M\times N}$, where $H(\lfloor \mathbf X\rfloor )<\infty$ and $\mathbf W$ has an absolutely continuous distribution with $h(\mathbf W)>-\infty$ and $H(\lfloor \mathbf W\rfloor )<\infty$. Furthermore, let $\mathbf U$ be  uniformly distributed on $B(\mathbf 0; 1)$ and  independent of $\mathbf X$ and $\mathbf W$. For $k\in \mathbb N$, we have
\ban	&\frac{1}{MN}\left | h(2^k\mathbf X +\mathbf W)-\mathbb E\lefto [\log \frac{1}{\mu(B(\mathbf X; 2^{-k}))}\right ]\right | \nonumber	\\ &\leqslant \frac{\max \{ h(\mathbf W- \mathbf U) -h(\mathbf W), h(\mathbf U-\mathbf W)-h(\mathbf U)\}}{MN} +\log 6 , \label{eq:rhsmax}	\ean
where $\mu$ is the distribution of $\mathbf X$ and   $B(\mathbf A;\varepsilon)$ denotes the $\ell^\infty$-ball with center $\mathbf A\in \mathbb R^{M\times N}$ and radius $\varepsilon$. \end{lemma}
\begin{IEEEproof}
Since $h(\mathbf U)>-\infty$ and $H(\lfloor \mathbf U\rfloor)<\infty$, we can apply the Ruzsa triangle inequality \eqref{eq:ruzsa} and find
\ban  \nonumber &\left | h(2^k \mathbf  X+ \mathbf W) -		h(2^k \mathbf  X+ \mathbf U)\right | \\ &\leqslant \max \{ h(\mathbf W- \mathbf U) -h(\mathbf W), h(\mathbf U-\mathbf W)-h(\mathbf U)\} .\label{eq:rhsfinite} \ean
 The RHS of \eqref{eq:rhsfinite} is finite by Proposition~\ref{prop:existence} and Corollary~\ref{cor:diffent}. By the scaling property of differential entropy, we have $h(2^k \mathbf  X+ \mathbf U)= h(\mathbf X+2^{-k}\mathbf U)+MNk \log 2$, and following the steps  \eqref{eq:prop10a}--\eqref{eq:lastline}, we find that
\ban &h(2^k \mathbf  X+ \mathbf U)\\ &=h(\mathbf X+2^{-k}\mathbf U)+MNk \log 2 \\ &\leqslant 
\mathbb E\lefto [\log \frac{1}{\mu (B(\mathbf X; 2^{-k}))} \right ]+ MNk \log 2  + MN \log (2^{-k+1})\\&= \mathbb E\lefto [\log \frac{1}{\mu (B(\mathbf X; 2^{-k}))} \right ] +MN \log 2.\label{eq:put}\ean 
For a bound in the other direction, we note that $B(\mathbf X+2^{-k}\mathbf U; 2^{-k})\subseteq B(\mathbf X; 2^{-(k-1)})$ and thus
\ba h(2^k \mathbf  X+ \mathbf U) &\stackrel{\eqref{eq:dens}}{=}\mathbb E\lefto [\log \frac{1}{\mu (B(\mathbf X+2^{-k}\mathbf U; 2^{-k}))} \right ]\\ &\,\, \geqslant  \mathbb E\lefto [\log \frac{1}{\mu (B(\mathbf X; 2^{-(k-1)}))} \right ].\ea
It follows that
\ban &\mathbb E\lefto [\log \frac{1}{\mu (B(\mathbf X; 2^{-k}))} \right ] -  h(2^k \mathbf  X+ \mathbf U) \\ &\leqslant  \mathbb E\lefto [\log \frac{1}{\mu (B(\mathbf X; 2^{-k}))} \right ] -\mathbb E\lefto [\log \frac{1}{\mu (B(\mathbf X; 2^{-(k-1)}))} \right ] \\ &\!\!\!\!\!\! \stackrel{\text{Lemma~\ref{lem:moreaccurate}}}{\leqslant} H([\mathbf X]_{k}) - H([\mathbf X]_{k-1}) +MN \log 3\\ &\leqslant H([\mathbf X]_{k} \! \mid \! [\mathbf X]_{k-1})+MN \log 3 \label{eq:lemchaina} \\ &\leqslant MN( \log 2 + \log 3)	\label{eq:lemchainb}, \ean
where in \eqref{eq:lemchaina} we applied the chain rule, and \eqref{eq:lemchainb} follows since given $[\mathbf X]_{k-1}$ there are at most $2^{MN}$ possible values for $[\mathbf X]_{k}$.
Putting together \eqref{eq:rhsfinite}, \eqref{eq:put}, and \eqref{eq:lemchainb} completes the proof.
\end{IEEEproof}

The next lemma, which is a straightforward extension of \cite[Lem.~4]{WSV11draft} to the vector case,  provides a sufficient condition for the iterated function system used in the  construction of the  random vector  in \eqref{eq:ifs} to satisfy the open set condition.

\vspace{.1cm}
\begin{lemma}\label{lem:openset}
Consider the iterated function system $\{F_1,... ,F_m\}$ with $F_i( x )=r x +  w_i$, for $ x\in\mathbb R^n$,  $r\in (0,1)$, and pairwise different vectors $w_1,...,w_m\in\mathbb R^n$. Let furthermore $\mathcal W:=\{w_1,... , w_m\}$. Then, the open set condition (see Definition~\ref{def:opensetcondition}) is satisfied if
\begin{align}	r\leqslant \frac{\mathsf m(\mathcal W)}{\mathsf m(\mathcal W)+\mathsf M(\mathcal W)}.	\label{eq:contract}\end{align}
\end{lemma}
\begin{IEEEproof}
The proof follows closely that of the corresponding result for the scalar case reported in \cite[Lem.~4]{WSV11draft}, and is provided in the Online Addendum \cite[Lem.~1]{SB14extended}  for completeness. 
\end{IEEEproof}

The following lemma provides an upper bound on the difference in entropy of two discrete random matrices. 
\vspace{.1cm}
\begin{lemma}[{\cite[Lem.~7]{WSV11draft}}]\label{lem:distanceentropy}
Let $\mathbf U$ and $\mathbf V$ be discrete random matrices in $\mathbb R^{M\times N}$ such that  $\|\mathbf U - \mathbf V \|_\infty\leqslant\varepsilon$ and $\mathsf m(\mathcal U)\geqslant\delta>0$ where $\mathcal U$ is the set of possible realizations of $\mathbf U$. Then,
\ba 	H(\mathbf U) - H(\mathbf V) \leqslant MN \log \lefto (1+\left\lfloor\frac{2\varepsilon}{\delta}\right \rfloor \right ).		\ea
\end{lemma}
\begin{IEEEproof}
By the chain rule we have 
\ba H(\mathbf U)-H(\mathbf V)\leqslant H(\mathbf U\! \mid\! \mathbf V) .\ea 
The result now follows from the fact that given $\mathbf V$ each of the $MN$ entries of $\mathbf U$ can  take on at most $1+\left\lfloor\frac{2\varepsilon}{\delta}\right \rfloor $ different values.
\end{IEEEproof}

 Next, we bound the error in the entropy of the quantized output signals which results from replacing the input distributions by their fractional parts, a crucial step in the proof of Theorem~\ref{thm:mainb}.\vspace{.1cm}
\begin{lemma}\label{lem:fract}
Consider the deterministic matrices $\mathbf H_1,... , \mathbf H_K\in \mathbb R^{M\times M}$ and let $\mathbf X_1^N,... ,\mathbf X_K^N$ be random matrices in $\mathbb R^{M\times N}$. 
For every $k\in\mathbb N$, we  have
\begin{align*}	\left |H\lefto (\left [\sum_{j=1}^K\mathbf H_{j}\mathbf X_j^N \right ]_k\right ) -H\lefto (\left [\sum_{j=1}^K \mathbf H_{j}\lefto (\mathbf X_j^N\right )\right ]_k\right )	\right | \\ \hspace{1cm}\leqslant \sum_{j=1}^K H\lefto (\floor{\mathbf X_j^N}\right ) +MN \log 2 ,\end{align*}
where $(\mathbf A):=\mathbf A -\lfloor \mathbf A \rfloor$ denotes the fractional part of the real matrix $\mathbf A$.
\end{lemma}
\begin{IEEEproof}
The proof follows closely that of the corresponding result for the scalar case reported in \cite[Lem.~10]{WSV11draft}, and is provided in the Online Addendum \cite[Lem.~2]{SB14extended}  for completeness. 
\end{IEEEproof}

The following   result  shows that $H(\lfloor \mathbf X\rfloor)$ is finite if all entries of $\mathbf X$ have  finite second moment.
\vspace{.1cm}
\begin{lemma}\label{lem:linearbound}
Let $\mathbf X^N$ be a random matrix in $\mathbb R^{M\times N}$ with columns $X^{(1)}, ... , X^{(N)}$ where $X^{(n)}=(X^{(n)}[1],... ,X^{(n)}[M]) \in \mathbb R^M$, for $n=1,... ,N$. If
\begin{align}	\sum_{m=1}^M\sum_{n=1}^N \mathbb E[(X^{(n)}[m])^2]\leqslant MN	\label{eq:apclem},\end{align}
then
\begin{align}	H(\lfloor \mathbf X \rfloor)	\leqslant \frac{MN}{2}\log\lefto(\frac{26}{3}\pi e\right ).	\end{align}
\end{lemma}
\begin{IEEEproof}
\begin{align}		&H(\lfloor \mathbf X \rfloor)\\ 	&\leqslant \sum_{m=1}^M\sum_{n=1}^NH(\lfloor X^{(n)}[m] \rfloor) 			\label{eq:bound1}\\
										&\leqslant MN\log 2 +\sum_{m=1}^M\sum_{n=1}^NH(\lfloor |X^{(n)}[m]| \rfloor) \label{eq:bound2}\\
										&\leqslant MN \log 2+\sum_{m=1}^M\sum_{n=1}^N \frac{1}{2}\log\lefto (2\pi e \lefto ( \mathbb E[(X^{(n)}[m])^2]+\frac{1}{12} \right )\! \right )\label{eq:bound3}	\\	
										&\leqslant \frac {MN}{2} \Bigg (\log 4 \nonumber \\  &\hspace{.8cm} +\log\lefto (2\pi e \left ( \frac{1}{MN}\sum_{m=1}^M\sum_{n=1}^N \mathbb E[(X^{(n)}[m])^2]+\frac{1}{12} \right )\right )\! \Bigg )\label{eq:bound4}\\	
										&\leqslant \frac{MN}{2}\log\lefto(\frac{26}{3}\pi e\right ) \label{eq:bound5},
\end{align}
where \eqref{eq:bound1} is by the chain rule, \eqref{eq:bound2} follows from 
\begin{align*} H(\lfloor X \rfloor) & \leqslant H(\lfloor X \rfloor,   \lfloor |X| \rfloor)\\ & =  H(\lfloor |X| \rfloor) + H(\lfloor X \rfloor\! \mid \!  \lfloor |X| \rfloor)\\ &\leqslant H(\lfloor |X| \rfloor) +\log 2  \end{align*}
for arbitrary   random variables $X$, in \eqref{eq:bound3} we used \cite[Eq.~(8.94)]{CT06}, \eqref{eq:bound4} is due to Jensen's inequality, and \eqref{eq:bound5} follows from \eqref{eq:apclem}.
\end{IEEEproof}

The following lemma  is a straightforward extension of \cite[Lem.~14]{WSV11draft} to the vector case.
\vspace{.1cm}
\begin{lemma}\label{lem:entgeneral}
Let $\mathcal V\subseteq \mathbb R^n$ be a  set such that $0<\mathsf m(\mathcal V),\mathsf M(\mathcal V)<\infty$ and let $r>0$ be such that 
\ban	r\leqslant \frac{\mathsf m(\mathcal V)}{\mathsf m(\mathcal V)+\mathsf M(\mathcal V)}.	\label{eq:conddist}\ean
Then, for every $\ell \in \mathbb N$ with $\ell\geqslant 1$, we have 
\ban \mathsf m(\mathcal V+r \mathcal V + \ldots + r^{\ell-1}\mathcal V) \geqslant r^{\ell-1}\mathsf m(\mathcal V). \label{eq:inj} \ean
Moreover, the mapping $\mathcal V^\ell \to  \mathcal V+r \mathcal V + \ldots + r^{\ell-1}\mathcal V$, $( v_1, ..., v_\ell)\mapsto v_1 + rv_2+ \ldots +r^{\ell-1}v_\ell$ is a one-to-one correspondence. 
\end{lemma}
\begin{IEEEproof} 
The proof follows closely that  of the corresponding result for the scalar case reported in \cite[Lem.~13]{WSV11draft}, and is provided in the Online Addendum \cite[Lem.~3]{SB14extended}  for completeness. 
\end{IEEEproof}

We next derive  an alternative limiting expression for $\Dof(\mathbf H)$ in terms of the entropy of quantized versions of the noiseless channel output vectors.  We restrict the corresponding statement in Proposition~\ref{prop:outsourced} below to Gaussian noise so that it can be applied directly in the proof of Theorem~\ref{thm:main}. However, in the proof of Theorem~\ref{thm:mainc}, we show that the statement continues to hold for more general noise distributions, specifically those satisfying the conditions of Theorem~\ref{thm:mainc}.
\vspace{.1cm}
\begin{proposition}\label{prop:outsourced}
The DoF of the channel \eqref{eq:channel} (with Gaussian noise) are given by
\ba &\Dof(\mathbf H)= \limsup_{k\to\infty}\lim_{N\to\infty} \sup_{\mathbf X_1^N,... , \mathbf X_K^N}\frac{1}{N k}   \\ &  \sum_{i=1}^K \left [  H\lefto (\left [\sum_{j=1}^K \mathbf H_{i,j}\mathbf X_j^N \right ]_k\right ) - H\lefto ( \left [\sum_{j\neq i}^K \mathbf H_{i,j}\mathbf X_j^N \right ]_k \right ) \right ],\ea
where $\mathbf X_i^N= \left (X_{i}^{(1)}\, ... \; X_{i}^{(N)}\right )$ and the supremum is taken over all independent $\mathbf X_1^N, ... ,\mathbf X_K^N$ satisfying
\begin{align}	\frac{1}{MN} \sum_{m=1}^M\sum_{n=1}^N \mathbb E \lefto [\left(X_{i}^{(n)}[m] \right)^2\right ]\leqslant 1 ,	\label{eq:apc} \quad \text{for all $i$}.\end{align}
\end{proposition}
\begin{IEEEproof}
We begin by arguing that in computing $\limsup_{\snr\to\infty}\frac{C_\text{sum}(\mathbf H;\, \snr)}{\frac{1}{2}\log\snr}$ we may restrict to an exponential sequence $\snr_k$. Specifically, let $\snr_k$ be an increasing sequence of positive real numbers such that $\snr_{k+1}\leqslant c \cdot\snr_k$, for all $k\in\mathbb N$ and some real constant $c>1$. For a given $\snr>\snr_0$, we can then find a  $k$ such that $\snr_k<\snr \leqslant \snr_{k+1}$. Since $C_\text{sum}(\mathbf H;\snr)$ is monotonic in $\snr$, we get
\begin{align}	\frac{C_\text{sum}(\mathbf H;\snr)}{\frac{1}{2}\log \snr} &\leqslant \frac{C_\text{sum}(\mathbf H;\snr_{k+1})}{\frac{1}{2}\log \snr_k}\\ & =\frac{C_\text{sum}(\mathbf H;\snr_{k+1})}{\frac{1}{2}\log \snr_{k+1}+	\frac{1}{2}\log \frac{\snr_k}{\snr_{k+1}}}\\ &\leqslant\frac{C_\text{sum}(\mathbf H;\snr_{k+1})}{\frac{1}{2}\log \snr_{k+1}-	\frac{1}{2}\log c} .\end{align}
This implies 
\begin{align}		\limsup_{\snr\to\infty}\frac{C_\text{sum}(\mathbf H;\snr)}{\frac{1}{2}\log\snr} \leqslant \limsup_{k\to\infty}\frac{C_\text{sum}(\mathbf H;\snr_k)}{\frac{1}{2}\log\snr_k}. \label{eq:eq}\end{align}
By virtue of $\snr_k$ being a sequence that tends to $\infty$, we also have
\begin{align}		\limsup_{\snr\to\infty}\frac{C_\text{sum}(\mathbf H;\snr)}{\frac{1}{2}\log\snr} \geqslant \limsup_{k\to\infty}\frac{C_\text{sum}(\mathbf H;\snr_k)}{\frac{1}{2}\log\snr_k}, \end{align}
which implies equality in \eqref{eq:eq}. From now on we restrict ourselves to the particular subsequence $\snr_k=4^k$.

As a second preliminary remark, we note that for independent $\mathbf X_1^N, ... , \mathbf X_K^N$  with $\mathbf X_i^N = ( X_i^{(1)} \, ... \;  X_i^{(N)} )$ and $X_i^{(n)}\in \mathbb R^M$, we have 
\begin{align*}	&I\lefto(\mathbf X_i^N;\mathbf Y_i^N\right )\\ &= I\lefto (\sum_{j=1}^K \mathbf H_{i,j} \mathbf X_j^N,\snr \right )- I\lefto ( \sum_{j\neq i}^K \mathbf H_{i,j} \mathbf X_j^N ,\snr \right ),\end{align*}
where, for a random matrix $\mathbf X$, we set $I(\mathbf X,\snr):=I(\mathbf X; \sqrt{\snr}\mathbf X+\mathbf W)$ with $\mathbf W$ consisting of i.i.d.\ zero-mean Gaussian random vectors, independent of $\mathbf X$ and with identity covariance matrix.
The limiting characterization of sum-rate capacity in Proposition~\ref{prop:limiting} thus yields
\begin{align} &\Dof(\mathbf H)= \limsup_{\snr\to\infty}\lim_{N\to\infty} \sup_{\mathbf X_1^N,... , \mathbf X_K^N}\frac{2}{N\log \snr}   \nonumber \\ & \sum_{i=1}^K   \left [ I\lefto (\sum_{j=1}^K \mathbf  H_{i,j} \mathbf X_j^N,\snr \right ) - \lefto ( \sum_{j\neq i}^K \mathbf  H_{i,j} \mathbf X_j^N ,\snr \right ) \right ],	\label{eq:limiting}\end{align}
where the supremum is taken over all independent $\mathbf X_1^N, ... ,\mathbf X_K^N$ satisfying the power constraint \eqref{eq:apc}.

Since $h(\mathbf W)=\frac{MN}{2}\log(2\pi e)>-\infty$, $H(\lfloor \mathbf W \rfloor )<\infty$ by Lemma~\ref{lem:linearbound}, and the RHS of \eqref{eq:rhsmax} is bounded by a numerical constant as  a consequence of $h(\mathbf W- \mathbf U),h(\mathbf W), h(\mathbf U-\mathbf W),h(\mathbf U)$  all being finite, it follows from Lemma~\ref{lem:neu} that for a random matrix $\mathbf X$ with $H(\lfloor \mathbf X\rfloor )<\infty$, we have
\begin{align}	\left | I(\mathbf X,\snr_k)-\mathbb E\lefto  [\log \frac{1}{ \mu \Big (B \Big (\mathbf X;2^{-k}\Big )\Big )}\right  ] \right |\leqslant MN c,	\end{align}
where $\mu$ is the distribution of $\mathbf X$, 
$c$ is a numerical constant, we used $I(\mathbf X,\snr_k)=h( 2^k\mathbf X+\mathbf W)-h(\mathbf W)$, and moved $h(\mathbf W)$ to the other side in inequality \eqref{eq:rhsmax}. By Lemma~\ref{lem:moreaccurate}, we also have
\begin{align}	0\leqslant H([ \mathbf X]_{k})-	\mathbb E\lefto [\log \frac{1}{ \mu \Big (B \Big (\mathbf X;2^{-k}\Big )\Big )}\right  ]	\leqslant	MN\log 3. \end{align}
Therefore,  we get for the mutual information terms in \eqref{eq:limiting} that
\ban &	-\frac{M\left ( c + \log 3\right )}{ \frac{1}{2}\log \snr_k}\label{eq:explicitboundl} \\ &\label{eq:explicitbound} \leqslant \frac{ I\lefto (\sum_j  \mathbf  H_{i,j} \mathbf X_j^N,\snr_k \right )-  H\lefto (\left [\sum_{j} \mathbf H_{i,j}\mathbf X_j^N \right ]_k\right )}{N\frac{1}{2}\log \snr_k}\\ &\leqslant \frac{M(c+ \log 3)}{\frac{1}{2}\log \snr_k} \label{eq:explicitboundu}
,\ean
for all independent $\mathbf X_1^N, ... , \mathbf X_K^N$ satisfying \eqref{eq:apc}. 
The lower bound \eqref{eq:explicitboundl} and the upper bound \eqref{eq:explicitboundu}  tend to zero for $k\to \infty$, and do so uniformly with respect to $N$ and the inputs $\mathbf X_1^N, ... , \mathbf X_K^N$. Therefore, we
 may replace the mutual information terms in \eqref{eq:limiting} by the corresponding entropy terms to get
\begin{align} & \Dof(\mathbf H) 
=\limsup_{k\to\infty}\lim_{N\to\infty} \sup_{\mathbf X_1^N,... , \mathbf X_K^N}\frac{1}{N k} \nonumber \\ 
&\sum_{i=1}^K \left [  H\lefto (\left [\sum_{j=1}^K \mathbf H_{i,j}\mathbf X_j^N \right ]_k\right )- H\lefto ( \left [\sum_{j\neq i}^K \mathbf H_{i,j}\mathbf X_j^N \right ]_k \right ) \right ],	\label{eq:limitent}\end{align}
where we used $\snr_k=4^k$.
\end{IEEEproof}

\subsection{Proof of Theorem~\ref{thm:mainb}}


First, we argue that we may assume 
\begin{align}		\sum_{m=1}^M\mathbb E[(X_{i}[m])^2]\leqslant M, \quad i=1,...,K,	\label{eq:apcspecial}	\end{align}
in addition to $H(\floor{X_i})<\infty$ and the existence of all information dimension terms appearing in \eqref{eq:dofallexist}. 
This follows, since first by Lemma~\ref{lem:expsub} we can compute the information dimension by
\ba d\lefto (\left [\sum_{j} \mathbf H_{i,j} X_j \right ]_k\right ) = \lim_{k\to \infty} \frac{H\lefto (\left [\sum_{j} \mathbf H_{i,j}X_j \right ]_k\right )}{k}, \ea
and thus by Lemma~\ref{lem:fract} and the assumption $H(\floor{X_i})<\infty$ the RHS of \eqref{eq:dofallexist} does not change when we replace the inputs by their respective fractional parts. The argument is concluded by noting that  the fractional parts always satisfy \eqref{eq:apcspecial} as a consequence of  $0\leqslant (x)<1$, for $x\in \mathbb R$.


Now let $X_1, ... , X_K$ be independent and such that the information dimension terms in \eqref{eq:dofallexist} exist and the power constraint \eqref{eq:apcspecial} is satisfied. For each $i=1,...,K$ and for given $N$, we take  the $N$ columns of $\mathbf X_i^N$ to be   i.i.d.\ copies of $X_i$. Since the $X_i$ satisfy \eqref{eq:apcspecial}, the $\mathbf X_i^N$ satisfy \eqref{eq:apc}, and by Proposition~\ref{prop:limiting}, we find that
\begin{align}	&C_\text{sum}(\mathbf H; \snr)\\	&\geqslant \lim_{N\to \infty}\frac{1}{N}\sum_{i=1}^K I( \mathbf X^N_i; \mathbf Y^N_i) \\
									&\geqslant \sum_{i=1}^K I( X_i; Y_i)  \label{eq:single}\\
									&=\sum_{i=1}^K\left [I\lefto (\sum_{j=1}^K \mathbf H_{i,j}  X_j,\snr \right )- I\lefto ( \sum_{j\neq i}^K \mathbf H_{i,j} X_j ,\snr \right )\right ]. \label{eq:ach}\end{align}
It therefore follows that 
\ban &\Dof(\mathbf H)\\&=\limsup_{\snr\to\infty}\frac{C_\text{sum}(\mathbf H; \snr)}{\frac{1}{2}\log \snr}\\ &\geqslant \limsup_{\snr\to\infty} \frac{1}{\frac{1}{2}\log \snr}\nonumber \\ &\hspace{1cm}  \sum_{i=1}^K\left [I\lefto (\sum_{j=1}^K \mathbf H_{i,j}  X_j,\snr \right )- I\lefto ( \sum_{j\neq i}^K \mathbf H_{i,j} X_j ,\snr \right )\right ] \\ &\!\!\! \stackrel{\text{Thm.~\ref{thm:guionnet}}}=\sum_{i=1}^K \left [ d \lefto (\sum_{j=1}^K \mathbf H_{i,j} X_j \right )-	 d \lefto ( \sum_{j\neq i}^K \mathbf H_{i,j} X_j \right )\right ],  \label{eq:infdimexist}\\ &=\dof(X_1, ... ,X_K ; \mathbf H)\ean
where we used that, by assumption, the information dimension terms in \eqref{eq:infdimexist} exist.
 \hfill \endIEEEproof

\subsection{Proof of Theorem~\ref{thm:main}}

As achievability, i.e., \begin{align}		\sup_{ X_1, ... , X_K}\dof(X_1, ... ,X_K ; \mathbf H)  \leqslant  \Dof (\mathbf H )	\end{align}
was established in Theorem~\ref{thm:mainb}, it remains to prove the converse.

The proof architecture is as follows. Starting from  Proposition~\ref{prop:outsourced} we have a limiting characterization of $\Dof(\mathbf H)$ based on terms that are related to  information dimension  through Lemma~\ref{lem:expsub}. The main task is to show that for  
arbitrary multi-letter inputs we can construct corresponding single-letter inputs that  achieve the same value in the supremization in Proposition~\ref{prop:outsourced} as the underlying multi-letter inputs. These  single-letter inputs are  obtained by discretizing the underlying multi-letter inputs and ``encoding'' the resulting discrete random vectors using self-similar distributions. 
The corresponding (noiseless) output distributions are then shown to also be self-similar. This means that we can compute the information dimension of the output distributions using Theorem~\ref{thm:infdimifs}. We will, however, have to ensure that the open set condition is satisfied, which will be accomplished by using a result from Diophantine approximation theory.
This result applies to almost all  channel matrices $\mathbf H$ only, with the consequence of the statement of Theorem~\ref{thm:main} being restricted to almost all channels. 
We emphasize that the idea for this construction was first proposed in \cite{WSV11draft} for the scalar case.

We begin by noting that Proposition~\ref{prop:outsourced}  implies that $\Dof(\mathbf H)$ can be expressed  as 
\ban &\Dof(\mathbf H)= \limsup_{k\to\infty}\lim_{N\to\infty} \sup_{\mathbf X_1^N,... , \mathbf X_K^N}\frac{1}{N k}\nonumber \\ & \sum_{i=1}^K \left [  H\lefto (\left [\sum_{j=1}^K \mathbf H_{i,j}\mathbf X_j^N \right ]_k\right )- H\lefto ( \left [\sum_{j\neq i}^K \mathbf H_{i,j}\mathbf X_j^N \right ]_k \right ) \right ], \label{eq:limitent2}\ean
where the supremum is taken over all independent $\mathbf X_1^N, ... ,\mathbf X_K^N$ satisfying the power constraint \eqref{eq:apc}.
We now want to apply Lemma~\ref{lem:fract} to find that we can replace the inputs in \eqref{eq:limitent2} by their fractional parts, which, in turn, means that it suffices to consider
  $\mathbf X_i^N$, $i=1,...,K$, that  take on values in $[0,1]^{M\times N}$ only. For this to work out it suffices to establish an upper bound on $\sum_{j=1}^K H\lefto (\floor{\mathbf X_j^N}\right )$ that is uniform with respect to the inputs $\mathbf X_i^N$ and grows at most linearly in $N$. Such an upper bound follows immediately from Lemma~\ref{lem:linearbound}, taking into account  the  power constraint \eqref{eq:apc}. 
We can therefore conclude that for every $\varepsilon>0$ and every $N_0>0$, there exist $N,k\geqslant N_0$ and independent $\mathbf X_1^N,... ,\mathbf X_K^N$ taking on values in $[0,1]^{M\times N}$ such that
\begin{align}	&\Dof(\mathbf H) \leqslant \varepsilon + \frac{1}{N k} \\ &
 \sum_{i=1}^K \left [ H\lefto (\left [\sum_{j=1}^K \mathbf H_{i,j}\mathbf X_j^N \right ]_k\right )- H\lefto ( \left [\sum_{j\neq i}^K \mathbf H_{i,j}\mathbf X_j^N \right ]_k \right )\right ].	\label{eq:dominate1}\end{align}
It remains to  single-letterize \eqref{eq:limitent2} by constructing  $\widetilde{{X}}_1,... , \widetilde{{X}}_K$ out of the $\mathbf X_1^N,... ,\mathbf X_K^N$ such that \linebreak $\dof(\widetilde{{X}}_1,... , \widetilde{{X}}_K; \mathbf H)$ upper-bounds the RHS of \eqref{eq:dominate1}. This then completes the proof. 

We begin the construction of the single-letter transmit vectors  with preparatory steps.
Consider monomials in the $(KM)^2$ channel coefficients, i.e.,  let $f_1,f_2, ...$ be the monomials\footnote{A ``monomial'' $f$ in $\ell$ variables is a function of the form $f(x_1,...,x_\ell)=x_1^{d_1}x_2^{d_2}\cdots x_\ell^{d_\ell}$  for non-negative integers $d_1,..., d_\ell$.} of all degrees\footnote{The ``degree'' of a monomial is to be understood as the sum of all exponents of the variables involved (sometimes called the total degree).} in $(KM)^2$ variables enumerated as follows:   $f_1,...,f_{\varphi(d)}$  are  the monomials of degree  not larger than $d$, where  
\ba {\varphi(d)}:=\binom{(KM)^2+d}{d} .\ea
As  $\varphi(d)$ and $\varphi(d+1)$ will occur frequently in the remainder of the proof, we simplify the notation by setting $L:=\varphi(d)$ and $L':=\varphi(d+1)$.
Since $L'=\frac{(KM)^2+d+1}{d+1}L$, we have $(L'-L)/L' \xrightarrow{d\to \infty} 0$. Suppose we are given $\varepsilon >0$ in \eqref{eq:dominate1}.  We choose $d$ large enough  so that
 \ban (L'-L)/L'< \varepsilon. \label{eq:deps}\ean 
We will construct single-letter transmit vectors with self-similar homogeneous distributions, with  the corresponding finite set $\mathcal W$ (cf.\ p.~\pageref{p:W}) given by a subset of the lattice of $\mathbb Z$-linear combinations of the monomials $f_1(\mathbf H), f_2(\mathbf H),...$  in the channel coefficients. In order to guarantee the open set condition for the resulting distributions of the noiseless output and noiseless interference, we will need the following two results from Diophantine approximation theory. 
\vspace{.1cm}
\begin{lemma}\label{lem:dio1}
 For $\mathbf A\in \mathbb R^{m\times n}$ define $w(\mathbf A)$ to be the supremum of all $w>0$ such that for arbitrarily large $Q>0$ the set
\ba 	\{  u \in \mathbb Z^{n}\! \setminus\! \{0\}	 \mid  |\mathbf A u |_{\mathbb Z}< Q^{-w}, \|  u\|_\infty \leqslant Q 	\}		\ea
is non-empty. If $g_1,...,g_k$ are $k$ pairwise distinct  monomials of positive degrees, then 
\ban 	w\lefto (\left (g_1(a)\, ... \, g_k(a)\right) ^T\right ) =\frac{1}{k} \quad \text{ and } \quad  w\lefto (\left (g_1(a)\, ... \, g_k(a)\right) \right ) =k	 \label{eq:obtained}\ean
for almost all $a\in \mathbb R^\ell$.
\end{lemma}
\begin{IEEEproof}
See \cite[Thm.~KM, p.~823]{BV10}.
\end{IEEEproof}
\vspace{.1cm}
\begin{lemma}\label{lem:dio2}
Let $\mathbf A\in \mathbb R^{m\times n}$, $\kappa:=2^{1-m-n}((m+n)!)^2$, and $\alpha,\beta>0$. If the set
\ba \left 	\{u\in\mathbb Z^m\!\setminus\! \{0\} \; \middle \vert\; |\mathbf A^Tu|_\mathbb Z< \frac{\kappa}{\alpha}, \|u\|_\infty\leqslant \beta  \right \} 		\ea
is empty, then for all $z\in \mathbb R^m$, the set
\ba \left 	\{v\in\mathbb Z^n\!\setminus\! \{0\} \; \middle \vert\; |\mathbf Av+z|_\mathbb Z\leqslant  \frac{\kappa}{\beta}, \|v\|_\infty\leqslant \alpha  \right \} 		\ea
is non-empty.
\end{lemma}
\begin{IEEEproof}
See \cite[Lem.~3, p.~756]{BL05}.
\end{IEEEproof}

We apply Lemma~\ref{lem:dio1} to find that for almost all matrices $\mathbf H\in \mathbb R^{KM\times KM}$
\ba 	w\lefto (\left (f_2(\mathbf H)\, ... \, f_{L}(\mathbf H)\right) ^T\right ) &=\frac{1}{L-1} \quad \text{ and }\\   w\lefto (\left (f_2(\mathbf H)\, ... \, f_{L'}(\mathbf H)\right) \right ) &=L'-1,	\ea
where $w(\cdot )$ was defined in Lemma~\ref{lem:dio1}; note that we intentionally left out the constant monomial $f_1$. Therefore, for every $s>0$ and almost all $\mathbf H$, there is a $Q_0$ such that for all $Q\geqslant Q_0$, we have
\ban \nonumber 	&\{  u \in \mathbb Z\! \setminus\! \{0\}	 \mid \\ &\hspace{.5cm}  |\left (f_2(\mathbf H)\, ... \, f_L(\mathbf H)\right) ^T u |_{\mathbb Z}<  Q^{-\frac{1}{L-1-s}}, \|  u\|_\infty \leqslant Q 	\}= \emptyset \label{eq:empty1} \\	
&\{  u \in \mathbb Z^{L'-1}\! \setminus\! \{0\}	 \mid \nonumber \\ &\hspace{.5cm}  |\left (f_2(\mathbf H)\, ... \, f_{L'}(\mathbf H)\right)  u |_{\mathbb Z}< Q^{-L'+1-s}, \| u\|_\infty \leqslant  Q 	\}= \emptyset	. 	\label{eq:empty2}\ean
We choose $s>0$ to be small enough for
\ban 	\left |\frac{L'-L+2s}{L'} - \frac{L'-L}{L'} \right |<\varepsilon	\label{eq:Leps}\ean
to hold. Since \eqref{eq:empty1} and \eqref{eq:empty2} hold, individually,  for almost all channel matrices $\mathbf H$ it follows that they also hold simultaneously for almost all $\mathbf H$. 
 In the remainder of the proof, we assume that  $\mathbf H$ satisfies both \eqref{eq:empty1} and \eqref{eq:empty2}. 
By the scaling invariance in  Lemma~\ref{lem:scaleinv}, we may furthermore assume that $\|\mathbf H \|_\infty \leqslant 1$.
Next, for a given finite set $\mathcal A\subseteq \mathbb R$, we define the quantizer to $\mathcal A$ as 
\ba \mathsf Q_{\mathcal A}(x) := \argmin_{a\in \mathcal A}|x-a| ,  \ea
with ties broken arbitrarily. For an arbitrary  real constant $T>0$, define the following sets 
\ban 	&\mathcal V:= \nonumber  \frac{1}{T}\Bigg \{\sum_{i=1}^{L}	u_if_i(\mathbf H) \; \Big \vert \\ &\; u_1,...,u_{L}\in\mathbb Z, |u_1|\leqslant LT, |u_2|,...,|u_{L}|\leqslant T \Bigg\}\subseteq [-2{L},2L],\label{eq:defV}\\
&\mathcal V':=\frac{1}{T}\Bigg\{\sum_{i=1}^{L'}	u_if_i(\mathbf H) \; \Big \vert \nonumber \\ & \; u_1,...,u_{L'}\in\mathbb Z, |u_1|,...,|u_{L'}|\leqslant LT \Bigg \}\subseteq [-L'L,L'L],	\label{eq:defV'}\ean
which will be used in our construction of the self-similar single-letter input distributions.
The choice of $\mathcal V$ and $\mathcal V'$ is adapted to the channel in the sense that if $x_1,...,x_K\in \mathcal V^M$ then $\sum_{j=1}^K\mathbf H_{i,j} x_j \in \mathcal V'^M$, since multiplication by a channel coefficient only increases the degrees of the involved monomials by $1$.

We next discretize the inputs $\mathbf X_1^N,... ,\mathbf X_K^N$  by quantizing their entries to the set $\mathcal V$ and characterize the resulting quantization error through \eqref{eq:empty1} and Lemma~\ref{lem:dio2}.
To  this end,  we set $Q=\left(\frac{T}{\kappa}\right ) ^{L-1-s}$ and apply Lemma~\ref{lem:dio2} with $\mathbf A=\left (f_2(\mathbf H), ... , f_L(\mathbf H)\right) $, $\alpha=T,\beta=Q,m=1, n=L-1$, and $\kappa=2^{1-L}(L!)^2$;  the  condition of   Lemma~\ref{lem:dio2} is satisfied by \eqref{eq:empty1} if we choose $T$ large enough for $Q\geqslant Q_0$ to hold. Lemma~\ref{lem:dio2} then says that for every $\theta\in [0,1]$, there exists a $v= (v_2\, ...\,  v_L)^T \in \mathbb Z^{L-1}\!\setminus\! \{0\} $ with $\|v\|_\infty\leqslant T$ such that
\ban |(\left (f_2(\mathbf H), ... , f_L(\mathbf H)\right)  v - T\theta |_{\mathbb Z} \leqslant \frac{\kappa^{L-s}}{T^{L-1-s}}. \label{eq:above} \ean
Since $\|\mathbf H\|_\infty \leqslant 1$ we have $f_i(\mathbf H)\leqslant 1$, for all $i$, and thus $|(\left (f_2(\mathbf H), ... , f_L(\mathbf H)\right)  v - T\theta | \leqslant LT$. Therefore, the integer that best approximates $(\left (f_2(\mathbf H), ... , f_L(\mathbf H)\right)  v - T\theta$ also has absolute value $\leqslant LT$. Denote   this best approximating integer by $v_1$. 
Since $f_1$ is the constant monomial, dividing both sides of \eqref{eq:above} by $T$ yields\ba \left |\frac{1}{T} \sum_{i=1}^L v_if_i(\mathbf H) -\theta \right | \leqslant \left (\frac{\kappa}{T}\right )^{L-s}. \ea
We thus  have
\ban \sup_{\theta\in[0,1]}|\mathsf Q_{\mathcal V}(\theta)-\theta|\leqslant \left (\frac{\kappa}{T}\right )^{L-s} . \label{eq:quanterror}\ean
Next, we lower-bound the minimum distance between elements of $\mathcal V'$ using \eqref{eq:empty2}.  Note that 
\ba &\mathcal V' - \mathcal V' =\\ & \frac{1}{T}\left \{\sum_{i=1}^{L'}	u_if_i(\mathbf H) \; \middle \vert \; u_1,...,u_{L'}\in\mathbb Z, |u_1|,...,|u_{L'}|\leqslant 2LT \right \}. \ea
We choose $T$ large enough for $2LT\geqslant Q_0$ to hold. By \eqref{eq:empty2} we then get that for all $u=(u_2\,...\,u_{L'})^T\in\mathbb Z^{L'-1}\! \setminus\!  \{0\}$ with $\|u\|_\infty\leqslant 2LT$
\ba |\left (f_2(\mathbf H), ... , f_{L'}(\mathbf H)\right)  u |_{\mathbb Z}\geqslant  (2LT)^{-L'+1-s}	\ea
must hold. Therefore, no matter how we choose the integer $u_1$, we have 
\ba \left |\sum_{i=1}^{L'}	u_if_i(\mathbf H) \right |\geqslant (2LT)^{-L'+1-s},	\ea
and thus 
\ban \mathsf m(\mathcal V')\geqslant\frac{(2LT)^{-L'+1-s}}{T}. \label{eq:mindist} \ean

Before getting into the juice of the argument, we need some preparatory material, which explains how to choose the parameter $N_0$ (from \eqref{eq:dominate1}) sufficiently large to obtain  bounds guaranteeing that the open set condition for the noiseless output distributions is satisfied. The reader can safely skip this part initially and consult it when needed later in the proof.  
First, note that the expression
\ban \frac{1}{k} \log\lefto (c_1+c_2 2^{c_3 k } \right )  \ean
for positive constants $c_1,c_2, c_3$ converges to $c_3$ as $k\to \infty$. We can therefore take $N_0$ large enough such that for all $k\geqslant N_0$  we have 
\ban 
\left |	\frac{1}{k} \log\lefto ( c_1+  c_2 2^{\frac{L'-L+2s}{L'}k} \right ) 	-\frac{L'-L+2s}{L'}\right |& <\varepsilon	 , \label{eq:logneu2}\ean
where $ c_1,  c_2$ will be specified below and $\varepsilon$ was chosen above. Finally, we set
\ban T:= 2^{\frac{k}{L'}}  \label{eq:defT}\ean
and pick $N_0$ large enough such that for all $k\geqslant N_0$,  in addition to \eqref{eq:logneu2}, we also have 
\ban 2^{-k} &\leqslant \frac{(2L)^{-L'+1-s}T^{-L'-s}}{16KML} .\label{eq:cond1} \ean

%

We are now ready to proceed to the finale of the proof and single-letterize \eqref{eq:dominate1}.
First, we set $\widetilde{\mathbf X}_i^N:= \mathsf Q_{\mathcal V} (\mathbf X_i^N)$, for $i=1,...,K$,  where $\mathsf Q_{\mathcal V}(\cdot )$ is applied entry-wise. Note that 
\ban &\left \|\left [\sum_{j=1}^K \mathbf H_{i,j}\mathbf X_j^N\right ]_k - \sum_{j=1}^K \mathbf H_{i,j}\widetilde{\mathbf X}_j^N\right \|_\infty \\  & \leqslant 2^{-k} + \left \|\sum_{j=1}^K \mathbf H_{i,j}(\mathbf X_j^N - \widetilde{\mathbf X}_j^N)\right \|_\infty \label{eq:mehrdetail0}\\ &\leqslant 2^{-k} + \sum_{j=1}^K \left \| \mathbf H_{i,j}(\mathbf X_j^N - \widetilde{\mathbf X}_j^N)\right \|_\infty\\ &\!\! \stackrel{\eqref{eq:quanterror}}{\leqslant}
 2^{-k}+K M\left (\frac{\kappa}{T}\right )^{L-s}. \label{eq:mehrdetail1}\ean
Applying Lemma~\ref{lem:distanceentropy} with $\varepsilon= 2^{-k}+KM \left (\frac{\kappa}{T}\right )^{L-s}$ and $\delta=2^{-k}$ which is the minimum distance of the value set for $\left [\sum_{j=1}^K \mathbf H_{i,j}\mathbf X_j^N\right ]_k$, we find
\ban 
	&\frac{1}{Nk}\left ( H\lefto (\left [\sum_{j=1}^K \mathbf H_{i,j}\mathbf X_j^N \right ]_k\right )-H\lefto (\sum_{j=1}^K \mathbf H_{i,j}\widetilde{\mathbf X}_j^N\right ) \right ) \\ &\leqslant 	\frac{M}{k}\log \lefto (1+ \frac{2\left ( 2^{-k}+KM \left (\frac{\kappa}{T}\right )^{L-s}\right )}{2^{-k}}\right ) \label{eq:collecterror0}\\ &=\frac{M}{k}\log \lefto (3+2KM \kappa^{L-s}2^k T^{-L+s}\right )\\ &\!\! \stackrel{\eqref{eq:defT}}{=}  \frac{M}{k}\log \lefto (3+2KM \kappa^{L-s}2^{\left (1-\frac{L-s}{L'}\right ) k} \right )\\
&\leqslant \frac{M}{k}\log \lefto (3+2KM \kappa^{L-s}2^{\frac{L'-L+2s}{L'} k} \right )\\
 &\!\! \stackrel{\eqref{eq:logneu2}}{\leqslant} M \left (\frac{L'-L+2s}{L'}+\varepsilon \right )\\ &\!\! \stackrel{\eqref{eq:Leps}}{\leqslant} M \left (\frac{L'-L}{L'}+2\varepsilon \right )\\ &\!\! \stackrel{\eqref{eq:deps}}{\leqslant} 3M\varepsilon. \label{eq:collecterror}
\ean
Similarly, again using Lemma~\ref{lem:distanceentropy} with $\varepsilon$ as before and $\delta=(2L)^{-L'+1-s}T^{-L'-s}$, we obtain by \eqref{eq:mindist} that
\ban 
	&\frac{1}{Nk}\left ( H\lefto (\sum_{j=1}^K \mathbf H_{i,j}\widetilde{\mathbf X}_j^N\right ) -H\lefto (\left [\sum_{j=1}^K \mathbf H_{i,j}\mathbf X_j^N \right ]_k\right )\right )\\ &\leqslant 	\frac{M}{k}\log \lefto (1+ \frac{2\left ( 2^{-k}+KM \left (\frac{\kappa}{T}\right )^{L-s}\right )}{(2L)^{-L'+1-s}T^{-L'-s}}\right )\\ 
	&=	\frac{M}{k}\log \! \bigg (1+2 (2L)^{L'-1+s}T^{L'+s} \phantom{3}\nonumber \\ &\hspace{1.8cm}   \times \left ( 2^{-k}+KM \left (\frac{\kappa}{T}\right )^{L-s}\right ) \bigg )\\
&\!\! \stackrel{\eqref{eq:defT}}{=}  	\frac{M}{k}\log \! \bigg (1 +2 (2L)^{L'-1+s}KM\kappa^{L-s} \phantom{3}\nonumber \\ &\hspace{1.8cm} \times \left (2^{\left (\frac{L'+s}{L'}-1\right ) k}+2^{\frac{L'-L+2s}{L'} k}\right ) \bigg) \\&\; {\leqslant}   \;	\frac{M}{k}\log \! \bigg  (1+2 (2L)^{L'-1+s}KM\kappa^{L-s} \phantom{3} \nonumber \\ &\hspace{1.8cm} \times  \left (2^{\frac{L'-L+2s}{L'} k}+2^{\frac{L'-L+2s}{L'} k}\right ) \bigg ) \label{eq:alsoused}\\
&\!\! \stackrel{\eqref{eq:logneu2}}{\leqslant} M \left (\frac{L'-L+2s}{L'}+\varepsilon \right )\\ &\!\! \stackrel{\eqref{eq:Leps}}{\leqslant} M \left (\frac{L'-L}{L'}+2\varepsilon \right )\\ &\!\! \stackrel{\eqref{eq:deps}}{\leqslant} 3M\varepsilon ,\label{eq:collecterror2}
\ean
where in \eqref{eq:alsoused} we used  $s/L' \leqslant (L'-L+2s)/L'$. In summary, we have shown that 
\ban \frac{1}{Nk}\left | H\lefto (\left [\sum_{j=1}^K \mathbf H_{i,j}\mathbf X_j^N \right ]_k\right )-H\lefto (\sum_{j=1}^K \mathbf H_{i,j}\widetilde{\mathbf X}_j^N\right )\right | \leqslant 3M\varepsilon. \label{eq:therefore}\ean

We next let $r=2^{-k}$ and set
\ban W_i:=\sum_{n=1}^N r^{n-1} \widetilde X_i^{(n)} , \quad i=1,...,K,\label{eq:ifstransmit2} \ean
where $\widetilde{\mathbf X}_i^N= (\widetilde X_i^{(1)}\, ... \; \widetilde X_i^{(N)})$. The single-letter transmit vectors are then constructed as
\ban \widetilde X_i:=\sum_{\ell=0}^\infty  r^{N\ell } W_{i,\ell}, \quad i=1,...,K, \label{eq:ifstransmit}\ean
where $\{ W_{i,\ell}\}_{\ell\geqslant 0}$ are i.i.d.\ copies of the discrete random vector $W_i$. By \eqref{eq:ifs} it follows that the $\widetilde X_i$ have self-similar homogeneous distributions with similarity ratio $r^N$. 
Note that although Proposition~\ref{prop:sing} requires the input distributions only to have a singular component, for simplicity, we take them to be purely singular.
The essence of this construction is that the noiseless channel output vectors   
\ba \sum_{j=1}^K \mathbf H_{i,j}\widetilde X_j = \sum_{\ell=0}^\infty r^{N\ell}  \sum_{j=1}^K \mathbf H_{i,j}W_{j,\ell},\quad \text{$i=1,...,K$},	\ea
  again have self-similar homogeneous distributions with similarity ratio $r^N$. Note that $W_i$ takes value in $\mathcal W:=\sum_{n=1}^Nr^{n-1}\mathcal V^M$. 
Since multiplication of an element in $\mathcal V$ by a channel coefficient increases the degrees of the involved monomials by $1$, we have  $\sum_{j=1}^K\mathbf H_{i,j}\mathcal V^M\subseteq \mathcal V'^M$. Therefore, it follows that $\mathsf m (\sum_{j=1}^K\mathbf H_{i,j}\mathcal V^M )\geqslant \mathsf m(\mathcal V'^M)=\mathsf m(\mathcal V')$. Moreover, by \eqref{eq:defV} and $\|\mathbf H \|_\infty\leqslant 1$ we have $\mathsf M(\sum_{j=1}^K\mathbf H_{i,j}\mathcal V^M)\leqslant 4KML$, and  thus get
\ba &\frac{\mathsf m(\sum_{j=1}^K\mathbf H_{i,j}\mathcal V^M)}{\mathsf m(\sum_{j=1}^K\mathbf H_{i,j}\mathcal V^M)+\mathsf M(\sum_{j=1}^K\mathbf H_{i,j}\mathcal V^M)}\\ &\geqslant \frac{\mathsf m(\sum_{j=1}^K\mathbf H_{i,j}\mathcal V^M)}{2\mathsf M(\sum_{j=1}^K\mathbf H_{i,j}\mathcal V^M)}\\ &\!\!\stackrel{\eqref{eq:mindist}}{\geqslant} \frac{(2L)^{-L'+1-s}T^{-L'-s}}{8KML}\\ &\!\!\stackrel{\eqref{eq:cond1}}{\geqslant} r. \ea
By Lemma~\ref{lem:entgeneral} this implies 
\ban 	\mathsf m \lefto (\sum_{j=1}^K\mathbf H_{i,j} \mathcal W \right ) &= \mathsf m \lefto (\sum_{j=1}^K\mathbf H_{i,j} \sum_{n=1}^Nr^{n-1}\mathcal V^M \right )  \label{eq:mindistpower1}\\ &= \mathsf m \lefto (\sum_{n=1}^Nr^{n-1} \sum_{j=1}^K\mathbf H_{i,j} \mathcal V^M \right )  \label{eq:mindistpower2}\\ &\geqslant r^{N-1} 	\mathsf m\lefto  (\sum_{j=1}^K\mathbf H_{i,j} \mathcal V^M \right ) .	 \label{eq:mindistpower}\ean
For the maximum distance we have $\mathsf M(\sum_{j=1}^K \mathbf H_{i,j}\mathcal W)=\mathsf M(\sum_{j=1}^K\mathbf H_{i,j} \sum_{n=1}^Nr^{n-1}\mathcal V^M)\leqslant 8KML$, by \eqref{eq:defV} and $\sum_{n=1}^N r^{n-1}\leqslant 2$. We  obtain
\ba 	r^N &\!\! \stackrel{\eqref{eq:cond1}}{\leqslant}r^{N-1}\frac{(2L)^{-L'+1-s}T^{-L'-s}}{16KML}\\ &\!\! \stackrel{\eqref{eq:mindist}}{\leqslant} r^{N-1} \frac{	\mathsf m\lefto  (\sum_{j=1}^K\mathbf H_{i,j} \mathcal V^M \right )}{16KML}\\ &\!\! \stackrel{\eqref{eq:mindistpower}}{\leqslant} \frac{	\mathsf m \lefto (\sum_{j=1}^K\mathbf H_{i,j} \mathcal W \right )}{	2\mathsf M \lefto (\sum_{j=1}^K\mathbf H_{i,j} \mathcal W \right ) } \\ &\leqslant \frac{	\mathsf m \lefto (\sum_{j=1}^K\mathbf H_{i,j} \mathcal W \right )}{\mathsf m \lefto (\sum_{j=1}^K\mathbf H_{i,j} \mathcal W \right )+	\mathsf M \lefto (\sum_{j=1}^K\mathbf H_{i,j} \mathcal W \right ) } .	\ea
It thus follows from Lemma~\ref{lem:openset} that the distribution of $\sum_{j=1}^K \mathbf H_{i,j} \widetilde X_j$ satisfies the open set condition. By Theorem~\ref{thm:infdimifs}, we then get
\ban d\lefto ( \sum_{j=1}^K \mathbf H_{i,j} \widetilde X_j\right )&=\frac{H\lefto (\sum_{j=1}^K \mathbf H_{i,j} W_j \right )}{N \log \frac{1}{r}}\\ &=\frac{H\lefto (\sum_{j=1}^K \mathbf H_{i,j} W_j \right )}{N k},		\label{eq:finalfinal1}\ean 
and Lemma~\ref{lem:entgeneral} yields
\ban H\lefto (\sum_{j=1}^K \mathbf H_{i,j} W_j \right )=H\lefto (\sum_{j=1}^K  \mathbf H_{i,j}\widetilde{\mathbf X}_j^N \right ), \label{eq:finalfinal2}\ean
resulting in 
\ban d\lefto ( \sum_{j=1}^K \mathbf H_{i,j} \widetilde X_j\right )=H\lefto (\sum_{j=1}^K  \mathbf H_{i,j}\widetilde{\mathbf X}_j^N \right ). \label{eq:finalfinal3}\ean 
We can apply  the  same  arguments to the quantity $\sum_{j\neq i}^K \mathbf H_{i,j}\mathbf X_j^N$ and use the result thereof along with \eqref{eq:finalfinal3} in the sum on the RHS of \eqref{eq:dominate1}. Collecting  the errors from \eqref{eq:collecterror} and \eqref{eq:collecterror2}, we finally get
\ba  &\Dof(\mathbf H)\\ &\leqslant 
\varepsilon +  6MK\varepsilon +  \sum_{i=1}^K \left [ d\lefto (\sum_{j=1}^K \mathbf H_{i,j}\widetilde{ X}_j \right )- d\lefto (\sum_{j\neq i}^K \mathbf H_{i,j}\widetilde{ X}_j \right )\right ]\\ &=\varepsilon +  6MK\varepsilon  + \dof(\widetilde{{X}}_1,... , \widetilde{{X}}_K; \mathbf H). \ea
Since $\varepsilon>0$ can be chosen arbitrarily small, this completes the proof.  
 \hfill \endIEEEproof

\subsection{Proof of Theorem~\ref{thm:maind}}

The proof is along the same lines as that of   Theorem~\ref{thm:main}. We therefore  detail only the steps that require modification. Specifically, we start from \eqref{eq:dominate1} which states that  for every $\varepsilon>0$ and every $N_0>0$, there exist $N,k\geqslant N_0$ and independent $\mathbf X_1^N,... ,\mathbf X_K^N$ that take value in $[0,1]^{M\times N}$ such that
\begin{align}	&\Dof(\mathbf H)\leqslant 
\varepsilon + \frac{1}{N k} \nonumber  \\ &\sum_{i=1}^K \left [ H\lefto (\left [\sum_{j=1}^K \mathbf H_{i,j}\mathbf X_j^N \right ]_k\right )- H\lefto ( \left [\sum_{j\neq i}^K \mathbf H_{i,j}\mathbf X_j^N \right ]_k \right )\right ].	\label{eq:dominate}\end{align}
Since we assume all entries of $\mathbf H$ to be rational, we may use the scaling invariance of $\Dof(\mathbf H)$ formalized in Lemma~\ref{lem:scaleinv} to argue that it suffices to consider $\mathbf H$ with integer entries only. This drastically simplifies controlling the distance properties of the noiseless outputs. Specifically, for $\mathbf H$ with all entries rational, we can employ codebooks built from integers,  which allows us to  control the distances of the noiseless outputs for \emph{all}  channel matrices $\mathbf H$.
More concretely, we replace the sets $\mathcal V$ and $\mathcal V'$ in \eqref{eq:defV}, \eqref{eq:defV'}  by 
\ban 	\mathcal V&:=2^{-(k-p)}\{0, 1, ..., 2^{k-p}\}, \label{eq:vneu1} \\ 	\mathcal V' &:=2^{-(k-p)}\{-KMH_{\text{max}}2^{k-p} ,-KMH_{\text{max}}2^{k-p}+1, ... , \nonumber \\  &\hspace{2cm} KMH_{\text{max}}2^{k-p}\}, \label{eq:vneu2}\ean
for $k\geqslant p$, where $H_{\text{max}}:=\| \mathbf H\|_\infty$ and  $p$ is a positive integer such that
\ban 	2^{-p}\leqslant \frac{1}{8KMH_{\text{max}}}		.	 \label{eq:scond} \ean
 Then, we have
\ban \sup_{\theta\in[0,1]}|\mathsf Q_{\mathcal V}(\theta)-\theta|&\leqslant 2^{-(k-p)}, \label{eq:quanterror2}\\  \mathsf m(\mathcal V')&\geqslant 2^{-(k-p)},\label{eq:quanterror3}\\ \mathsf M(\mathcal V')&\leqslant 2KM H_{\text{max}}.\label{eq:quanterror4}\ean
Setting $\widetilde{\mathbf X}_i^N:= \mathsf Q_{\mathcal V} (\mathbf X_i^N)$, for $i=1,...,K$, we get  $\left \|\left [\sum_{j=1}^K \mathbf H_{i,j}\mathbf X_j^N \right ]_k - \sum_{j=1}^K \mathbf H_{i,j}\widetilde{\mathbf X}_j^N\right \|_\infty\leqslant 2^{-k}+2^{-(k-p)}KMH_{\text{max}}$ by \eqref{eq:quanterror2} (cf.\ \eqref{eq:mehrdetail0}--\eqref{eq:mehrdetail1}). Through  application of  Lemma~\ref{lem:distanceentropy} and using \eqref{eq:quanterror3}, we  find that
\ban 
	&\frac{1}{Nk}\left | H\lefto (\left [\sum_{j=1}^K \mathbf H_{i,j}\mathbf X_j^N \right ]_k\right )-H\lefto (\sum_{j=1}^K \mathbf H_{i,j}\widetilde{\mathbf X}_j^N\right ) \right |\\ &\leqslant \frac{M}{k}\log \lefto  (1+2 \frac{2^{-k}+2^{-(k-p)}KMH_{\text{max}}}{2^{-k}}\right )\\ &\leqslant \frac{M}{k}\log (3+2KMH_{\text{max}} 2^{p}). \label{eq:smallenough} \ean
Choosing $N_0$ sufficiently large, we ensure that \eqref{eq:smallenough} is as small as desired for all $k\geqslant N_0$. The construction of the single-letter transmit vectors now follows the exact same steps as in the proof of Theorem~\ref{thm:main}. 
The only element that differs from what was done there is the choice of the sets $\mathcal V$ and $\mathcal V'$ in \eqref{eq:vneu1}, \eqref{eq:vneu2}. We therefore only need to verify the open set condition for the iterated function system constructed according to \eqref{eq:ifstransmit2} and \eqref{eq:ifstransmit}  using $\mathcal V$ and $\mathcal V'$ in \eqref{eq:vneu1}, \eqref{eq:vneu2}. 
First, we note that since the entries of the matrices  $\mathbf H_{i,j}$  are integers, we have  $\sum_{j=1}^K\mathbf H_{i,j}\mathcal V^M\subseteq \mathcal V'^M$ which implies $\mathsf m (\sum_{j=1}^K\mathbf H_{i,j}\mathcal V^M )\geqslant \mathsf m(\mathcal V'^M)=\mathsf m(\mathcal V')$ and $\mathsf M (\sum_{j=1}^K\mathbf H_{i,j}\mathcal V^M )\leqslant \mathsf M(\mathcal V'^M)= \mathsf M(\mathcal V')$. We set  $r:=2^{-k}$. Using \eqref{eq:quanterror3} and \eqref{eq:quanterror4} we find  that
\ban &\frac{\mathsf m(\sum_{j=1}^K\mathbf H_{i,j}\mathcal V^M)}{\mathsf m(\sum_{j=1}^K\mathbf H_{i,j}\mathcal V^M)+\mathsf M(\sum_{j=1}^K\mathbf H_{i,j}\mathcal V^M)}\\ &\geqslant \frac{\mathsf m(\sum_{j=1}^K\mathbf H_{i,j}\mathcal V^M)}{2\mathsf M(\sum_{j=1}^K\mathbf H_{i,j}\mathcal V^M)}\\ &\geqslant \frac{\mathsf m(\mathcal V')}{2\mathsf M(\mathcal V')}\\&\!\!\!\!\!\!\! \stackrel{\eqref{eq:quanterror3},\eqref{eq:quanterror4}}{\geqslant} \frac{2^{-(k-p)}}{4KMH_{\text{max}}}\\ &\!\! \stackrel{\eqref{eq:scond}}{\geqslant} r. \label{eq:esta}\ean
It thus follows by Lemma~\ref{lem:entgeneral} that  for $\mathcal W:=\sum_{n=1}^Nr^{n-1}\mathcal V^M$ (cf.\ \eqref{eq:mindistpower1}--\eqref{eq:mindistpower})
\ban 	\mathsf m \lefto (\sum_{j=1}^K\mathbf H_{i,j} \mathcal W \right ) &\geqslant r^{N-1} 	\mathsf m\lefto  (\sum_{j=1}^K\mathbf H_{i,j} \mathcal V^M \right ).	\label{eq:implies1} \ean
Moreover note that
 $\sum_{j=1}^K \mathbf H_{i,j}\mathcal W\subseteq \sum_{n=1}^Nr^{n-1}\mathcal V'^M$ and therefore  $\mathsf M(\sum_{j=1}^K \mathbf H_{i,j}\mathcal W)\leqslant 4KMH_{\text{max}}$, since $\sum_{n=1}^N r^{n-1}\leqslant 2$. This implies
\ban 	r^N &\!\! \stackrel{\eqref{eq:scond}}{\leqslant} r^{N-1}\frac{2^{-(k-p)}}{8KMH_{\text{max}}}\\ &\!\! \stackrel{\eqref{eq:quanterror3}}{\leqslant} r^{N-1} \frac{	\mathsf m\lefto  (\sum_{j=1}^K\mathbf H_{i,j} \mathcal V^M \right )}{8KMH_{\text{max}}}\\ &\!\! \stackrel{\eqref{eq:implies1}}{\leqslant} \frac{	\mathsf m \lefto (\sum_{j=1}^K\mathbf H_{i,j} \mathcal W \right )}{	2\mathsf M \lefto (\sum_{j=1}^K\mathbf H_{i,j} \mathcal W \right ) } \\ &\leqslant \frac{	\mathsf m \lefto (\sum_{j=1}^K\mathbf H_{i,j} \mathcal W \right )}{\mathsf m \lefto (\sum_{j=1}^K\mathbf H_{i,j} \mathcal W \right )+	\mathsf M \lefto (\sum_{j=1}^K\mathbf H_{i,j} \mathcal W \right ) } . \label{eq:estb}	\ean
All the remaining steps follow exactly the proof of Theorem~\ref{thm:main}.
\endIEEEproof


\addtolength{\textheight}{-18.6cm}

\subsection{Proof of Theorem~\ref{thm:mainc}}

We first note that the limiting characterization in Proposition~\ref{prop:limiting} holds for general noise distributions satisfying the assumptions of Theorem~\ref{thm:mainc}. Next, we conclude that Proposition~\ref{prop:outsourced} also  extends to noise distributions that satisfy the assumptions of Theorem~\ref{thm:mainc}, as the only conditions on the noise distributions used in the proof of Proposition~\ref{prop:outsourced} are those needed by Lemma~\ref{lem:neu}, which, in turn, precisely equal the assumptions in Theorem~\ref{thm:mainc}. This establishes the claim.
%
 \hfill \endIEEEproof


\begin{thebibliography}{10}
\providecommand{\url}[1]{#1}
\csname url@samestyle\endcsname
\providecommand{\newblock}{\relax}
\providecommand{\bibinfo}[2]{#2}
\providecommand{\BIBentrySTDinterwordspacing}{\spaceskip=0pt\relax}
\providecommand{\BIBentryALTinterwordstretchfactor}{4}
\providecommand{\BIBentryALTinterwordspacing}{\spaceskip=\fontdimen2\font plus
\BIBentryALTinterwordstretchfactor\fontdimen3\font minus
  \fontdimen4\font\relax}
\providecommand{\BIBforeignlanguage}[2]{{%
\expandafter\ifx\csname l@#1\endcsname\relax
\typeout{** WARNING: IEEEtran.bst: No hyphenation pattern has been}%
\typeout{** loaded for the language `#1'. Using the pattern for}%
\typeout{** the default language instead.}%
\else
\language=\csname l@#1\endcsname
\fi
#2}}
\providecommand{\BIBdecl}{\relax}
\BIBdecl

\bibitem{SB12Allerton}
D.~Stotz and H.~B\"olcskei, ``Degrees of freedom in vector interference
  channels,'' \emph{Proc. 50th Ann. Allerton Conf. on Communication, Control,
  and Computing}, pp. 1755--1760, {Oct.} 2012.

\bibitem{SB14extended}
------, ``ADDENDUM to ``{D}egrees of freedom in vector interference
  channels'','' \emph{available online:
  \burl{http://www.nari.ee.ethz.ch/commth/research/downloads/dof_addendum.pdf}},
  {Feb.} 2016.

\bibitem{WSV11draft}
Y.~Wu, S.~Shamai~(Shitz), and S.~Verd\'u, ``A formula for the degrees of
  freedom of the interference channel,'' \emph{IEEE Trans. Inf. Theory},
  vol.~61, no.~1, pp. 256--279, {Jan.} 2015.

\bibitem{GS07}
A.~Guionnet and D.~Shlyakhtenko, ``On classical analogues of free entropy
  dimension,'' \emph{Journal of Functional Analysis}, vol. 251, pp. 738--771,
  {Oct.} 2007.

\bibitem{KM12}
I.~Kontoyiannis and M.~Madiman, ``Sumset and inverse sumset inequalities for
  differential entropy and mutual information,'' \emph{IEEE Trans. Inf.
  Theory}, vol.~60, no.~8, pp. 4503--4514, {Aug.} 2014.

\bibitem{CJ08}
V.~R. Cadambe and S.~A. Jafar, ``Interference alignment and degrees of freedom
  of the {K}-user interference channel,'' \emph{IEEE Trans. Inf. Theory},
  vol.~54, no.~8, pp. 3425--3441, Aug. 2008.

\bibitem{MMK08}
M.-A. Maddah-Ali, A.~S. Motahari, and A.~K. Khandani, ``Communication over
  {MIMO X} channels: Interference alignment, decomposition, and performance
  analysis,'' \emph{IEEE Trans. Inf. Theory}, vol.~54, no.~8, pp. 3454--3470,
  {Aug.} 2008.

\bibitem{MM09}
A.~S. Motahari, S.~O. Gharan, M.-A. Maddah-Ali, and A.~K. Khandani, ``Real
  interference alignment: Exploiting the potential of single antenna systems,''
  \emph{IEEE Trans. Inf. Theory}, vol.~60, no.~8, pp. 4799--4810, {Aug.} 2014.

\bibitem{EO09}
R.~H. Etkin and E.~Ordentlich, ``The degrees-of-freedom of the {K}-user
  {Gaussian} interference channel is discontinuous at rational channel
  coefficients,'' \emph{IEEE Trans. Inf. Theory}, vol.~55, no.~11, pp.
  4932--4946, {Nov.} 2009.

\bibitem{CJ09}
V.~R. Cadambe and S.~A. Jafar, ``Parallel {Gaussian} interference channels are
  not always separable,'' \emph{IEEE Trans. Inf. Theory}, vol.~55, no.~9, pp.
  3983--3990, {Sep.} 2009.

\bibitem{Jaf11}
S.~A. Jafar, ``Interference alignment --- {A} new look at signal dimensions in
  a communication network,'' \emph{Foundations and Trends in Communications and
  Information Theory}, vol.~7, no.~1, 2011.

\bibitem{Rey59}
A.~R\'enyi, ``On the dimension and entropy of probability distributions,''
  \emph{Acta Mathematica Hungarica}, vol.~10, pp. 1--23, {Mar.} 1959.

\bibitem{Wu11}
Y.~Wu, ``Shannon theory for compressed sensing,'' \emph{Ph.D. dissertation,
  Princeton University}, Sep. 2011.

\bibitem{PR80}
A.~Peled and A.~Ruiz, ``Frequency domain data transmission using reduced
  computational complexity algorithms,'' \emph{Proc. IEEE Int. Conf. on
  Acoustics, Speech, and Sig. Proc.}, vol.~5, pp. 964--967, {Apr.} 1980.

\bibitem{GK12}
A.~El~Gamal and Y.-H. Kim, \emph{Network Information Theory}.\hskip 1em plus
  0.5em minus 0.4em\relax Cambridge, UK: Cambridge University Press, 2012.

\bibitem{Dur10}
R.~Durrett, \emph{Probability: Theory and Examples}.\hskip 1em plus 0.5em minus
  0.4em\relax New York, NY: Cambridge University Press, 2010.

\bibitem{Tay97}
J.~Taylor, \emph{An Introduction to Measure and Probability}.\hskip 1em plus
  0.5em minus 0.4em\relax New York, NY: Springer, 1997.

\bibitem{WV10}
Y.~Wu and S.~Verd\'u, ``R\'enyi information dimension: Fundamental limits of
  almost lossless analog compression,'' \emph{IEEE Trans. Inf. Theory},
  vol.~56, no.~8, pp. 3721--3748, {Aug.} 2010.

\bibitem{Hut81}
J.~E. Hutchinson, ``Fractals and self similarity,'' \emph{Indiana University
  Mathematics Journal}, vol.~30, pp. 713--747, 1981.

\bibitem{BHR05}
C.~Brandt, N.~Viet~Hung, and H.~Rao, ``On the open set condition for
  self-similar fractals,'' \emph{Proc. of the American Mathematical Society},
  vol. 134, no.~5, pp. 1369--1374, {Oct.} 2005.

\bibitem{GH89}
J.~S. Geronimo and D.~P. Hardin, ``An exact formula for the measure dimensions
  associated with a class of piecewise linear maps,'' \emph{Constructive
  Approximation}, vol.~5, pp. 89--98, {Dec.} 1989.

\bibitem{You82}
L.-S. Young, ``Dimension, entropy and {Lyapunov} exponents,'' \emph{Ergod. Th.
  {\&} Dynam. Sys.}, vol.~2, pp. 109--124, {Mar.} 1982.

\bibitem{SB14}
D.~Stotz and H.~B\"olcskei, ``Explicit and almost sure conditions for {$K/2$}
  degrees of freedom,'' \emph{Proc. IEEE Int. Symp. on Inf. Theory}, pp.
  471--475, Jun. 2014.

\bibitem{Ahl71}
R.~Ahlswede, ``Multi-way communication channels,'' \emph{Proc. of the 2nd Int.
  Symp. on Inf. Theory}, pp. 23--52, Sep. 1971.

\bibitem{WV12a}
Y.~Wu and S.~Verd\'u, ``Functional properties of minimum mean-square error and
  mutual information,'' \emph{IEEE Trans. Inf. Theory}, vol.~58, no.~3, pp.
  1289--1301, {Mar.} 2012.

\bibitem{Ung02}
G.~Ungerb{\"o}ck, ``Huffman shaping,'' in \emph{Codes, Graphs, and Systems},
  R.~E. Blahut and R.~Koetter, Eds.\hskip 1em plus 0.5em minus 0.4em\relax
  Kluwer Academic Publishers, {Mar.} 2002, pp. 299--313.

\bibitem{Dur11}
G.~Durisi, V.~I. Morgenshtern, H.~B\"olcskei, U.~G. Schuster, and
  S.~Shamai~(Shitz), ``Information theory of underspread {WSSUS} channels,'' in
  \emph{Wireless Communications over Rapidly Time-Varying Channels},
  F.~Hlawatsch and G.~Matz, Eds.\hskip 1em plus 0.5em minus 0.4em\relax
  Academic Press, 2011, pp. 65--116.

\bibitem{BCT11}
G.~Bresler, D.~Cartwright, and D.~Tse, ``Feasibility of interference alignment
  for the {MIMO} interference channel,'' \emph{IEEE Trans. Inf. Theory},
  vol.~60, no.~9, pp. 5573--5586, {Sep.} 2014.

\bibitem{Ash90}
R.~B. Ash, \emph{Information Theory}, ser. Dover Books on Mathematics.\hskip
  1em plus 0.5em minus 0.4em\relax Mineola, NY: Courier Dover Publications,
  1990.

\bibitem{Rud87}
W.~Rudin, \emph{{Real and complex analysis}}, 3rd~ed.\hskip 1em plus 0.5em
  minus 0.4em\relax McGraw-Hill, 1987.

\bibitem{BC15}
S.~Bobkov and G.~Chistyakov, ``Entropy power inequality for the {R\'enyi}
  entropy,'' \emph{IEEE Trans. Inf. Theory}, vol.~61, no.~2, pp. 708--714,
  {Feb.} 2015.


\bibitem{CT06}
T.~M. Cover and J.~A. Thomas, \emph{Elements of Information Theory},
  2nd~ed.\hskip 1em plus 0.5em minus 0.4em\relax New York, NY:
  Wiley-Interscience, 2006.

\bibitem{BV10}
V.~Beresnevich and S.~Velani, ``An inhomogeneous transference principle and
  {D}iophantine approximation,'' \emph{Proc. London Math. Soc.}, vol. 101,
  no.~3, pp. 821--851, {Mar.} 2010.

\bibitem{BL05}
Y.~Bugeaud and M.~Laurent, ``On exponents of homogeneous and inhomogeneous
  {D}iophantine approximation,'' \emph{Moscow Math. Journal}, vol.~5, no.~4,
  pp. 747--766, {Oct.} 2005.

\end{thebibliography}
\end{document}